\documentclass[11pt]{article}
	\pdfoutput=1
\usepackage[T1]{fontenc}

\usepackage{listings}

\usepackage{datetime}
\usepackage{comment}					% to comment large parts of text

\usepackage[usenames,dvipsnames]{xcolor}
\usepackage[setpagesize=false,pagebackref=false, 
linktocpage, bookmarksopen=true, colorlinks=true, 
linkcolor=Maroon,citecolor=Maroon,urlcolor=Maroon]{hyperref}

\usepackage[parsep]{collref}				% collect references in groups

\usepackage{amsmath, amssymb,amsthm}
\usepackage{stackrel}
\numberwithin{equation}{section}
\usepackage{bm,environ,mathrsfs,array,arydshln}
\usepackage{booktabs,float,slashed}
\usepackage{appendix}
\usepackage[mathcal]{euscript}
\usepackage{tensor} 						% Ratcliffe package to write tensors
\usepackage{mathabx}
\usepackage[vcentermath]{youngtab}
\usepackage{simpler-wick}

\usepackage{graphicx,epsfig,epic}
\usepackage{tikz}
\usepackage{tikz-feynman} 
\allowdisplaybreaks

\usepackage{framed}						% for shaded equations \begin{shaded} + \end{shaded} or \bs + \es
\definecolor{shadecolor}{rgb}{0.9996078, 0.984314, 0.960784}
\definecolor{framecolor}{rgb}{0,0,0}
\definecolor{TFTitleColor}{RGB}{1,1,1}
\newenvironment{frshaded}{%
    \MakeFramed {\FrameRestore}}%
    {\endMakeFramed}

\definecolor{myred}{RGB}{233, 33, 45}

\newcommand{\bs}{\begin{frshaded}}			% framed with background in shadecolor 
\newcommand{\es}{\end{frshaded}\noindent}

\def\ba#1\ea{\begin{align}#1\end{align}}		        %  clever way to bypass the known problem...
\newcommand{\be}{\begin{equation}}
\newcommand{\ee}{\end{equation}}
\newcommand{\bea}{\begin{equation} \begin{aligned}} 
\newcommand{\eea}{\end{aligned} \end{equation}}
\newcommand{\mc}{\mathcal }

\newcommand{\wt}{\widetilde}

\newcommand{\la}{\label}
\newcommand{\eps}{\varepsilon}

\DeclareMathOperator{\tr}{\text{tr}}
\DeclareMathOperator{\vol}{vol}

\newcommand{\N}{\mathcal N}

\newcommand{\sql}{\sqrt\l}
\renewcommand{\l}{\lambda}
\newcommand{\gym}{g_{\scalebox{0.45}{\text{YM}}}}

\newcommand{\LDR}{\ell}

\newcommand{\G}{\Gamma}

\def \np {\newpage}
\def \ed {\np \small
\baselineskip 11pt
\bibliography{2loop}
\small
\bibliographystyle{JHEP-v2.9}
\end{document}
}
\def \iffa  {\iffalse}
\def \te {\textstyle}
\newcommand{\rf}[1]{(\ref{#1})}
\def\ov{\over}
\def \ci {\cite}
\def \foot {\footnote}

\def\la{\label}\def \a {\alpha}
\def\foot{\footnote}
\def \adss {AdS$_5 \times S^5$\ }
\def \adsc {AdS$_4\times { \rm CP}^3$ } 

\def \s {\sigma} \def \del {\partial} 
\def \ha {\tfrac{1}{2}}

\def \bb {{\rm b}}
\def \OO {{\cal O}}
\def \cc {{\rm c}}
\def \no {\nonumber}
  
 \def \g {\gamma}
 
\def \adst {AdS$_2$ }

  \def \rmb {{_{\rm b}}}   \def \rmf {{_{\rm f}}}  \def \rmbf {{_{\rm bf}}}
   \def \b {\beta } 
\def \m {\mu} \def \n {\nu} \def \L {\mc L}

\def \G  {\Gamma}

   \def \rk {\tilde  \rfe}

\def \Ge {\gamma_{\rm E}} \def \rG {{\rm G}}
\def \CP {{\rm CP}}
\def \ZZ {{\mathbb Z}}

\def \rk {{\rm k}}
\def \rE {{\rm E}}

\def \cJ {{\mathcal J}}
\def \hn {\hat \nabla}
\def \vp {\varphi}
\def \Gb {\bar \Gamma}
\def \DF {\mathfrak D}
\def \Dsl {\slashed{\mathfrak D}}
\def \rT {{\rm T}}
\def \hb {{\rm h}}
\def \cX {\mathcal X}
\def \X {{\rm X}}
\def \bth {\bar\theta}
\def \ii {{\rm i}}
\def \cL {\mc L}
\def \vol {{\rm vol}} \def \AdS {{\rm AdS}}
\newcommand{\al}[1]{\begin{align}{#1}\end{align}}
  \def \td {\tilde }
 \def \rT {{\rm T}}
\def \rM {{\rm  M}}
\def \m {\mu}  \def \n {\nu}
\def \rF {{\rm F}}

\def \ru {{\rm u}}
 \def \rM  {{\rm M}}
\def \cZ {{\cal Z}}

\def \rc {{\rm c}}

\def \rB {{\rm B}}
\def \cc {{\rm c}}

\def \rt {{\rm t}}

\def \adst {AdS$_2$}   % needed in the abstract
\begin{document}
\begin{titlepage}
\begin{tabbing}
\hspace*{10.5cm} \=  \kill % set the tabbings
\>  Imperial--TP--2026--AAT--04 \\
\> %none
\end{tabbing}

%\centerline{\small\today\ -- \currenttime}

\vspace*{5mm}
\begin{center}
{\Large\sc  \bf
On 2-loop   AdS$_2$ 
superstring  partition function \vskip 3pt
  in  AdS$_5 \times S^5$ and   AdS$_4\times {\rm CP}^3$ }
\vspace*{6mm}

Arkady A. Tseytlin\footnote{Also at  ITMP of MSU  and Lebedev Inst. \ \  E-mail: \texttt{tseytlin@ic.ac.uk}}

\vspace*{4mm}
{\small
Abdus Salam Centre for Theoretical Physics,\\ Imperial College London,  SW7 2AZ, U.K.
%\vskip 0.2cm {\small E-mail: \texttt{tseytlin@ic.ac.uk}
}
\vspace*{0.7cm}
\end{center}
\begin{abstract}
We revisit the 2-loop computation of the partition function of the GS superstring expanded near the AdS$_2$ surface in AdS$_5\times S^5$ and AdS$_4\times {\rm CP}^3$, dual to  $\frac{1}{2}$-BPS circular Wilson loop in ${\cal N}=4$ SYM and ABJM theories. In \ci{Beccaria:2026ffm} it was done in AdS$_5\times S^5$ using dimensional regularization and a UV pole was found. Since dimensional regularization is potentially problematic for the GS string, here we use a heat-kernel cutoff instead. Then the contribution of the local path integral measure, proportional to $\delta(0)$, can no longer be ignored. We show that it cancels the $\delta(0)$ terms coming from the quartic interactions in the action, leaving the square of the difference ${\rm Str}\,E$ between the regularized numbers of bosonic and fermionic modes. The 2-loop coefficient is thus UV finite, the pole of \ci{Beccaria:2026ffm} being an artefact of dimensional regularization. If all fields are regularized with the same smearing kernel, ${\rm Str}\,E=0$ and the 2-loop coefficient $f_2$ vanishes, in apparent tension with $f_2=\frac{3}{32\pi^2}$ from the localization expression in SYM theory. In the AdS$_4\times{\rm CP}^3$ case the same cancellation leaves a finite term proportional to the square of the coincident value of the massless 2d fermion Green's function, giving, ironically, $f_2=\frac{3}{32\pi^2}$ instead of  $f_2=0$  required by localization. This term is the level-zero (string) part of the result 
of  \ci{Beccaria:2026zjj}
 for the AdS$_2\times S^1$ M2 brane in AdS$_4\times S^7/\mathbb Z_k$, where it is cancelled by the higher $S^1$ modes, which also remove a finite mass shift of the massless fermion that would be in conflict with the AdS$_2$ supersymmetry. This suggests that the type II string path integral measure should contain an extra contribution, supplied in the IIA case by the parent M2 brane; in the AdS$_5\times S^5$ case it is still to be identified and may resolve the remaining 2-loop mismatch with localization.
\end{abstract}

\iffa
We revisit the 2-loop computation of the partition function of the GS superstring expanded near the AdS$_2$ surface in AdS$_5\times S^5$ and AdS$_4\times {\rm CP}^3$, dual to  $\frac{1}{2}$-BPS circular Wilson loop in ${\cal N}=4$ SYM and ABJM theories. In arXiv:2601.08809 it was done in AdS$_5\times S^5$ using dimensional regularization and a UV pole was found. Since dimensional regularization is potentially problematic for the GS string, here we use a heat-kernel cutoff instead. Then the contribution of the local path integral measure, proportional to $\delta(0)$, can no longer be ignored. We show that it cancels the $\delta(0)$ terms coming from the quartic interactions in the action, leaving the square of the difference ${\rm Str}\,E$ between the regularized numbers of bosonic and fermionic modes. The 2-loop coefficient is thus UV finite, the pole of arXiv:2601.08809 being an artefact of dimensional regularization. If all fields are regularized with the same smearing kernel, ${\rm Str}\,E=0$ and the 2-loop coefficient $f_2$ vanishes, in apparent tension with $f_2=\frac{3}{32\pi^2}$ from the localization expression in SYM theory. In the AdS$_4\times{\rm CP}^3$ case the same cancellation leaves a finite term proportional to the square of the coincident value of the massless 2d fermion Green's function, giving, ironically, $f_2=\frac{3}{32\pi^2}$ instead of  $f_2=0$  required by localization. This term is the level-zero (string) part of the result 
of  arXiv:2609.14497 for the AdS$_2\times S^1$ M2 brane in AdS$_4\times S^7/\mathbb Z_k$, where it is cancelled by the higher $S^1$ modes, which also remove a finite mass shift of the massless fermion that would be in conflict with the AdS$_2$ supersymmetry. This suggests that the type II string path integral measure should contain an extra contribution, supplied in the IIA case by the parent M2 brane; in the AdS$_5\times S^5$ case it is still to be identified and may resolve the remaining 2-loop mismatch with localization.
\fi

%\vskip 3.5cm
\end{titlepage}

%%%%%%%%%%%%%%%%%%%%%%%%%%%%%%%%%%%%%%%%%%%%%%%%%%%%%%%%%

{\small
\makeatletter
\newcommand*{\toccontents}{\@starttoc{toc}}
\makeatother
\toccontents
}

\

\iffa

\setcounter{footnote}{0}
\begin{flushright}
\small
{\it ``Imagination is more important than knowledge.\ \ \ \ \ \ \ \ \ \ \ \ \ \ 

Knowledge is limited. Imagination encircles the world.'' \ \ \ 

\ \ A.~Einstein\foot{Interview with G.\,S.~Viereck,\  \emph{The Saturday Evening Post}, 26 October 1929.}\  
    (with unintended  doubt in the power of %relevance
      AI) \ \ \ }
% interview with G.\,S.~Viereck,\\
% \emph{The Saturday Evening Post}, 26 October 1929
\end{flushright}

\fi

\section{Introduction and summary}

The aim of this paper is to revisit the computation of the 2-loop  correction to the partition function of the superstring expanded near the \adst\ minimal surface in \adss and \adsc   which is  dual to the 
expectation value of the $\ha$ BPS circular Wilson loop in the planar limit of $\N=4$ SYM and ABJM theories respectively, clarifying and elaborating on the discussion in the two recent papers \ci{Beccaria:2026ffm} and \ci{Beccaria:2026zjj}.

\subsection*{\adss}

{\bf 1.\ } The AdS/CFT correspondence relates the expectation value of the $\ha$ BPS circular Wilson loop in $\N=4$ SYM theory, known exactly from localization \ci{Erickson:2000af,Drukker:2000rr,Pestun:2007rz}, to the partition function of the \adss superstring on the disk with the \adst\ minimal surface ending on the circle at the boundary \ci{Berenstein:1998ij,Drukker:2000ep}. In the planar limit $\langle W\rangle = {2N\ov \sql} I_1 ( \sql)$, which, expanded at large 't Hooft coupling $\l= N \gym^2$ and expressed in terms of the string tension 
$T={L^2\ov 2\pi \a'} = {\sql \ov 2\pi} $ and the string coupling $   g_s ={\gym^2\ov 4 \pi} $
gives
\be \la{1}
\langle W\rangle =\te   \sqrt\frac{2}{\pi}\frac{N}{\l^{3/4}}e^{\sql}
 \te   \big[1-\frac{3}{8\,\sql}+...   \big]
   = {1\ov  \sqrt{2\pi} g_s }  \sqrt {T \ov 2\pi} e^{2\pi T}\te  \big[ 1-  {3\ov 16 \pi} T^{-1}   +... \big] \ .
    %\qquad \ \ \ c_1 = {1\ov \sqrt{ 2 \pi}} \ .  
    \ee
The exponent is the classical string action on \adst\ with the regularized $\AdS_2$ volume and the prefactor is the 1-loop correction \ci{Drukker:2000ep,Giombi:2020mhz}. Then writing  as in \ci{Beccaria:2026ffm} 
\ba \la{2}
  &   \langle W\rangle = Z_1 \,  e^{-F} \ , \ \ \ \ \    F=  - f(T) V \ ,   \qquad V\equiv \vol(\AdS_2)=-2\pi \ , \qquad \te Z_1= {1\ov  g_s }  \sqrt {T \ov 2\pi}\ , \\
  &\la{3}\te   f(T)  =  f_0 T + f_1 +  f_2  T^{-1} + ... \ , \qquad f_0=-1 \ , \qquad f_1= {1\ov 4 \pi}   \log ({2\pi}) \ , \qquad 
  \ \ \  f_2= \frac{3}{32\pi^{2}}\ ,
\ea
the subleading term in \rf{1} should correspond to the 2-loop coefficient $f_2$ in the free energy of the \adss string expanded near the \adst\ minimal surface.

In \ci{Beccaria:2026ffm} the 2-loop correction to the \adss string free energy was computed directly from the GS action \ci{Metsaev:1998it} expanded to quartic order near the \adst\ surface in the static gauge and the ``adapted'' $\kappa$-symmetry gauge. The important simplification is that in this gauge there are no cubic interaction  vertices, so that the 2-loop contribution is given just by the ``double-bubble'' diagrams, i.e. by the expectation value $\langle \L_4\rangle$ of the quartic term in the Lagrangian. All the contractions reduce to the coincident values of the \adst\ propagators and of their derivatives, which are constants, so that
$ \langle S_4\rangle = V \langle \L_4\rangle  .$

The computation of $\langle \L_4\rangle$ in \ci{Beccaria:2026ffm} was done in dimensional 
 regularization, i.e.  %doing the tensor and spinor algebra in 2d but
  continuing the propagators to AdS$_{1+d}$ with $d=1-2 \eps$. It was found that the result has the form
\be \la{4}
\te  f_2=  \rc_\rmb  \,\rG^{2} +\,  \rc_{\rmbf} \,  \rG\, {\rm S} + \rc_\rmf \,{\rm S}^{2}\ , \qquad
 \rc_\rmb = \frac{9(1-d)}{2(1+d)}\ ,  \qquad    \rc_{\rmbf} = \frac{24(1-d)}{(1+d)^{2}} \ , \qquad
\rc_\rmf = \frac{2(1-d)(5-3d)}{(1+d)^{2}}\ .
\ee
Here $\rG$ and ${\rm S}$ are the coincident-point values of the massive bosonic and fermionic AdS$_{1+d}$ propagators, each having a $1\ov \eps$ pole, while all the coefficients are proportional to $1-d = 2 \eps$. As a result, taking $\eps \to 0$ the double poles cancel
(separately in each sector) 
 but a single pole survives, so that
\be \la{102}\te
 f_2 = - {11\ov 32\pi^2 \eps} +   f_{2, \rm fin} \ .
\ee
$f_{2, \rm fin}$ depends on the finite parts of the coincident propagators, i.e. on the choice of a scheme. This divergent result is puzzling, precluding a comparison with the gauge theory coefficient  in \rf{3}.

As was noted in \ci{Beccaria:2026ffm}, the use of dimensional regularization in the GS string case is potentially problematic: the GS action involves the $\epsilon^{\m\n}$ symbol in the fermionic WZ term and in the $\kappa$-symmetry projectors, and the global \adst\ supersymmetry of the gauge-fixed action \ci{Drukker:2000ep} is not manifestly preserved by it. 

It was also observed in \ci{Beccaria:2026ffm} that if one formally sets $1+d =2$ in the coefficients in \rf{4} before expanding the Green's functions near $d=1$ (``hard'' dimensional reduction regularization, in which all index contractions are first done in 2 dimensions and only then the result is regularized) the 2-loop correction just vanishes.\foot{One  could interpret  the  UV  pole   present, in particular, 
   in the bosonic  $\rG^{2}$ sector that originates just  from 
 Nambu part of the   GS string action     as  a reflection of formal  2-loop non-renormalizability of the  Nambu  action 
 in dimensional  regularization that was seen already  in flat target space case in 
  \ci{Beccaria:2025xry}.  Thus the issue  of UV pole is not specific to GS string. 
  Continuation away from 2 dimensions  appears to  break the integrability of the Nambu action,
requiring  addition of  counterterms to restore it. 
Supersymmetry of the GS action appears to translate
 this property to the other two sectors, but does 
   not imply the cancellation of the pole (that the 2-loop  UV pole  survives in the GS  string case in flat space case 
   was seen  also in  \ci{Beccaria:2025xry}).
   As we shall see below, in 2d  cutoff regularization  with the measure contributions
    included  all   log  UV  divergent terms   cancel    sector by sector (cf. \rf{37},\rf{999}), 
 just like in the ``hard'' dimensional reduction  prescription applied to \rf{4}. 
}

%v2
Dimensional regularization  automatically sets $\delta(0)$ 
 terms to zero, 
but   if one  does the calculation  directly in 2d with an  explicit  UV cutoff $\Lambda$   such terms   will  need to be accounted  for.
As was  discussed  in Appendix B of \ci{Beccaria:2026ffm}, in this case all propagator-squared terms (that  produce $\log^2 \Lambda$ 
terms)  automatically  cancel in the 2-loop  correction, separately   in the bosonic, mixed and fermionic sectors, 
consistently  with the structure of \rf{4}. However,  the $\delta(0)$
  or ``equation-of-motion''  (EOM) 
  type terms (originating  from kinetic operators acting on the corresponding Green's function) 
  survive and may still  produce (in the case of curved \adst\ world sheet)  UV divergent $\log \Lambda$ terms. 

{\bf 2.\ } 
The aim of the present paper is to carefully 
redo this 2-loop computation using  a power-like (proper-time or heat-kernel) cutoff directly in 2 dimensions.
One may also use the $\zeta$-function regularization, which removes the power divergences but keeps the potential logarithmic ones.

We shall find that the resulting 2-loop coefficient $f_2$ is UV finite, so that the pole in \rf{102} is an artefact of the dimensional reduction regularization. It is produced by the evanescent coefficients $\sim  1-d$ in \rf{4} multiplying the poles of the product of 
coincident propagators, i.e. by the terms which in a 2d cutoff  regularization are absent.  
% are cancelled by the path integral measure. %%% CHANGED
 
Moreover, in the natural scheme in which all the fields are regularized in the same way the coefficient $f_2$ vanishes, just like in the ``hard'' dimensional reduction prescription applied to \rf{4}. 
A possible interpretation of this result in relation to \rf{3} will be discussed in section \ref{s5}.

The use of a dimensionful  2d cutoff brings in a new element: one needs to account for the path integral measure contribution. In a generic 2d regularization the $\delta(0)$
 terms
 %($\delta(0) \equiv \delta^{(2)}(0)$ is the coincident value of the regularized 2d delta function) 
 or, equivalently, the ``equation-of-motion'' terms (where the kinetic operators act on the corresponding Green's functions) do not vanish. Indeed, $\delta(0) \equiv \delta^{(2)}(0)$ contains a quadratically divergent part and, in a curved background like \adst, also a finite part (see, e.g., \ci{Randjbar-Daemi:1987rfs}).

The expectation value $\langle \L_4\rangle$ in the $\zeta$-function regularization in 2d was  discussed in Appendix B of \ci{Beccaria:2026ffm} and was shown to be expressed in terms of such EOM  terms. However, the contribution of the path integral measure, which is also proportional to $\delta(0)$, was not computed there.\foot{Its possible presence was acknowledged in footnote 30 in \ci{Beccaria:2026ffm}.} Our aim will be to complete 
 the computation in Appendix B of \ci{Beccaria:2026ffm} by including the 2-loop contribution of the measure.
%, which will turn out to cancel that of the action, i.e. of $\langle \L_4\rangle$.

The reason why the measure contribution is nontrivial is that the string action has the structure of a Nambu action in curved space, so that the interaction terms contain powers of 
 derivatives,  $\cL_4 \sim x^2 (\del x)^2 + (\del x)^4 + ...$. While by the definition of the Feynman path integral the measure is trivial in the phase-space formulation, once the canonical momenta are integrated out to get the coordinate-space form of the action one gets an extra local factor that depends on the fields and their first derivatives; for the first-order GS fermions a similar factor arises from the second-class constraints. %%% CHANGED (fermionic part added)

Let us denote by $E_\vp$ the regularized value of $\delta(0)$ for one of the 8+8 string fluctuation fields $\vp=\{x,y; \theta\}$, i.e. the coincident-point value of the kinetic operator applied to the corresponding Green's function $G_\vp(\xi, \xi')$. We will refer to 
 $E_\vp$    as the EOM term for $\vp$. 
 Let $\rG_\vp$ be the coincident-point  value $G_\vp(\xi,\xi)$ of the Green's function itself. 
 The regularized $\delta(0)$ appearing in the measure will be denoted by $\rE$.

As was observed in Appendix B of \ci{Beccaria:2026ffm}, all terms in $\langle \L_4\rangle$ have the structure $E_\vp\, \rG_{\vp'}$ or $E_\vp\, E_{\vp'}$ (where $\vp$ and $\vp'$ stand for some of the 
 fluctuation fields). Thus $\langle \L_4\rangle$ is proportional to the EOM terms and formally vanishes if they are set to zero, in line with the vanishing in the hard dimensional reduction prescription mentioned above.
  It was anticipated there that the sum of the $E_\vp\, E_{\vp'}$ terms may vanish due to 2d 
  supersymmetry, while the $E_\vp\, \rG_{\vp'}$ terms may be cancelled by the contribution of the measure.

{\bf 3.\ }  This is essentially what we shall find below, with one refinement. 
The phase-space path integral measure of the gauge-fixed string theory turns out to contain two $T^{-1}$ (2-loop) contributions: 
(i) $ E \langle \cJ \rangle$ type  terms,  where $E$ stands for  particular $E_\vp$ 
and $\cJ$ is a local combination quadratic in the fields and their derivatives, and 
(ii)  field-independent $E^2$ terms. 
Combined together they cancel {\it all} the terms in $\langle \L_4\rangle$ containing the coincident propagators, i.e. the $E_\vp\, \rG_{\vp'}$ ones, for any choice of the regularization of the individual fields, provided each term in the measure is multiplied by the $\delta(0)$ of the field whose integration produced it (see Appendix \ref{apA3}). 

What remains then is the finite square of the  difference between the regularized 
numbers of bosonic and fermionic modes per unit \adst\ volume,
\be \la{1061}
 f_2 = \tfrac18 \big( {\rm Str}\, E \big)^2 \ , \qquad \qquad {\rm Str}\, E \equiv  3 E_x + 5 E_y - 8 E_\theta \ .
\ee
In ${\rm Str}\, E$ the divergences cancel between the 8 bosonic and 8 fermionic degrees of freedom,
so that, provided the $\delta(0)$ in each term of the measure is the one of the corresponding
field, 
the 2-loop coefficient is UV finite for any choice of the regularization of the individual
fields within the class of isometry-preserving regulators with a common cutoff scale (i.e. with the
same leading short-distance behaviour $E_\vp = {1\ov 4\pi t} + \dots$ for all fields, cf. \rf{233}),
 and its value is determined by the single regularization-dependent finite constant
${\rm Str}\, E$. If instead the same $\delta(0)=\rE$ is used in all the measure terms while the fields
in the action are regularized differently, uncancelled logarithmic divergences remain and they 
 cancel  under a linear condition on $E_\vp,\rE$  satisfied, e.g.,  for the common smearing,  see \rf{313}.
% (continue with the rest of the sentence: "and its value is determined by ...")
% and in section 2.2, prescription (iii), ADD one sentence:
We assume that the regularizing kernels of all the fields are built from their
\adst\ Laplacians with a common cutoff parameter $t$, so that the leading term ${1\ov 4\pi t}$ of
$E_\vp$ is the same for all fields; for kernels with different cutoff scales the power divergences in
${\rm Str}\, E$ in \rf{999} would not cancel.

${\rm Str}\, E$ vanishes, in particular, if all the fields are regularized in the same way (i.e. with the same ``smearing'' operator),
\be \la{103}
E_x=E_y=E_\theta = \rE \ ,
\ee
so that the $\delta(0)$'s in the action automatically coincide with the $\delta(0)$'s in the measure. 
Below  we shall often assume such  common smearing, in which case the measure may be written 
in terms of  a single $\rE$ constant. 

The more general case will be considered in Appendix \ref{apA3}. 
The condition ${\rm Str}\, E=0$, satisfied by the common smearing \rf{103}, is required by the
2d Lorentz covariance of the measure: with the field-attributed $\delta(0)$'s the measure density
contains a non-covariant term proportional to ${\rm Str}\, E$ which, while not contributing to the
free energy, would break the Lorentz invariance of the 1-loop effective action
(see Appendix \ref{apA3}). In this case the total 2-loop correction vanishes,
%The assumption of a common smearing \rf{103}  might be  a consequence of the underlying 2d supersymmetry of the gauge-fixed string action. In this case the total 2-loop correction vanishes,
\be \la{5}
f_2 = 0 \ ,
\ee
as in the ``hard'' dimensional reduction interpretation of \rf{4}.

{\bf 4.\ }  The vanishing of $f_2$  appears to be 
 inconsistent with the gauge-theory result in \rf{3}, and in section 5 we discuss two possible interpretations 
 of this result. 
 One is  to draw analogy   with the  ABJM case, where  the string computation  may
  be compared not with the  Wilson loop at fixed $N$ but with its ``grand-canonical'' counterpart as 
  suggested by the ensemble conjecture \ci{Gautason:2025plx,vanMuiden:2026nsp,Bobev:2026gir,vanMuiden:2026lno},
   which in the M2 brane case explains the vanishing of the 2-loop correction found in  \ci{Beccaria:2026zjj}.\foot{We thank Jesse van Muiden for suggesting this possibility.}  
   
   The other is that the type II GS string path integral measure  may  contain an
    extra finite contribution not accounted so far  in  our  computation. 
    An indication that the standard definition of the  GS string measure  may be 
     incomplete  comes  already at the  1-loop order: 
     the string partition function on \adst\ has a universal logarithmic UV divergence proportional to the Euler number of the disk
     \ci{Drukker:2000ep}, $B_2 = {1\ov 4\pi}\int d^2\xi\sqrt g\, R^{(2)}=1$, which must be removed by hand to reproduce the prefactor in \rf{1} \ci{Giombi:2020mhz}. 
     
     At the same time,  in the \adsc case
      it is removed automatically if the string  partition function 
       is defined as the limit of the M2 brane one:  the   higher  $S^1$   M2 brane  modes contribute 
       $2 \zeta(0) = -1$ to the coefficient of the UV divergent factor 
       $\log \Lambda$ \ci{Giombi:2023vzu}. 
       An  analogous 2-loop 
        result  is established below in the \adsc case, where the parent M2 brane  path integral  is available.
        For the  \adss  GS theory  we  do not know  such ``parent'',   so  the question remains open.

Let us note that the fact that the 2-loop expectation value $\langle \L_4\rangle$ is proportional to the EOM terms appears to be related to the special geodesic embedding of the AdS$_p$ brane world volume into the AdS$_D$ target space factor. Indeed, the same was observed in \ci{Beccaria:2025ahf} in the case of the M2 brane wrapping AdS$_3$ in AdS$_7$ of AdS$_7 \times S^4$ and in \ci{Tseytlin:2026dhy} in the case of the M2 brane wrapping $S^3$ in $S^7$ of AdS$_4 \times S^7$. Let us also mention that another example in which the contribution of a measure (in that case the collective-coordinate one, required due to the presence of zero modes) cancels the 2-loop contribution coming from the action is the M2 brane partition function expanded near $S^3/\ZZ_k$ in $\AdS_4 \times S^7/\ZZ_k$ \ci{Tseytlin:2026dhy}.

\subsection*{\adsc}

{\bf 5.\ }  The $\ha$ BPS circular Wilson loop in the ABJM theory is dual, in the type IIA limit of large $N$ and $k$ with $\l={N\ov k}$ fixed, to the type IIA fundamental string on \adst\ in the \adsc background. The localization expression of \ci{Drukker:2010nc} in the planar limit may be written as  \ci{Beccaria:2026zjj}
\al{
 \langle W \rangle =  & \frac{N}{4\pi\l}e^{\pi\sqrt{2(\l- {1\ov 24})}}  +  \mc O(N^{0})
= \frac{N}{4\pi\l}e^{\pi\sqrt{2\l}}   \te  \big[1-\frac{\pi}{24\sqrt 2}\frac{1}{\sql}+\dots    \big]+...\no \\= &Z_{1}
\ e^{2\pi \rT_1  }  \big[ 1 + 0 \cdot  \rT_1^{-1}   + \OO(\rT_1 ^{-2})
 \big]   + ... \ , \la{118} \\
&\rT_1=  \tfrac{1}{\sqrt 2}\sqrt{\l- \te {1\ov24}} \ , \qquad \qquad Z_{1}=\frac{\sqrt{ \rT_1} }{\sqrt{2 \pi} \ g_{\rm s} } \ , \qquad \qquad  T_1 = \sqrt{\tfrac{\l}{2}} \ , \la{41}
}
where $\rT_1$ is the string tension expressed in terms of the shifted coupling (cf. \ci{Bergman:2009zh}) and $T_1$ is its ``unshifted'' value. %   ($g_s$ is assumed to be $\sqrt \pi (2 \lambda)^{5/4}/N$).
 The direct 1-loop GS string computation \ci{Kim:2012tu,Giombi:2020mhz} requires an extra $\sqrt{ T_1}$ type measure contribution to reproduce $Z_{1}$ in \rf{41}. This factor is generated automatically if the string  partition function  is defined as the $k\to \infty$ limit of the M2 brane 1-loop expression \ci{Giombi:2023vzu}.

Thus, in the same normalization as in \rf{2},\rf{3} (with $T\to \rT_1$ and $F = - f V = 2\pi f$) the ABJM 
gauge theory prediction is that there should be no 2-loop \adsc string correction if the string result is expressed in terms of $\rT_1$ in \rf{41}, i.e.
\be \la{42}
f_2 = 0 \ .
\ee
The 2-loop correction for the M2 brane in AdS$_4 \times S^7/\ZZ_k$ wrapping $\AdS_2$ and the Hopf fiber of $S^7/\ZZ_k$ was shown in \ci{Beccaria:2026zjj} to vanish, in agreement with the localization result in the grand-canonical ensemble interpretation. Expanding the 3d fields in the Fourier modes on $S^1$ one finds $8+8$ towers of \adst\ fields, the $n=0$ level being the IIA string fluctuations \ci{Sakaguchi:2010dg,Kim:2012tu,Giombi:2020mhz}. The IIA string action is the double dimensional reduction \ci{Duff:1987bx} of the M2 brane action, and the static and $\kappa$-symmetry gauge-fixed quartic action of the IIA string expanded near \adst\ is obtained from the one of \ci{Beccaria:2026zjj} by restricting to the $S^1$-independent ($n=0$) modes. The relation between the tensions and the 2-loop string and M2 brane coefficients is \ci{Beccaria:2026zjj}
\be \la{43}
\rT_1 =\te  { {\pi\ov 2k}}  \rT_2 \ , \qquad \qquad   f_{2} = \frac{1}{4k} f_{2\rM}(k) \Big|_{k\to \infty}  \ .
\ee
The M2 brane free energy in the normalization of \ci{Beccaria:2026zjj} has the expansion $F_{\rM} = \rT_2 f_{0\rM}(k) + f_{1\rM}(k) + \rT_2^{-1} f_{2\rM}(k) + ...$, and the string $f_2$ coefficient is defined as in \rf{2},\rf{3}.\foot{In \ci{Beccaria:2026zjj} the string coefficient was defined as $f_{2s} = [{\pi \ov 2k} f_{2\rM}]_{k\to \infty}$, which is the coefficient of $\rT_1^{-1}$ in $F$ itself, so that $f_{2s} = 2\pi f_2$.}

{\bf 6.\ }  The computation of the 2-loop correction to the type IIA string partition function on \adst\ in \adsc follows the same pattern as in the \adss case up to one special contribution. The 2d fluctuation spectrum here \ci{Sakaguchi:2010dg,Kim:2012tu,Giombi:2020mhz} contains, in addition to the  bosons  $x$ and $w^a$ 
 and the massive fermionic fields $\psi$ also  a massless 2d fermion $\chi$ (paired by the preserved supersymmetry with one of the massless ${\rm CP}^3$ scalars $w^a$). The coincident-point value $\rG_\chi$ of its Green's function is finite and is determined by the choice of the boundary conditions at the \adst\ boundary \ci{David:2023btq,Beccaria:2026zjj},
\be \la{104}
\rG_\chi =\te  \pm {1\ov 4\pi} \ .
\ee
As in the \adss case the 2-loop contribution of the action combined with the 2-loop contribution of the measure results in the cancellation of all of 
 log divergent terms    and we  end up with  % EOM $E$-terms, i.e. in the general case
\be 
f_2 = \tfrac18 ( {\rm Str}\, E )^2 + \tfrac32  \rG_\chi^2\ , \qquad \qquad
{\rm Str}\, E = 2 E_x + 6 E_w - 6 E_\psi - 2 E_\chi\ .
 \ee
 This is similar to the \adss  expression in  \rf{1061}  
but  contains an extra  finite  non-EOM term $\rG_\chi^2$  coming from  $\langle \cL_4\rangle$.
Then   for the  choice of the  scheme-dependent constant ${\rm Str}\, E =0$ 
 which is realised, in particular, for the  common smearing as in \rf{103} 
\be \la{6}
f_2 = \tfrac32  \rG_\chi^2 = \tfrac{3}{32\pi^2}\ .
\ee
We will explain below how this 
residual contribution is consistent with the formal truncation to the massless (zero KK level) 
sector of the M2 brane result of \ci{Beccaria:2026zjj}: there the same term appears as the $n=0$ contribution to a ``perfect square''  result
  $\langle \cL_4\rangle =-\frac38 \cX^2$,  
 and is cancelled by the regularized contribution of   the sum over all  non-zero KK levels. 
 The direct 2d string computation has no analog of the latter, and thus gives \rf{6} rather than the vanishing result \rf{42} required by the localization (which is found if the string expression is defined  from 
  the M2 brane one according to \rf{43}).

The contribution of the non-zero KK levels can be identified, from the 2d point of view, 
with a finite local term of the measure type, a finite coefficient times a
 mass-type bilinear of the massless fermion plus a constant (cf. \rf{430}). 
 That  accounting  for this term is  crucial 
  is  demonstrated   by a 2-point function  test  in Appendix \ref{s44}: in the  direct 
   2d  type IIA 
   string computation the massless fermion $\chi$ 
   acquires a finite 1-loop mass while its scalar superpartner does not, in conflict with the residual \adst\ 
   supersymmetry.
    It is precisely the term induced by the KK tower that removes this mass shift 
    (with the choice of plus sign in  \rf{104} for which the M2 brane  contribution in \ci{Beccaria:2026zjj}    vanishes). 
    
    We thus have  evidence that the M2 brane path integral, where the issue of the local measure is less ambiguous than in the string case (cf.  Appendix \ref{apB}), supplies an extra 
     contribution to the type IIA GS string path integral measure, which restores the agreement with the localization prediction \rf{42}. 
       One may speculate that an analogous contribution is present in the type IIB string in \adss  where no ``parent'' 
     path integral is  known  at present. 
     %Curiously, the correction required there to match localization result, 
   %  $\Delta f_2 = \tfrac{3}{32\pi^2}$, is equal and opposite to the one required in the 
  %   IIA case, $\Delta f_2 = - \tfrac{3}{32\pi^2}$.

{\bf 7.\ }  The rest of the paper is organized as follows. In section 2 we first  give a general discussion  of  the contribution
 of the path integral measure in a power-cutoff regularization  (starting with
  a bosonic sigma model and then considering the  GS string in a curved background in the static and $\kappa$-symmetry gauges). We then find  the 2-loop measure  contribution 
  for the \adss string expanded near \adst. 
  
  In section 3 we combine this with the contribution of the quartic action (correcting some issues 
  in the  computation in  Appendix B of \ci{Beccaria:2026ffm}) 
  and  derive the expression in \rf{1061}. %show that the total expression vanishes  under the assumption of  the common smearing. 
    In section 4 we repeat  similar  analysis for the \adsc string,
   using the zero-level truncation of the expansion of the M2 brane action in \ci{Beccaria:2026zjj}, and then compare with the 2-loop M2 brane result. 
   Section 5 presents a summary and some concluding remarks.

%%% CHANGED: organization of the appendices (A-F)
Appendix \ref{apA} contains some details of the construction of the measure. In Appendix \ref{apA3} we define the measure with field-specific regularized $\delta(0)$'s and show that the 2-loop coefficient is then given by the square of the regularized supertrace of the identity operator \rf{1061}. 

Appendix \ref{apA2} contains comments on the relation of the canonical measure to the invariant measure on the supercoset. In Appendix \ref{apB} we discuss the measure in the M2 brane case and explain why it was not important in the 3d computations of \ci{Beccaria:2025ahf,Beccaria:2026zjj}. Appendix \ref{apa2} contains the derivation of the coincident limits of the fermionic propagator. 

In Appendix \ref{s44} we discuss the 1-loop corrections to the 2-point functions. In the \adsc case 
 the finite mass shift of the massless fermion 
  is cancelled by the term supplied by the KK modes of the M2 brane.
  In the \adss  case  we find 
  the absence of a mass shift (as opposed to a wave-function renormalization) for the massive scalar 
   in contrast to the finite mass counterterm found in dimensional regularization in Appendix C of \ci{Beccaria:2026ffm}.
We also compare with the recent 1-loop computation of the tilt-operator 2-point function (the
    Bremsstrahlung coefficient) in \ci{Castellani:2026wwh}, and point out that the evanescent terms
   needed there to match the localization result play the role of the missing measure factor $\sqrt T$.

%%%%%%%%%%%%%%%%%%%%%%%%%%%%%%%%%%%%%%%%%%%%%%%%%%%%
\section{Path integral measure in  power-cutoff regularization\la{s2}}

The free energy of a 2d theory with quartic interactions at the 2-loop order is given by the expectation value of the quartic action in the free theory, plus the contribution of the path integral measure. In dimensional regularization the latter is absent: the measure contributes terms proportional to $\delta(0)$, which is  set to zero. In a  generic 2d regularization $\delta(0)$ is a non-zero quantity, with a power-divergent $\sim \Lambda^2$ and, in a curved 2d background, also a finite part, and the measure 
 may contribute  at the same order as $\langle \L_4 \rangle$.

In this section we review how the measure is determined   and what it contributes, first for a  sigma model and  its generalization, then for the GS string in flat space, and finally for the GS string in a curved background in the static and $\kappa$-symmetry gauges. We then determine 
 the relevant part of the  measure for the \adss string expanded near the \adst\  minimal surface and compute its 2-loop contribution.

The starting point is the phase-space path integral. For a quantum-mechanical 
 system with coordinates $z^I$, momenta $p_I$ and Hamiltonian $H(z, p)$ the Euclidean partition function is
\be \la{20}
Z = \int \prod_\xi dz^I\, dp_I \ \exp\big\{ \int d \xi \, \big[ i\, p_I \del_0 z^I - H(z, p)\big] \big\} \ ,
\ee
with the standard  Liouville measure $\prod_\xi dz^I\, dp_I$ which is invariant under canonical transformations. If there are no constraints this is the complete definition: the measure in phase space is trivial. The familiar Lagrangian path integral, with its non-trivial measure, is obtained by integrating out the momenta.
This   construction  has a  straightforward generalization to a  2d   field theory.

Consider a 2d sigma model with the Euclidean action
\be \la{21}
S= \ha  T \int d^2 \xi\, \sqrt g \, g^{\mu \nu } \, G_{IJ}(X)\, \del_\m  X^I \del_\n  X^J \ .
\ee
Expanding near some   background point (with $G_{IJ} =\delta_{IJ} + ...$)
 and rescaling the fluctuation   fields  $z^I$ by $T^{-1/2}$   we  may set 
%with the fields rescaled, $X^I = T^{-1/2} z^I$, so that
\be \la{201}
T\,   G_{IJ}(X) \del^\m X^I \del_\m X^J  \ \to \      \del^\m z^I \del_\m z^I +    T^{-1} h_{IJKL}\,  z^K z^L  \del^\m z^I \del_\m z^J    + ... \ .
\ee
The momenta enter the Hamiltonian quadratically, $H= \ha G^{IJ}(X) p_I p_J + ... $, and the Gaussian integral over them gives the Lagrangian path integral with the familiar covariant 
measure \ci{Lee:1962vm, Salam:1970fso,Gervais:1976ws}.%Salomonson:1976yn.
 Expanding  in  loop  counting parameter $T^{-1}$  we get\foot{We absorb the curvature scale into $T$   to make it dimensionless.}
\ba \la{22}
\prod_\xi \big[ \det G_{IJ}(X(\xi))\big]^{1/2} &= \exp\big[ \ha \delta(0) \int d^2 \xi \sqrt g\, \log\det G_{IJ}(X)\big] \no \\
& = 1 +  \ha  T^{-1} \delta(0) \int d^2 \xi \sqrt g\, h_{IIKL}\, z^K z^L  + \OO(T^{-2}) \ .
\ea
Here $\delta(0)\equiv \delta^{(2)}(\xi,\xi)$.
% arises as the number of integration variables per unit area, i.e. it is the regularized coincident value of the delta function. 
Thus at order $T^{-1}$  the measure  contains  a term proportional to 
$\delta(0)$  and an integral of  a local combination quadratic in the fields. 
Its expectation value  is  $\sim  T^{-1} \delta(0) \rG_z$, where $\rG_z\equiv G_z(\xi,\xi) 
$ is the coincident value of the $z$-propagator.
It thus  contributes at the same order  as the 2-loop expectation value of the quartic interaction term  in \rf{201}, i.e.
$T^{-1} \langle h_{IJKL}z^Kz^L
 \del z^I \del z^J\rangle \sim  \rG_z\,  \delta(0) $, since the contraction of the two derivatives at coincident points produces
 % (up to a mass term) 
  again $\delta(0)$.

This is the standard mechanism by which the measure cancels the power divergences in the perturbation theory of sigma models in an explicit cutoff regularization \ci{Charap:1970xj,Honerkamp:1971sh,Honerkamp:1971xtx} (see also \ci{Friedan:1980jm,Andreev:1990iv}). In dimensional regularization where  $\delta(0)=0$, the measure contribution  is trivial and all  power divergences in the loop 
diagrams are also absent.

In a flat 2d space with a cutoff $\Lambda\to \infty $ one has $\delta(0)= {1 \ov 4 \pi}\Lambda^2$ and the measure terms are purely power divergent. In a curved 2d space there are also additional finite terms. Using the proper-time (heat-kernel) cutoff, $\delta(0)$ is represented by the coincident value of the heat kernel of the corresponding Laplacian,
\al{ \la{233}
&\delta(0)\equiv \delta^{(2)}(\xi,\xi) \ \  \to \ \   \rE \equiv  e^{ - t (-\nabla^2 + m^2 )}\delta^{(2)}(\xi,\xi')\Big|_{\xi'\to\xi}  \ , \\
&  \rE = \te  {1\ov 4\pi t} + {1\ov 4\pi} \big[ {1\ov 6} R^{(2)}(\xi)  - m^2\big] + \OO(t) \ , \qquad\qquad   t = \Lambda^{-2} \to 0\ .\la{23}
}
It thus  has a finite part  depending on  the curvature  and mass. In the $\zeta$-function regularization (used, e.g.,  in Appendix B of \ci{Beccaria:2026ffm}) the  power divergence is dropped  but the  finite part  remains
\be \la{202}
\delta(0)\ \to \  \zeta(0;\xi,\xi) = \te {1\ov 4 \pi} \big[{1\ov 6} R^{(2)} - m^2 \big] \ .
\ee
In \adst\   where  $R^{(2)}=-2$ these finite parts are non-zero and differ from field to field if each $\delta(0)$ 
 is regularized with its own kinetic operator with the corresponding mass.
 %r1
 One may also use a  common ``massless''  smearing for all the fields
   which will be one of the regularization options discussed below.\foot{Note that for the fermions ``massless''  smearing will be assumed 
   to be the one with $e^{t \hat \nabla^2}$   i.e. with the spinor derivative Laplacian,   while 
     the square of the   massless Dirac operator  is  $- \hat \nabla^2 + {1\ov 4} R^{(2)}$.}
  %(i.e. with $m^2$ in \rf{23} being the mass of the field). 
  % In the  problem we will  discuss below   it will be  important to use a common smearing operator for all the fields.

When the Lagrangian is not quadratic in the velocities  % (as, e.g., in  the Nambu  string case  in the static gauge) 
 the  momentum integral in the analog of \rf{20} is no longer  Gaussian, but it can still be 
 computed  perturbatively in $T^{-1}$, expanding around the  Gaussian point. This is the case, e.g.,  for the string in static gauge, where the Nambu action expanded in powers of the fields contains $(\del z)^4+...$ terms. Let $u^I $ be the velocity in a local orthonormal frame on the 2d surface, so that the Euclidean Lagrangian is 
 \be 
 \L = \ha u^2 + ... + T^{-1} \L_4(u, ...) + ...  \ , \qquad \qquad   u^I=\del_0 z^I \ . \la{2233} 
 \ee
  Shifting the momenta by their saddle-point values,  i.e. $p = i u + \td p$, the quartic term becomes $T^{-1}\L_4(u - i \td p, ...)$, and the  integral over the  momentum  fluctuations $\td p_I$  with  % in a cell of proper area $w$ 
     $ \langle \td p_I(\xi)\,   \td p_J(\xi') \rangle =\delta^{(2)}(\xi,\xi')  \, \delta_{IJ} 
$  gives 
\be \la{2031}
 \big\langle \L_4(u - i \td p, ...)\big\rangle = \L_4(u,...) - \tfrac{1}{2}  \delta(0)  \Delta_u \L_4 + \tfrac{1}{ 8} [\delta(0)]^2 \Delta_u^2 \L_4 \ , \qquad \qquad \Delta_u \equiv \sum_I {\del^2 \ov \del u^I \del u^I} \ .\ee
 Here $\delta(0)$  corresponds to the field  whose momentum was integrated over (cf. Appendix \ref{apA3}). 
%where the derivatives with respect to the velocities are taken at fixed values of the other arguments. 
%This  contributes  to the measure   with the factor $- wT^{-1} $, and the summation over the 
%cells turns $w^{-1}$ into $\delta(0)$.
Thus  at order $T^{-1}$ one finds two types of  contributions   to the measure. The first is   
\be \la{24}
M_1=  \delta(0)\, \cJ_{\rm mom} \ , \qquad \qquad \cJ_{\rm mom} \equiv  \ha \Delta_u \L_4 = \ha \sum_I {\del^2 \L_4 \ov \del u^I \del u^I }\ ,
\ee
which is the first  term in the 
expansion of $\ha \log \det { \del^2 \L\ov  \del u \del u}$ (the generalization of \rf{22}). $ \cJ_{\rm mom} $
  is a local combination quadratic in the fields.   The subscript ``mom'' indicates that   this  term comes from the integration over the momenta.
The second contribution  is $[\delta(0)]^2$  times 
  a field-independent constant   coming from the quartic in momenta part of the Hamiltonian (the last term in \rf{2031}) 
\be \la{203}
M_2= c\, [\delta(0)]^2 \ , \qquad \qquad c \equiv  - \tfrac18 \Delta_u^2 \L_4 \ .
\ee
 This  term is field-independent but it  still 
  contributes to the 2-loop free energy: in a power-cutoff regularization  $[\delta(0)]^2$ contains, in addition
   to the quartic and quadratic divergences, also a finite part  (cf. \rf{23}). 

A further contribution arises when the system has second-class constraints, as is the case for fermions whose Lagrangian is linear in the time derivatives,
\be \la{2041}
 \L_{\rm f} = a_\a \dot \theta^\a + ... \ , \qquad \qquad   a_\a = [\theta^T A(\theta,z)]_\a \ , \qquad \dot \theta = \del_0 \theta \ , 
\ee
with the dots standing for terms without $\dot   \theta$.
     Starting with \rf{2041}  the  corresponding canonical momenta $\pi_\a$ are then not independent: there are second-class constraints
\be \la{204}
\phi_\a \equiv \pi_\a - a_\a (\theta, z) \approx 0 \ . %\qquad \qquad a_\a = {\del^R \L \ov \del \dot  \theta^\a} \ .
\ee
%where $a_\a$ is the coefficient of $\dot \theta^\a$ in the Lagrangian (the right derivative).
 The phase-space path integral \rf{20} for a system with second-class constraints contains the factor 
  \ci{Senjanovic:1976br,Fradkin:1975cq} (see also \ci{Faddeev:1969su,Henneaux:1992ig})
 \be \prod_\xi \delta(\phi)\, \sqrt {{\rm sdet} \{\phi, \phi\}}\ .\la{4455}\ee
 For the Grassmann-odd constraints 
\be \la{205}
 \{\phi_\a, \phi_\b\} = \mc C_{\a\b}(\theta,z) = \del^L_\a a_\b + \del^L_\b a_\a \ , \qquad \ \ \ 
\ee
where $ \del^L_\a =  {\del^L  \ov \del \theta^\a} $   and $\mc C_{\a\b}$ is 
 the (symmetric) matrix in  the kinetic term  in    \rf{2041}.   
 The factor \rf{4455},\rf{205}  is  thus the fermionic analog of \rf{22}: a  Jacobian related to   the  
  field-dependent  coefficient in  the first-order fermionic kinetic term.
 %  Here the right-derivative convention means $\delta \L = a_\m\, \delta \dot\theta^\a + ...$; the left-derivative convention differs by signs which must be converted consistently and cannot change the determinant. 
 
   Since $a_\a$ in \rf{2041} may depend also on the spatial derivatives $\del_1 \theta$, the matrix $\mc C$ is a priori a differential operator in the spatial direction. Its variation with respect to that dependence produces terms proportional to derivatives of the spatial delta function, 
   whose traces vanish at the order considered here, so that only the algebraic matrix trace  remains. 
  Expanding  $\mc C$ in  $T^{-1}$  we then get an extra contribution to the  $T^{-1}$ term in the 
  measure   factor  (in addition to the ``bosonic''
  ones in \rf{24} and \rf{203}) 
  \al{M_3=  \delta(0) \, \cJ_{\rm sc}  \ , \qquad \qquad \cJ_{\rm sc} = - \ha \tr \big( \mc C_0^{-1} \mc C_2 \big) \ ,\qquad \ \ \ \ 
  \mc C  = \mc C_0 + T^{-1} \mc C_2(\theta, z) + ...\ . \la{25}}
The subscript ``sc''   indicates   the second-class constraint origin of the  term $ \cJ_{\rm sc} $
which  is again a local combination quadratic in the fields, now containing both 
bosonic terms (from the dependence of the fermion kinetic term on the bosonic fields, e.g. through the induced metric) and fermionic bilinears (from the four-fermion terms containing $\dot \theta$ factor).

 When $\cJ_{\rm sc}$ depends on the bosonic momenta (through the $\del_0 z$ velocity dependence of $\mc C_2$) the constant  $c$  in \rf{203} receives an additional contribution $- \ha \Delta_u \cJ_{\rm sc}$.
Altogether, the  leading  2-loop  term in the 
 measure  factor  in the Lagrangian path integral  then takes the  following form
\al{\la{206}
& \mc M = 1 +  T^{-1}   \int d^2 \xi \sqrt g\, \Big[ \delta(0)\, \cJ  +  c\,[ \delta(0)]^2 \Big] + \OO(T^{-2})\ ,\\
&  \qquad \ \  \cJ = \cJ_{\rm mom} + \cJ_{\rm sc} \ , \qquad  \qquad c = - \tfrac18 \Delta_u^2 \L_4 - \ha \Delta_u \cJ_{\rm sc} \ .\la{2006}
}
Here  $\cJ_{\rm mom}$ includes, in general,  also the velocity
 Hessian of the mixed boson-fermion quartic terms. % (which gives fermionic bilinears).
$\delta(0)$   factor in each term in \rf{206}  corresponds  in general to a  particular field   whose integration produced it. 
%r1
Below in sections \ref{s23} and \ref{s24}  we will consider a particular prescription
   when  all  such $\delta(0)$  in the  measure are  regularized in the same way, i.e. set equal to the same $\rE$. 
   The   general definition of the measure will be  discussed in Appendix \ref{apA3}. 

%21
Let us note that one could in principle avoid the explicit construction of the
measure by doing the 2-loop computation directly in the phase-space form
\rf{20}, where the bosonic measure is flat and the 2-loop correction is the
expectation value of the  corresponding quartic Hamiltonian $H_4(p,z,...)$. The measure terms are
then hidden in the momentum propagators: with $p = i \dot z + \td p$ one has then 
$\langle p_I(\xi)\, p_J(\xi') \rangle = \delta_{IJ}\big[ K_I(\xi,\xi') -
\del_0 \del'_0 G_I(\xi,\xi')\big]$. The contact term $K_I$, which
carries the smearing kernel of the field whose momentum is contracted, is
precisely what  leads to the  $\delta(0)\, \cJ_{\rm mom}$ and the $[\delta(0)]^2$
terms  after the integration over the momenta. 

In this sense the
phase-space form fixes unambiguously the $\delta(0)$ attached  to each term
of the measure (cf. Appendix \ref{apA3}). It does not, however, remove the
fermionic second-class  constraint factor $(\det \mc C)^{-1/2}$  in \rf{4455}--\rf{25}, which is field dependent
in any variables. Also, it singles out the time direction, so that the
covariant \adst\  regularization  is not directly available in this case. 
We therefore use the phase-space form of the path integral only to derive the
configuration-space measure, and regularize  it covariantly afterwards.

Let us stress %add a remark that 
that the role of the right canonical measure  choice is  more generally: it is what makes the
partition function invariant under local  field redefinitions  and  change of  the gauge.\foot{This is the path-integral form of the equivalence theorem for local field redefinitions
\ci{Chisholm:1961tha,Kamefuchi:1961sb,Salam:1970fso}, in which the ultralocal Jacobian
is  what compensates the change of the contact terms (see
\ci{Lee:1962vm,Gervais:1976ws,Salomonson:1976yn} for the quantum-mechanical and
\ci{Fradkin:1973wke,Fradkin:1984ai} for the field-theory discussions of the role of the local
measure).  In dimensional regularization both  contributions  vanish \ci{tHooft:1973wag}.
 The
gauge independence of the reduced phase-space path integral was  discussed in 
\ci{Faddeev:1969su,Kallosh:1972ap,Fradkin:1975cq,Batalin:1977pb,Henneaux:1992ig}.} 
Under a
local redefinition of the fluctuation fields, $\vp \to \vp + T^{-1} \rF (\vp, \del \vp)$,
the action changes at order $T^{-1}$ by $\rF \cdot \delta S_2/\delta \vp$, whose expectation
value is a pure contact term 
$
 \langle \int d^2 \xi \sqrt g\ \rF \cdot  {\delta S_2 \ov \delta \vp} \rangle
 = \int d^2 \xi \sqrt g\  E_\vp \cdot \langle {\del \rF\ov \del \vp} \rangle
 + \dots 
$
%(the contraction of $\vp_A$ in $f_A$ with the field equation of $\vp_A$ gives the
%regularized $\delta(0)$ of that field, $E_A$; the 
(with the fermionic components entering with the opposite sign) 
where dots stand for the terms with derivatives
of the delta-function, which vanish in a symmetric regularization. 
At the same time, 
 the measure
changes by the Berezinian of the redefinition 
$
 J =1 + T^{-1}  \delta(0) \int d^2 \xi \sqrt g\ {\rm Str}\, {\del \rF \ov \del \vp} + ... \ . 
$
%with the same relative sign for bosons and fermions as in \rf{2061}. 
The two cancel  due 
to the  identity $\int [d\vp]\, {\delta \ov \delta \vp} ( \rF\, e^{-S}) = 0$, 
 provided the $\delta(0)$ in $J$ for the field $\vp$ is the same regularized
quantity $E_\vp$ that appears in the contact term.

This is the precise sense in
which the $\delta(0)$'s in the measure are associated with  the individual fields
(cf. Appendix \ref{apA3}): with any other regularization of the Jacobian the result would
depend on the choice of variables already at order $T^{-1}$. 
An example   of this is the
redefinition \rf{4174} below relating two forms of the massless four-fermion terms in section
\ref{s4}, whose Berezinian cancels the change of $\langle \L_4 \rangle$. 

The same
reasoning applies to a change of the gauge: two algebraic gauges are related by a field
redefinition together with the ratio of the ghost  determinants, both ultralocal,
and the  theorem of    \ci{Faddeev:1969su}     guarantees that the reduced Liouville measure is the same in both.
Thus  the gauge-fixed partition function is gauge independent once the $\delta(0)$'s in
the Jacobians and in the contact terms of the action are regularized in the same way. 
The measure
prescription we use here  is thus not an additional assumption but the condition for the
2-loop result to be a property of the theory rather than of the parametrization.
%(though it fixes only the 
 What it
does not fix is the regularization-dependent  constant  ${\rm Str}\, E$  in \rf{1061},\rf{999} 
discussed in section
\ref{s3} and Appendix \ref{apA3}.

%, which is a property of the regulator and not of the variables.

\subsection{GS string  in  static and $\kappa$-symmetry gauges \la{s23}}

For the GS string  in flat space 
the fermions $\theta$ do have a first-order kinetic term, $\L_{\rm f} = \bar\theta \Gamma^a \del_a \theta + ...$  as in \rf{2041}, 
 so that the  above construction  applies. In flat target space the GS string in the light-cone gauge is a free theory and the measure is trivial. In the static gauge with the corresponding $\kappa$-symmetry gauge, as used in \ci{Beccaria:2025xry}, the action does contain quartic terms of the type $(\del z)^4$, $(\del z)^2 \theta \del \theta$ and $\theta\del\theta\theta\del\theta$, and the above construction produces the measure \rf{206}. In the dimensional regularization used in \ci{Beccaria:2025xry} it is again trivial. In a power-cutoff regularization its role would be to cancel the power divergences of the 2-loop diagrams.

Let us note that in  general in  the GS string 
 case the leading  fermionic kinetic term is of the first-order type  \rf{2041}, but the full action, being built from
  the induced metric ${\rm h}_{ij} = \Pi^a_i \Pi^a_j$ with $\Pi^a_0 = \del_0 X^a + \bth \G^a \del_0 \theta + ...$,  also contains terms quadratic in $\dot  \theta$ at order $T^{-1}$. 
  In the gauge-fixed actions used below such terms are absent to the quartic order.
  In the adapted $\kappa$-symmetry gauge the two-derivative four-fermion structures combine into the 2d determinant combination \rf{a41}, which is linear in $\dot  \theta$, while the mixed terms contain only one Killing-spinor derivative and are linear in $\dot  \theta$ by construction (this was verified explicitly for both backgrounds; see section \ref{s24} and Appendix \ref{apA}). We shall therefore assume the form \rf{2041}, with no $\dot  \theta^2$ terms.\foot{If the quartic action did contain a term $\tfrac{1}{2} T^{-1} \dot\theta^T B(\theta, z)\, \dot\theta$,
the fermionic Hessian $\del^2 \L/\del\dot\theta \del\dot\theta = T^{-1} B$ would be non-zero. Since
$B \sim \theta\theta$ is nilpotent, such a term does not change the constraint structure at the level
of the Grassmann-even parts, but it cannot be absorbed into the first-order treatment of this section:
the factor $(\det \mc C)^{-1/2}$ is built from the part of the action linear in $\dot\theta$, and there
is no expansion of the momentum integral around $B=0$. At order $T^{-1}$ such a term can, however, be
removed by the velocity-dependent field redefinition $\theta \to \theta - {1\ov 4T} A_0^{-1} B\, \dot\theta$,
which cancels it against the variation of the free term $\theta^T A_0 \dot\theta$ and changes the rest of
the action only at order $T^{-2}$. Its Jacobian contributes, in a symmetric regularization, only the
ultralocal term $-{1\ov 4T}\,\delta(0)\, \tr[A_0^{-1} \del_\theta B\, \dot\theta]$, which would then have to
be added to $\cJ_{\rm sc}$ in \rf{25}. For the gauge-fixed actions used in this paper no such terms are
present at the quartic order. The two-derivative four-fermion structures combine into the determinant
combination \rf{a41}, which is linear in $\dot\theta$, in accordance with the counting of the physical
fermionic degrees of freedom fixed by the $\kappa$-symmetry.}

Expanding the GS action in a curved background near a classical solution and fixing the static gauge on the bosons and the adapted $\kappa$-symmetry gauge on the fermions, one obtains the 2d Lagrangian for the $8+8$ physical fluctuations of the following structure   \ci{Beccaria:2026ffm} %\foot{We use the \adss case as an example; the discussion in the  \adsc case is analogous.}
\ba
   \cL = & (\del x^i)^2  + m^2_{\rmb}  x^2 + (\del y^a)^2  +  \bar \theta ( \slashed \nabla + m_{\rmf}) \theta  + T^{-1} \big[ (\del x)^4  + x^2 (\del x)^2 + x^4  + (\del x)^2 (\del y)^2  \no  \\
 &  +   (\del y)^4 + y^2 (\del y)^2 + (\del x  \del x  + x^2) (\theta \nabla \theta + \theta^2 ) + ...
     + \theta  \theta \theta  \nabla \theta   +   \theta\nabla  \theta\theta  \nabla \theta + ...     \big]  + \OO ( T^{-2}) \ . \la{26}
   \ea
 The physical phase space consists of the gauge-fixed coordinates $z^I= (x^i,y^a)$  and $ \theta^\a$ and 
their momenta $p_I $ and $ \pi_\a$ with the Liouville measure, the fermionic second-class constraints being imposed with the factor $(\det \mc C)^{-1/2}$ as above.  %This is the standard reduced-phase-space quantization of a constrained system.
 The bosonic momenta are integrated out perturbatively in $T^{-1}$, giving the measure \rf{206} with $\cJ= \cJ_{\rm mom} + \cJ_{\rm sc}$.

Let us comment on the role of the gauge-fixing determinants. The static gauge $X^0=\xi^0, \, X^1=\xi^1$ on the diffeomorphism constraints and the algebraic $\kappa$-symmetry gauge on the first-class half of the fermionic constraints are algebraic gauges, and the corresponding Faddeev-Popov determinants are ultralocal. 
%\be \la{207}
%\prod_\xi \det M(\vp(\xi)) = \exp \Big[ \delta(0) \int d^2 \xi \sqrt g\, \log \det M(\Phi)\Big] \ .
%\ee
%(for the odd $\kappa$-symmetry generators the determinant is the graded one, i.e. the corresponding factor enters with the inverse power). In dimensional regularization they are trivial. 
In a 2d regularization they are of the same $\delta(0)$ type as the measure terms above, and one may wonder whether they should be added. In fact, they  are already accounted for   in the above construction. 
 %The answer is that in the canonical construction (i)--(iii) they are already accounted for. 
 According to the  theorem of   \ci{Faddeev:1969su} (see also \ci{Henneaux:1992ig}), for first-class constraints $\beta_i$ with algebraic gauges $\nu_i$ the factor $\prod \delta(\beta) \delta(\nu ) \det \{ \nu, \beta\}$ in the Liouville measure of the parent phase space is precisely the Liouville measure of the reduced physical phase space. Thus the Liouville measure of the gauge-fixed system  already contains the static-gauge and $\kappa$-symmetry ghost determinants (the second-class fermionic constraints being treated separately by the factor \rf{4455}).\foot{The reduction theorem of  \ci{Faddeev:1969su}  is formal, i.e. it assumes an ultralocal relation between the parent and the reduced measures. In a regularized theory, the passage from a parent (e.g. covariant or BRST) formulation to the reduced one may leave finite ordering terms of the type discussed in \ci{Bastianelli:1991be,Bastianelli:2006rx} for a particle in curved space.
 % and the rotation from the original GS fermions to 2d spinors may carry a non-ultralocal Jacobian \ci{Fujikawa:1979ay}. 
  We shall return to this point in sections \ref{s4} and \ref{s5} and in Appendix \ref{apA2}.
}

Equivalently, the configuration-space form \rf{206} of the measure is the combined effect of the gauge-fixing Jacobians, of the second-class constraints and of the momentum integration; multiplying it by a separate $\kappa$-symmetry Jacobian would be double counting.\foot{The algebraic $\kappa$-symmetry Jacobian  alone would contribute a term $\delta(0) \int d^2\xi\sqrt g [ 2 (\del z)^2 + {1\ov 12} \bar \theta \Gb \theta]$ for the \adss case,  producing a non-cancelling logarithmic UV divergence at 2 loops.  In the canonical construction its bosonic part is cancelled identically by the change of variables from the original $\kappa$-gauge-fixed GS fermions to the 2d spinors with the canonical kinetic term.} %with its content reappearing in the quartic action.}

 The  factor $\delta(0)$ in the measure  is represented by the coincident value of the same smearing kernel  (cf. \rf{233})  that regularizes the  contributions  of the fluctuation  fields. We shall assume  that 
  the scalar contributions  are smeared with the kernel $e^{t \nabla^2}$ of the \adst\  Laplacian $-\nabla^2$ 
   and the spinor ones  with the kernel $e^{t \hn^2}$ of the spinor Laplacian  (the  square of the 
   2d Dirac operator is $-\slashed \nabla^2 = - \hn^2  + {1\ov 4} R^{(2)}$). 
%%% CHANGED 21/10/26 (field-attributed measure): sect. 2.3, smearing paragraph: field attribution of delta(0)
   The two kernels  are not one identical operator, but their coincident values differ only by terms vanishing for $t\to 0$, see \rf{218} below. Then $\delta(0)\to \rE$ with $\rE$ given by \rf{23} with $m^2=0$. 
   
   More precisely, each term in the measure comes with the $\delta(0)$ of the field whose integration produced it (the $I$-th term of $\cJ_{\rm mom}$ with the $\delta(0)$ of $z^I$, each term of the fermionic trace in $\cJ_{\rm sc}$ with that of the corresponding fermionic component, etc.). In the common smearing all of these coincide with $\rE$
    up to terms vanishing for $t \to 0$, and this is what we use  below. % in the main text.
     The more general form of the measure  which is required when
      the fields are smeared with different kernels
       is discussed in Appendix \ref{apA3}.
   %, and $\delta(0)^2 \to \rE^2$ is the product of two coincident kernels.

%Two further comments are in order. First, it is not true that ultralocal $\delta(0)$ factors give only power divergences and can %thus be ignored in a power-cutoff regularization. 

\iffa   %????????????????

The  term $\delta(0)   \cJ$ with a quadratic in fields  factor $\cJ$ in \rf{206}  contributes $\rE\,   \langle \cJ\rangle$ to the
 2-loop  free energy coefficient  (see \rf{225} below).   $\langle \cJ \rangle$ contains the coincident propagators $\rG_\vp$
 of the fluctuation fields $\vp=(z,\theta)$. Thus the  products
$
 \rE\, \rG_\vp = {1\ov 4\pi t} \rG_\vp  + ...
$
are  divergent as  $\Lambda^2 \log \Lambda^2 $ as $\rG_\vp$  contain log divergences $\sim \log t=-2\log \Lambda$. 
The  products of $\rE$ with EOM parts,  $\rE\,  E_\vp$     have power divergent plus  finite parts. 

 %Note that the finite part of a product is not the product of the finite parts: with
%\be \la{2081}
% \rE(t) = {1\ov 4\pi} \big( t^{-1} + a + b\, t + ... \big) \ , \qquad \quad
 %{\rm FP}\, [\rE(t)^2] = {a^2 + 2 b \ov 16 \pi^2} \ , \qquad \quad ({\rm FP}\, \rE)^2 = {a^2\ov 16\pi^2} \ ,
%\ee
%where FP denotes the finite part as $t \to 0$. This is why the action and 

Let us note that the limit $t\to 0$   should    be  applied  only  after taking  the products  of the smeared   quantities. 
%The measure are to be combined at non-zero cutoff, before the cutoff is removed, and why equality of the leading divergences alone would not be sufficient. 
The $\delta(0)$ terms thus contribute also  finite parts, and the cancellation found below is  a 
 non-trivial one: it holds only because the coefficients in the canonical measure are  correlated with those in the quartic term in the  string action.

\fi

\subsection{Measure contribution  for the \adss string \la{s24}}

Let us now apply the above construction to the \adss GS string expanded near the \adst\  minimal surface  starting with 
the quartic action in  \ci{Beccaria:2026ffm}.  We use the Euclidean \adst\  of unit radius with the metric $g_{\m\n}$
with curvature  $R^{(2)}=-2$ and regularized volume $V=-2\pi$.

The  gauge-fixed  GS  string   action  fluctuation field content in the two  cases  considered in this paper 
is summarized in the following table: %(all massive bosons have $m^2=2$; the fermionic masses are the coefficients of $\Gb$ in the corresponding Dirac operators):
\begin{center}
\begin{tabular}{@{}lccccc@{}}
\toprule
background & $m^2_{\rmb}=2$   bosons & $m^2_{\rmb}=0$ bosons & fermions & real components & $m_\rmf$\\
\midrule
\adss &  3  real $x^i$  &  5 real  $y^a$  & $\theta$ & $16$ & $+1$\\[3pt]
\adsc &  1  complex  $x$  &  3  complex $w^a$   & $\psi$, $\chi$ & $12$, $4$ & $-1$, $0$\\
\bottomrule
\end{tabular}
\end{center}
% The physical fluctuations in the gauge-fixed action 
%  are 3  scalars $x^i$ with $m^2_\rmb = 2$ (transverse directions in AdS$_5$), 5 massless scalars $y^{a}$ ($S^5$ directions) and
 %  16 real fermionic components $\theta$ which may be organized as 8   Majorana   2d fermions 
%    with the mass term  
    % (we use the notation of section 2 of , with the Euclidean continuation of the fermionic terms described in section \ref{s3}). 
   The quadratic Lagrangian in \adss  case is   \ci{Beccaria:2026ffm}
\al{ \la{27}
&\L_2 = \ha ( \del x\cdot \del x) + x^2 + \ha  (\del y\cdot \del y) + \ha \bth \Dsl \theta \ , \\
& \Dsl \equiv \g^\m \DF_\m = \hat {\slashed  \nabla} + \Gb \ , \qquad \DF_\m \equiv \hat  \nabla_\m + \ha \g_\m \Gb  \ ,\qquad 
\Gb = \G_{123}, \ \ \  \Gb^2=-1, \ \  \{\Gb, \g_\m\}=0 \ .\la{207}
}
where $\g_\m$ ($\m=0,1$) are the two  gamma matrices along the \adst\ directions (they are  10d   matrices $\G_0$ and $\G_4$
in the labelling of \ci{Beccaria:2026ffm}), and $\Gb = \G_{123}$ involves the three transverse AdS$_5$ directions.
Here $\DF_\m$ is the Killing-spinor derivative, $\bth = \theta^T C$ with $C$ the charge conjugation matrix, 
$\gamma_\mu = \Gamma_\mu  \  (\mu=0,1)$,  
$\g_{\m\n} = \ha [\g_\m, \g_\n]$, and all fermionic traces below are over the 16-dimensional space of the physical fermionic components. 
 The 16 real fermionic components $\theta$ may be organized as 8   Majorana   2d fermions 
    with the mass term  determined by $\bar \Gamma$. 
 The square of the Dirac operator is
\be \la{2009}
 - \hat {\slashed \nabla}^2 = - \hn^2 + \tfrac14 R^{(2)} = - \hn^2 - \tfrac12 \ , \qquad \qquad - \Dsl^2 = - \hn^2 + \tfrac12 \ ,
\ee
where $\hn^2$ is the spinor Laplacian. 

The detailed form of the \adss 
quartic Lagrangian in \rf{26}
\be \la{101}
\cL=\cL_2 +T^{-1}  \cL_4 + ... \ , \qquad \qquad 
\L_4 = \L_{4\rm b} + \L_{2\rm b,2 f} + \L_{4\rm f} \ , \ee
 is given in \rf{31}--\rf{33} below.
Applying the   procedure of constructing the leading 
 2-loop   contribution to the measure described above    to this   Lagrangian  %$\cL=\cL_2 +T^{-1}  \cL_4 + ...$
  one finds for \rf{206},\rf{2006}
    (see Appendix \ref{apA}) 
\al{ \la{28}
&\mc M = 1 +    T^{-1}   \int d^2 \xi \sqrt g\, \Big[ \delta(0)\, \cJ  - 6\, [\delta(0)]^2 \Big] + \OO (T^{-2}) 
\ , \qquad \qquad  \  \cJ = \cJ_{\rm b} + \cJ_{\rm f}  \ ,
\\
& \cJ_{\rm b} = \cJ_{\rm mom, b}  +  \cJ_{\rm sc, b} =  - \ha (\del x \cdot \del x +  \del y\cdot \del y) - \tfrac{13}{4} x^2 - \tfrac{5}{4} y^2 \ , \la{29} \\
& \cJ_{\rm f}  = \cJ_{\rm mom, f} +  \cJ_{\rm sc, f} =    - \bth \Dsl \theta - \tfrac{1}{8}  \bth \Gb \theta \ . \la{209}
}
 $\cJ_{\rm b}  $ is the sum of the velocity Hessian of $\L_{4\rm b}$  in  \rf{24}
(here $\del_0$  and $\del_1$ have  local frame indices)
\be \la{211}
 \cJ_{\rm mom, b} = - \tfrac{5}{2} u^2 + \tfrac{3}{2} s^2 + \tfrac{3}{4} x^2 - \tfrac{5}{4} y^2 \ , \qquad \ \ u^2 = \del_0 z^I \del_0 z^I \ , \ \ \ s^2 = \del_1 z^I \del_1 z^I \ ,\  \ \ \   z^I= (x^i, y^a) \ , 
\ee
%(in a local orthonormal frame, summed over all 8 bosons), 
and of the bosonic part of the second-class  constraint contribution \rf{25}, 
coming from the dependence of the fermion kinetic term in the mixed part of the quartic Lagrangian 
$\L_{2\rm b, 2f}$ on the bosonic fields,
\be \la{212}
 \cJ_{\rm sc, b} = 2u^2 - 2 s^2 - 4 x^2 \ .
\ee
Note  that  while   $\cJ_{\rm mom, b} $ and $ \cJ_{\rm sc, b}$  are not  Lorentz covariant,
their sum in  \rf{29} is:\foot{Recall that we  are using Euclidean signature, i.e.
$\del^\m z \del_\m z = (\del_0 z)^2 + (\del_1 z)^2$.}
this is a non-trivial check of the relative normalization of the bosonic   $\L_{4\rm b} $  \rf{31}
and the mixed quartic $\L_{2\rm b,2 f}$ \rf{32} interaction  terms.
The cancellation of the non-covariant terms $-2(u^2-s^2)$ in \rf{211} and $+2(u^2-s^2)$ in \rf{212}
relies on both being multiplied by the same $\delta(0)$. If, as discussed in Appendix
\ref{apA3}, each term of the measure is multiplied by the regularized $\delta(0)$ of the field
whose integration produced it, then \rf{211} comes with the bosonic $E_I$ and \rf{212} with
the fermionic $E_\a$, and the sum is covariant only if ${\rm Str}\, E=0$, see \rf{6002} below.
This will provide the main argument for the choice ${\rm Str}\, E=0$.

%Note  that  while   $\cJ_{\rm mom, b} $ and $ \cJ_{\rm sc, b}$  are not  Lorentz covariant,  their sum in  \rf{29} is:\foot{Recall that we  are using Euclidean signature, i.e. $\del^\m z \del_\m z = (\del_0 z)^2 + (\del_1 z)^2$.}
 % this is a non-trivial check of the relative normalization of the bosonic   $\L_{4\rm b} $  \rf{31} and the mixed quartic $\L_{2\rm b,2 f}$ \rf{32} 
% interaction  terms.

$\cJ_{\rm f}$ in \rf{209} is the sum of the velocity Hessian of the mixed term $\L_{2\rm b, 2f}$ and of the
 fermionic part of the second-class constraint contribution, coming from the $\del_0 \theta$-dependence of the  four-fermion term 
 $\L_{4\rm f}$
\be \la{2121}
 \cJ_{\rm mom, f} = \bth \g_1 \DF_1 \theta - \bth \g_0 \DF_0 \theta - \tfrac14 \bth \Gb \theta \ , \qquad \qquad
 \cJ_{\rm sc, f} = - 2\, \bth \g_1 \DF_1 \theta + \tfrac18 \bth \Gb \theta \ .
\ee
 Here $-\tfrac14 \bth\Gb\theta$ comes from the term $\tfrac18 (\del x\cdot \del x + 2 x^2 - \del y \cdot \del y) \bth \Gb\theta$ in $\L_{2\rm b, 2 f}$  in \rf{32}.
 The three mass-type four-fermion structures in $\L_{4\rm f}$ in \rf{33} 
  give contributions to the constraint matrix proportional to $\tfrac{7}{48}$, $\tfrac{1}{16}$ and $-\tfrac{5}{24}$, whose sum vanishes. 
  The two-derivative terms  combine so that no $(\del_0\theta)^2$ term appears  (see Appendix \ref{apA}).  Thus 
the   fermionic action   has   the structure of  \rf{2041} for    which  the above discussion leading to   \rf{25}--\rf{2006} applies.
The  sum \rf{209}  of  $ \cJ_{\rm mom, f} $  and  $ \cJ_{\rm sc, f}$   is   again Lorentz covariant.
 
 %non-covariant derivative combinations in \rf{2121} cancel in the sum, and in any case have vanishing expectation value on the homogeneous background. 
 
 Finally, the constant $c=-6 $  in \rf{28} 
   comes, according to \rf{206},  from the sum of  contributions of 
    (i) the quartic-in-momenta part of $\L_{4\rm b}$ in \rf{31}, 
   i.e. 
   $- \tfrac18 \Delta_u^2 \L_{4\rm b} = 10$, and  of  (ii)  the momentum dependence of $\cJ_{\rm sc, b}$ in \rf{212}, i.e.
    $- \ha \Delta_u \cJ_{\rm sc,b} = - 16$.  %  so that $c={80\ov 8} - {32\ov 2} =-6$. 

\

The next  step is to evaluate the expectation value   $\langle {\mc M }\rangle $   of \rf{28}   using the propagators  corresponding to 
the fields in \rf{27}. We  shall  assume   heat-kernel   regularization    using  ``massless''  boson and fermion 
smearing kernels (we suppress  $\delta^{(2)}(\xi,\xi')$ factors,  \rf{233},\rf{213})
%\subsection{Evaluation \la{s25}}
%To evaluate the 2-loop contribution of \rf{28} we need the coincident values of the propagators and of their derivatives. We regularize the propagators by the heat-kernel smearing with the kernels
\al{ \la{214}
&\langle x^i(\xi)\,  x^j(\xi')\rangle = \delta^{ij} G_x(\xi,\xi')  \ , \qquad \langle y^a(\xi)\,   y^b(\xi') \rangle = \delta^{ab} G_y(\xi,\xi')
 \ , \qquad \langle \theta(\xi)\,   \bth(\xi')  \rangle = S(\xi,\xi')\ ,  \\
&\qquad  G_x = (-\nabla^2 + 2)^{-1} K \ , \qquad \ \  G_y = (-\nabla^2)^{-1} K \ , \qquad \ \  S = \Dsl^{-1} K_\theta \ ,\la{2104} \\
&\qquad \qquad  K \equiv e^{t \nabla^2}     \ , \qquad \qquad K_\theta \equiv  e^{t \hn^2} \ , \qquad \qquad t = \Lambda^{-2} \to 0 \ . \la{213}
}
The regularized propagators satisfy the field equations
 with the smeared source, $(-\nabla^2 + 2) G_x = K$, etc. 
 
 Let  $[\,\dots\,]$  denote  the coincident-point value, i.e.  $[f(\xi,\xi')] \equiv f(\xi,\xi)$, and   define 
% the constants (\adst\  is a homogeneous space)
 \al{ \la{215}
&  [G_x]\equiv \rG_x  \ , \qquad [G_y]\equiv  \rG_y  \ , \qquad   [S] \equiv  - \Gb \, \rG_\theta \ ,  \\
%and of the regularized $\delta$-functions (``equation of motion'' or $E$-terms)
& \la{216}
E_x = E_y \equiv [K] = \rE \ , \qquad \qquad \qquad E_\theta \equiv [K_\theta] \  . % \ \ ({\rm per \ spinor\ component}) \ ,
}
$E$'s  are thus ``equation-of-motion'' terms ($E_\theta$ is defined per spinor component).
All the products of regularized quantities below are formed at finite $t$, and the limit $t\to0$ is taken at the end; in particular,  $[\delta(0)]^2\to\rE(t)^2$, whose finite part is not the square of the finite part of $\rE$.

Here we are using the common (massless) smearing \rf{213} for all the bosons and for all
the fermions, so that $\rE$ in \rf{216} is the same quantity as in \rf{233} with $m=0$, and the
$\delta(0)$ in  the measure in \rf{28} can be identified with $E_x = E_y = \rE$ and  also (up to
terms vanishing for $t \to 0$, see \rf{218} below)  with $E_\theta$. This is what allows us to
write the measure contribution \rf{225} below with a single   $\rE$.\foot{We assume that the regularizing kernels 
for  all the fields are built from their
\adst\ Laplacians with a common cutoff parameter $t$, so that the leading term ${1\ov 4\pi t}$ of
$E_\vp$ is the same for all fields.  For kernels with different cutoff scales the power divergences in
${\rm Str}\, E$ in \rf{999} would not cancel.}

 If instead  different  fields were
smeared with their   corresponding  kernels (with different masses), 
each term of the measure would have to be multiplied by the
 regularized $\delta(0)$ of the field whose integration produced it (cf. the remark  before %at the end of 
 section
\ref{s23}). The corresponding general form of the result is given in Appendix \ref{apA3}.

%where $\rE$ is the $\delta(0)$ of the measure in \rf{28}, which for the common smearing of (iii) coincides with $E_x$ and $E_y$. 

Explicitly, one   finds 
(see Appendix B of \ci{Beccaria:2026ffm} and \ci{Camporesi:1991nw,Camporesi:1995fb} for the spectral representations)
\ba
& \rG_x = \tfrac{1}{4\pi} ( \LDR -2) + \OO(t \LDR) \ , \qquad \rG_y = \tfrac{1}{4\pi} \LDR + \OO(t \LDR) \ , \qquad \rG_\theta = \tfrac{1}{4\pi} ( \LDR - 1) + \OO(t \LDR)  \ ,\qquad \ \ \ \  t\to 0 \ ,  \la{2170} \\
&\qquad \qquad \qquad     \LDR \equiv  - \log t + \Ge  = 2 \log \Lambda + \Ge \ , \la{217} \\
& \rE = \tfrac{1}{4\pi} \big( \tfrac{1}{t} - \tfrac{1}{3} + \tfrac{t}{15} + ... \big) \ , \qquad
  E_\theta = \tfrac{1}{4\pi} \big( \tfrac{1}{t} - \tfrac{1}{3} + \tfrac{t}{40} + ... \big) \ , \qquad \ \ E_\theta - \rE = - \tfrac{t}{96 \pi} + \OO(t^2) \ . \la{218}
\ea
The finite parts of $\rG_x, \rG_y, \rG_\theta$ are the same as in  \ci{Beccaria:2026ffm} (with ${1\ov \eps'} \to \LDR$). 
Since  the massless 
scalar and spinor heat kernels have the same first two Seeley-DeWitt
 coefficients ($\tfrac{1}{4\pi}$ and $\tfrac{1}{4\pi}\tfrac{1}{6}R^{(2)}$)  one concludes 
  that  $E_\theta$ and $\rE$ differ only at subleading   order in $t\to 0$.
 
 Note that if  instead each field were smeared with its own kinetic operator (i.e. $K_x = e^{-t(-\nabla^2+2)}$, etc.) the finite parts of the $E_\vp$ with  $\vp=(x,y, \theta)$ would be the $\zeta$-function regularization $\zeta(0)$ values 
  in Appendix B of \ci{Beccaria:2026ffm}, i.e. 
\be \la{219}
  (E_x, E_y, E_\theta)_{\rm fin} =\te {1\ov 4 \pi}  (-\tfrac{7}{3}, - \tfrac{1}{3}, - \tfrac{5}{6}) \ .
\ee
The two regularizations are related by a rescaling of the smearing kernels, $K \to e^{-t m^2} K$ for the scalars (with $m^2 = 2$ and $0$) and $K_\theta \to e^{-t/2}K_\theta$ for the spinor (whose kinetic operator squared is $-\hn^2 + \ha$ in \rf{2009}).
This  changes only the finite parts of the $E_\vp$; we shall comment on the corresponding change in the result below and in Appendix \ref{apA3}.

The coincident limits of the derivatives of the propagators follow from the homogeneity and isotropy of \adst\  and from the field equations. For the scalars $z= (x,y) $  one  has  $[\del_\m G_z]=0$ and
\be \la{220}
[\nabla_\m \nabla'_\n G_z] = \ha g_{\m\n}\, [-\nabla^2 G_z]  = \ha g_{\m\n} ( E_z - m^2_z \rG_z ) \ , \qquad\langle \del x\cdot \del x\rangle = 3 ( E_x - 2 \rG_x) \ , \ \ \ \ \langle \del y\cdot \del y\rangle = 5 E_y \ ,
\ee
where the prime denotes the derivative with respect to the second argument.
 For the fermions (see Appendix B of \ci{Beccaria:2026ffm} and Appendix \ref{apa2} below)
\ba
&\qquad \ \ \  [\hat \nabla_\m S] = \ha \g_\m (E_\theta - \rG_\theta) = - [\hat \nabla'_\m S] \ , \qquad \qquad\qquad  [\DF_\m S] = \ha E_\theta\, \g_\m \ , \la{221} \\
& [\hat \nabla_\m\hat  \nabla'_\n S] = \big[ \big( - \ha E_\theta + \tfrac14 \rG_\theta\big) g_{\m\n} + \tfrac14 \rG_\theta \g_{\m\n}\big] \Gb \ , \qquad \qquad
[\DF_\m S \overleftarrow{\DF}{}'_\n ] = \rk_\theta\, \g_{\m\n}\, \Gb \ , \qquad \rk_\theta = \ha E_\theta \ . \la{222}
\ea
In particular, the Killing-spinor derivative of the propagator at coincident points is a pure $E$-term, as it should be, since $\g^\m \DF_\m S = K_\theta$ is the field equation.\foot{\la{af222}The coefficient $\rk_\theta$ of the antisymmetric part of the mixed Killing-spinor second derivative is the one that appears in (B.6) of \ci{Beccaria:2026ffm} (where it was assumed to vanish). Its correct value is
 $\rk_\theta = \ha E_\theta$, see Appendix \ref{apa2}.
 The dependence of the 2-loop coefficient on $\rk_\theta$, and the fact that its UV finiteness
requires $2\rk_\theta = E_\theta$, is discussed in the footnote \ref{af36}. % after \rf{36}.
 For a fermion of general mass $m$
one finds (cf. \rf{49},\rf{410})
%\be \la{362}
$ [\DF_\m S(m) \overleftarrow{\DF}{}'_\n ] = \tfrac{m^2-1}{4}\, \rG_\theta(m)\, g_{\m\n} \Gb
 + \rk_\theta(m)\, \g_{\m\n} \Gb $  with  % \ , \qquad \qquad
$ \rk_\theta(m) = \tfrac{m}{2}\, E_\theta + \tfrac{1-m^2}{4}\, \rG_\theta(m) $. 
Here $\DF_\m \equiv \DF^{(m)}_\m = \hat \nabla_\m + \tfrac 12 m  \g_\m \Gb$ 
which is the  analog  of the Killing-spinor derivative in \rf{207} for a fermion of  generic mass $m$. 
Thus 
$\rk_\theta = \pm \ha E_\theta$ for $m = \pm 1$.
Note that for the massless fermion
of section \ref{s4} the antisymmetric coefficient   %in \rf{222}
 is not a contact term but the finite
coincident constant itself, $\rk_\chi = \tfrac14 \rG_\chi$. This is the origin of the
finite $\rG_\chi^2$ term in \adsc: the same structure that must reduce to a contact term
for the massive fermions  and is then removed by the measure,   happens to survive
 for the massless one.
    }
 Then
\be \la{223}
\langle \bth \g_\m \DF_\n \theta\rangle = - 8 E_\theta \, g_{\m\n} \ , \quad \langle \bth \Dsl \theta\rangle = - 16\, E_\theta  \ , \quad \langle \bth \Gb\theta\rangle = - 16\, \rG_\theta \ ,  \quad \langle x^2\rangle = 3 \rG_x \ , \quad \langle y^2\rangle = 5 \rG_y \ ,
\ee
where the minus signs in the fermionic expectation values are due to the fermionic statistics. Note that
the expectation value of the quadratic Lagrangian \rf{27}  is of course a pure  EOM  term
\be \la{224}
 \langle \L_2\rangle = \ha (3 E_x + 5 E_y - 16 E_\theta)\ . 
\ee
Using \rf{220},\rf{223} in \rf{29},\rf{209} we find that the   contribution of  the expectation value of the measure in \rf{28} to 
the 2-loop coefficient defined in \rf{2},\rf{3}  is then  
%for the expectation value of the measure density  $\cJ$ in \rf{28} is 
\be \la{225}  f_{2  \, \mc M} =  \rE \, \langle \cJ\rangle - 6 \rE^2 \  , \qquad \qquad 
\langle \cJ \rangle = - \tfrac{3}{2} E_x - \tfrac{5}{2} E_y + 16 E_\theta - \tfrac{27}{4} \rG_x - \tfrac{25}{4} \rG_y + 2 \rG_\theta \ .
\ee
The resulting  expression for the 2-loop coefficient defined in \rf{2},\rf{3}  is then  given by  the sum of 
 the contribution of the quartic action computed in the next section   and of $ f_{2  \, \mc M}$.

 \iffa 
  the  contribution of the  2-loop 
  measure  term in  $\langle \mc M\rangle $  \rf{28}  with  $\langle \cJ \rangle $   given by \rf{225}
\be \la{226}
f_2= f_{2 \, S}  + f_{2  \, \mc M} \ , \qquad \qquad 
f_{2 \, S} = - \langle \L_4\rangle\ , \qquad \ \ \ \ 
f_{2  \, \mc M} =  \rE \, \langle \cJ\rangle - 6 \rE^2 \ .
\ee
\fi

%%%%%%%%%%%%%%%%%%%%%%%%%%%%%%%%%%%%%%%%%%%%%%%%%%%%
\section{2-loop correction  in \adss   \la{s3}}

%\subsection{Quartic action}

The expansion of the \adss GS action near the \adst\ minimal surface in the static gauge and the adapted $\kappa$-symmetry gauge was given in section 2 of \ci{Beccaria:2026ffm} in Lorentzian signature.
% see (2.6)--(2.10) and (2.44) there. 
 We continue it to the Euclidean signature by rotating the time derivative together with the time-like 2d gamma matrix and rescaling the fermions, $\theta \to e^{i\pi/4} \theta$, $\bth \to e^{i\pi/4}\bth$, so that $ i S_{\rm L} \to - S_{\rm E}$.
 This fixes the relative sign of the bosonic and fermionic terms in the Euclidean action
\be \la{30}
 S= T S_0 + S_2 + T^{-1} S_4 + ... \ , \qquad S_2= \int d^2\xi \sqrt g\, \L_2 \ , \qquad S_4 = \int d^2\xi\sqrt g\, \L_4 \ .
\ee
  $\L_2$  is given by \rf{27} and $\cL_4$ by  (below $z=(x^i,y^a)$, all  2d  indices  are contracted with the \adst\ metric)
\ba
& \ \ \ \ \  \L_{4}= \L_{4\rm b} + \L_{2\rm b, 2f} + \L_{4\rm f} \ ,\la{300} \\
 \L_{4\rm b}&=\tfrac18(\del_\m z\cdot\del^\m z)^2-\tfrac14(\del_\m z\cdot\del_\n z)(\del^\m z\cdot\del^\n z)+\tfrac14 x^2 (\del _\m x\cdot \del^\m  x)-\tfrac14 y^2 (\del _\m y\cdot \del ^\m y)+\tfrac12(x^2)^2 \ ,\la{31}\\
 \L_{2\rm b, 2f}&=\tfrac14 A^{\m\n}\,\bth\g_\m\DF_\n\theta+\tfrac18\big(\del_\m x\cdot \del^\m  x+2x^2-\del_\m  y \cdot \del^\m y \big)\, \bth \Gb\theta \ , \la{32}  \\
& \qquad  A^{\m\n}=x^2g^{\m\n}-\del^\m z\cdot\del^\n z+\tfrac12g^{\m\n}\big(\del x\cdot \del x+\del y\cdot \del y\big) \ ,\la{302}\\
 \L_{4\rm f}&=\tfrac1{96}\, (\bth\Gb\theta)(\bth \Dsl \theta) +\tfrac1{192}(\bth\g^\m\G^{ij}\theta)(\bth\G_{ij}\Gb\DF_\m\theta)\no\\
 &\quad -\tfrac1{192}(\bth\g^\m\G^{ab}\theta)(\bth\G_{ab}\Gb\DF_\m\theta)
 +\tfrac1{16} (\bth \Dsl \theta)^2-\tfrac1{16}(\bth\g^\m\DF^\n\theta)(\bth\g_\n\DF_\m\theta) \ . \la{33}
\ea
Here $\G^{ij}$ and $\G^{ab}$ are the antisymmetrized products of the three AdS$_5$ and five $S^5$ transverse gamma matrices. The terms in $\L_4$ of \ci{Beccaria:2026ffm} antisymmetric in the transverse indices and the terms containing $\epsilon^{\m\n}$ have vanishing expectation values on the homogeneous isotropic background and are omitted (cf. (2.46) of \ci{Beccaria:2026ffm}). 
%There are no cubic terms. The Lagrangian \rf{27},\rf{31}--\rf{33} is the one for which the measure \rf{28}--\rf{209} was constructed.

\subsection{Expectation value of the quartic Lagrangian}

Using the coincident limits \rf{220}--\rf{223} and the Majorana fermion 
Wick  contraction rule (see \ci{Beccaria:2026ffm}    and Appendix \ref{apa2}) one finds   from \rf{31}--\rf{33} 
  %for the three parts of $\L_4$ 
\ba
\langle \L_{4\rm b} \rangle &= -\tfrac34 E_x^2-\tfrac54 E_y^2 +\tfrac{21}4 E_x \rG_x-\tfrac{25}4 E_y \rG_y \ , \la{34} \\
\langle \L_{2\rm b, 2 f}\rangle & = -12\, E_\theta\, \rG_x+\big(-6E_x+10E_y\big)\rG_\theta  \ , \la{35} \\
\langle \L_{4\rm f}\rangle & = 8\, E_\theta^2 - 2\, E_\theta \rG_\theta \ . \la{36}
\ea
Some comments are in order. In \rf{34}, the $\rG_x^2$ terms cancel 
 which is the statement that there is no double-logarithmic divergence, i.e.
  the counterpart of the cancellation of the double poles in \rf{4}. 
  In \rf{35}, $\langle A^\m{}_\m\rangle = 6 \rG_x$ (the contributions of 
  derivative terms in $\langle A^{\m\n}\rangle $ cancel), while the  expectation value 
  of the  bosonic factor in the  second term in \rf{32} is proportional to the bosonic equations of motion. 
  In \rf{36}, the three one-derivative structures in \rf{33} contribute $E_\theta \rG_\theta$ terms that cancel in the sum
   (this is the counterpart of the vanishing of the second line in (B.8) of \ci{Beccaria:2026ffm}).
    The two-derivative structures  in \rf{33} give $8 E_\theta^2$ from the ``direct'' contractions and  $-E_\theta \rG_\theta$ each from the extra ``Majorana'' contraction, which is where the antisymmetric coefficient $\rk_\theta = \ha E_\theta$ in \rf{222} enters.\foot{\la{af36}Let us note that 
    if one  keeps the antisymmetric coefficient $\rk_\theta$  general  then one finds that  it enters only in 
%\be \la{361}
$ \langle \L_{4\rm f}\rangle = 8\, E_\theta^2 - 4\, \rk_\theta\, \rG_\theta $.
%and, since $\rk_\theta$ enters nowhere else (the measure trace of $\cJ_{\rm sc,f}$
%involves only the single-derivative coincident limits), 
As a result, the total 2-loop coefficient
found below acquires the extra term $2 (2 \rk_\theta - E_\theta)\, \rG_\theta$ relative
to \rf{313}. 
This is a propagator ``tadpole'' term  divergent as $t^{-1} \log t$. Thus  
 the UV finiteness of $f_2$ requires $2 \rk_\theta = E_\theta$. 
 In fact, $\rk_\theta$ is not an adjustable parameter: its value $\rk_\theta = \ha E_\theta$ is derived in
Appendix \ref{apa2} (see also the footnote  \ref{af222} for the general-mass case, where
$\rk_\theta(m) = \tfrac m2 E_\theta + \tfrac{1-m^2}{4} \rG_\theta(m)$, so that for the massless
fermion of section \ref{s4} it is the finite constant $\tfrac14 \rG_\chi$ rather than a contact term).
With
$\rk_\theta = 0$  as  was assumed in (B.13) of \ci{Beccaria:2026ffm} the 2-loop
coefficient would not be UV finite in cutoff regularization.
}

Summing up \rf{34}--\rf{36}  we thus get 
\ba
\langle \L_4\rangle = & -\tfrac14 (3 E_x^2  + 5  E_y^2  - 32  E_\theta^2)
 +\big(\tfrac{21}4 E_x -12 E_\theta\big) \rG_x-\tfrac{25}4 E_y \rG_y
 + \big(-6 E_x+10 E_y - 2 E_\theta \big) \rG_\theta \la{37}\ . 
\ea
Each  term in \rf{37} contains at least one $E$-factor, i.e.  there are no products of two propagators at coincident points. \foot{Let us compare \rf{37} with  the expression in Appendix B of \ci{Beccaria:2026ffm}. 
  There  $E_x = 2 (\hb_x + \rG_x)$, $E_y = 2 \hb_y$ and $E_\theta = 2 \hb_\theta$ are the same
   $E$-terms as here (as follows by comparing  (B.4),(B.5) of \ci{Beccaria:2026ffm} with \rf{220},\rf{221}), while the fermionic coincident constant $\rG_\theta$ there is minus the one here (the two differ by the sign conventions for the mass term).
    $\langle \L_4\rangle$ there is defined with the Lorentzian sign, i.e. it is minus the Euclidean one here. Taking these notational differences into account, (B.13) of \ci{Beccaria:2026ffm} agrees with \rf{37} up to two corrections. First, (B.13) was derived assuming that the antisymmetric coefficient $\rk_\theta$ vanishes. With the corrected value $\rk_\theta = \ha E_\theta$ of \rf{222}, the last term in (B.8) of \ci{Beccaria:2026ffm}, $- \tfrac14 \rG_\theta \rk_\theta N_\theta$ with $N_\theta = 16$, gives the extra contribution $- 2 E_\theta \rG_\theta$ in \rf{36},\rf{37}. Second, the mixed terms (the $E_\theta \rG_x$ and the $E_{x,y}\rG_\theta$ ones) in (B.13) should enter with the opposite overall sign relative to the bosonic ones. This sign is fixed by the Euclidean continuation and is consistent with the dimensional reduction formula for the mixed contribution in (4.13) of \ci{Beccaria:2026ffm}. With these two corrections the result of Appendix B of \ci{Beccaria:2026ffm} coincides with \rf{37}.}
Thus $\langle \L_4\rangle$ contains   only 
 the ``equation of motion'' terms (as  anticipated in Appendix B of \ci{Beccaria:2026ffm}), which is  the  counterpart 
 of  
the vanishing  of \rf{4}  in ``hard'' dimensional  reduction    regularization.

\subsection{Total 2-loop coefficient}

The total  2-loop coefficient $f_2$ in  \rf{2},\rf{3} 
   is the sum of the contributions of the quartic  Lagrangian  and of the measure \rf{225}
\be \la{39}
f_2 = - \langle \L_4\rangle + \rE \, \langle \cJ \rangle - 6 \rE^2 \ . 
\ee
%with $\langle \L_4\rangle$ given by \rf{37} and $  \langle \cJ \rangle$ by \rf{225}. 
Let us introduce the differences between the $E$-terms of the individual fields and the regularized  $\delta(0)\equiv \rE$  in the measure
part 
\be \la{310}
\rho_x = E_x - \rE \ , \qquad \rho_y = E_y - \rE \ , \qquad \rho_\theta = E_\theta - \rE \ .
\ee
Substituting \rf{37} and \rf{225} into \rf{39} one finds that all the terms containing $\rE$  and $E_\vp$ 
  enter only   via $\rho_\vp$ 
\be \la{313}
f_2 =  \tfrac14 (3 \rho_x^2 + 5  \rho_y^2 - 32 \rho_\theta^2)  + \big( - \tfrac{21}4 \rho_x + 12 \rho_\theta\big) \rG_x + \tfrac{25}{4} \rho_y \rG_y + \big( 6 \rho_x - 10 \rho_y + 2 \rho_\theta \big)\rG_\theta \ .
\ee
%This is the main result of this section. The complete 2-loop coefficient is a polynomial in the differences \rf{310} between the regularized $\delta(0)$'s of the individual fields and the $\delta(0)$ of the measure, containing terms quadratic in the differences and terms linear in a difference times a coincident propagator; every term contains at least one difference. 
Note that in general all  $\rho_\vp$  are UV finite  (cf. \rf{218})
 while  all  $\rG_\vp$  in \rf{2170} contain only  log UV divergences, so all power divergences  cancel.

%%%%%%%%%%%%%%%%%%%%%
\iffa 
The cancellation works as follows: the coefficient of $\rE E_\vp$ in $\rE\langle \cJ \rangle$ is $-2$ times the coefficient of $E_\vp^2$ in $-\langle \L_4\rangle$, which completes the squares, while the $\rE^2$ terms cancel as $(\tfrac34 + \tfrac54 - 8) + (-\tfrac32 - \tfrac52 + 16) - 6 = 0$; and the coefficient of each $\rE \rG_\vp$ in $\rE\langle \cJ\rangle$ is minus the total coefficient of $E \rG_\vp$ in $-\langle \L_4\rangle$ ($-\tfrac{21}4 + 12 - \tfrac{27}{4} = 0$, $\tfrac{25}{4} - \tfrac{25}{4}=0$, $6 - 10 + 2 + 2 = 0$), which turns the propagator ``tadpoles'' $E\rG$ into $\rho \rG$. No finite part of any of the coincident propagators enters on its own, and no power divergence survives. This is an algebraic identity, not a numerical coincidence, and it holds for any smearing of the propagators.
\fi
%%%%%%%%%%%%%

For the common  choice of  smearing   in \rf{213},\rf{216}  
one has $\rho_x=\rho_y=0$ and $\rho_\theta  = - {t \ov 96 \pi} + \OO(t^2) $  (see \rf{218}) so that
\be \la{314}
f_2 \big|_{t\to 0} % = - 8 r^2 + 12 r \rG_x + 2 r \rG_\theta 
= -\te  {1\ov 192 \pi^2} t ( 7 \LDR - 13) + \OO(t^2 \LDR)\ \ \to \ \ 0 \ ,
\ee
i.e. the 2-loop correction to the \adss string free energy vanishes in the limit of the removed cutoff.

%r1
For a general choice of the $E_\varphi$ the cancellation holds separately for each propagator-dependent structure within each sector, provided each measure term is multiplied by the $\delta(0)$ of the field it originates from (Appendix \ref{apA3}): the tadpoles $E_x\rG_x$, $E_y\rG_y$ of  \rf{34}  cancel against $\sum_IE_I\langle\cJ^{(I)}_{\rm mom,b}\rangle$, those of \rf{35}
 against $\sum_\alpha E_\alpha\langle\cJ^{(\alpha)}_{\rm sc,b}\rangle+\sum_IE_I\langle\cJ^{(I)}_{\rm mom,f}\rangle$, and that of \rf{36}
  against $E_\theta\langle\cJ_{\rm sc,f}\rangle$  (see \rf{a51}). 
 This is the power-cutoff counterpart of the proportionality of each of the three coefficients in \rf{4} to $1-d$.
  The purely $E^2$ terms in \rf{37},\rf{39}, however, cancel (for the common smearing) only in the sum over the three sectors.
  %\foot{For  the   choice of the  common smearing \rf{216}   the bosonic, mixed and fermionic sectors (each including its own measure part and, for the bosonic and the mixed sectors, their shares $+10 \rE^2$ and $-16 \rE^2$ of the constant term $-6\rE^2$ in \rf{39})
%    give $8 \rE^2$, $-16 \rE^2$ and $8\rE^2$ respectively. These numbers are fixed by the multiplicities of the fields only.}

%(ii) %{\it Relation to the hard dimensional reduction.}
Let us note that doing the  2-loop 
 computation as above (i.e. staying in 2  dimensions) and then formally  setting all $\delta(0)$ terms  to zero   amounts  to 
 setting $E_\vp \to 0$ and $\rE \to 0$ in \rf{37}--\rf{39}   which kills the  contributions  of the 
 measure and  of $\langle \L_4\rangle$. This is  essentially equivalent to  the hard dimensional reduction prescription 
   \ci{Beccaria:2026ffm}, which thus gives the same result $f_2=0$ as the  above power-cutoff regularization
  with the  common smearing   choice with  the measure  contribution   included. 
    In the ``soft'' version of dimensional reduction used in the main part of \ci{Beccaria:2026ffm} the evanescent coefficients $ 1-d$ in \rf{4} multiply the double poles of the products of two propagators, leaving the single pole in $f_2$ and a convention-dependent finite part. The structure \rf{37} shows that this pole is an artefact of
     the dimensional  continuation: it is produced by terms that are identically cancelled by the measure in a power cutoff 2d  regularization.

 %{\it Dependence on the smearing.}
 \

 Let us now consider a  more general case   when one does not assume 
   the  common ``massless'' smearing   in \rf{2104}--\rf{216}.\foot{For example, 
  one  may use   the  one  adapted to the 
  kinetic operator of each  field (with the corresponding value of mass  as in \rf{233}). 
  With this particular choice   the finite parts of $E_\vp$  will be   given by unequal  $\zeta(0)$ values in \rf{219}.}
  In this case   the log UV divergence  in \rf{313}  will   not cancel unless there is  a  linear relation between the 
  $\rho_x,\rho_y,\rho_\theta$, i.e. between $E_x,E_y, E_\theta$ and $\rE$. 
  %%%%%%%%%%%%%%%%%%%%%%%%%%%%%%%%
\iffa   If one  takes the $\delta(0)$ of the measure to be $E_\theta$, one finds from \rf{313} with $\rho_x = - {3 \ov 8\pi}$, $\rho_y = {1\ov 8 \pi}$, $\rho_\theta=0$
\be \la{315}
f_2 = - {1\ov 16 \pi^2} \big( 3 \LDR - \tfrac14\big) \ ,
\ee
i.e. a logarithmic divergence survives. This is the signal of the mismatch between the regularizations of the different fields and of the measure. The two prescriptions are related exactly, at finite cutoff, by a change of the Gaussian covariances: for two sets of propagators $C$ and $C'$ one has, for any polynomial $P$ in the fields,
\be \la{3151}
 \langle P \rangle_C = \big\langle e^{\mc W_{C - C'}} P \big\rangle_{C'} \ ,
\ee
where $\mc W_{C-C'}$ performs one Wick contraction with the difference of the covariances (with the Grassmann signs included); for a quartic $P$ the exponential terminates after two contractions and generates local quadratic and field-independent terms. %%% CHANGED 21/10/26 (field-attributed measure): sect. 3.3, remark (c): replaced the reference to the old Appendix A.3 by the supertrace result
\fi
%%%%%%%%%%%%%%%%%%%%
This is
%Applied to the complete action-minus-measure insertion (including the measure trace), this converts \rf{315} into \rf{314}. A numerical comparison using changed propagators but unchanged insertions, as in \rf{315}, is a different computation. In fact,
% the logarithm in \rf{315} is 
 a reflection   of a ``hybrid'' prescription in which the fields in the action are smeared with different kernels while the measure  is assumed 
 to contain the same 
   regularized $\delta(0)$ value $\rE$. 
 
 In fact, for the general smearing a consistent  (``adapted measure'') procedure 
 discussed   in Appendix \ref{apA3} 
  is to  define  each term of the measure  using  the particular  $\delta(0)$   corresponding to 
  the field whose integration 
 produced it (so that each  $\delta(0)$ in the measure  is given by  one of $E_\vp$).
 Then  
  all the propagator-dependent  $E_\vp \rG_{\vp'}$   terms in $f_2$ (cf. \rf{313}) 
  cancel automatically  for
   arbitrary  smearing 
 kernels, i.e. for  arbitrary  $E_x,E_y,E_\theta$. 
 
 Remarkably, in this case 
 the 2-loop coefficient  is given just by the square of the regularized ``supertrace'' of the  regularized identity  operator 
\be \la{999}
 f_2 = \tfrac18  \big( {\rm Str}\, E\big)^2 \ , \qquad \qquad {\rm Str}\, E \equiv \sum_{\rm b} E_\vp - \ha \sum_{\rm f} E_\vp
  = 3 E_x + 5 E_y - 8 E_\theta \ .
\ee
The combination ${\rm Str}\, E$   is   UV finite\foot{Note that ${\rm Str}\, E$ is different from the  expectation value of the quadratic Lagrangian in \rf{224}   which 
 counts the Gaussian integration
variables of the configuration-space path integral (half a contact term per second-order
boson, minus half per first-order fermionic component). ${\rm Str}\, E$
instead counts the variables of the phase-space path integral from which the measure
descends: one momentum integration per boson and %, through $(\det \mc C)^{-1/2}$, half
half a unit per fermionic component, i.e. $8$ against $8$, which is why its power-divergent
part cancels.} % The two differ by half the bosonic contact sum,
%$\langle \L_2 \rangle = {\rm Str}\, E - \ha \sum_{\rm bosons} E_I$.}
%This expression  is automatically UV   finite.
%Its  value is thus determined by the   single constant ${\rm Str}\,E$.
and is a single constant that   thus determines $f_2$.

Note that the measure contributions  cancel log  divergent terms  in \rf{37}   sector by sector, 
 just like in the ``hard'' dimensional reduction  prescription applied to \rf{4}.

For the  common smearing $E_x=E_y=E_\theta$   we find  that    ${\rm Str}\, E=0$. %  again that   $f_2=0$.
If     each field   is smeared,  e.g.,   with its own  (massive) kinetic operator  then according to \rf{233},\rf{23} 
  the combination ${\rm Str}\, E$  will be  given by  the sum of the corresponding  second  Seeley 
   coefficients. This sum  does not  vanish 
   (which  is, in fact, the origin of the 
    non-cancelling    log UV divergence  at the 1-loop order    \ci{Drukker:2000ep,Giombi:2020mhz})   giving ${\rm Str}\, E = 
   - {1\ov 4 \pi} [ 3 \cdot  2 + 5 \cdot  0   -  8\cdot ( {1\ov 4} R^{(2)} + 1) ]= - {1\ov 2 \pi}$.
    In  this case we get $f_2 = {1 \ov 32 \pi^2}$.
    To match  $f_2 = {3 \ov 32 \pi^2}$   in \rf{3}  would require ${\rm Str}\, E =\pm  {\sqrt 3 \ov 2 \pi}$.

      %which is   similar to  the value in \rf{3}  up to the  factor of 3. 
 %\foot{Let us note that the $\zeta$-function evaluation of $\langle \L_4\rangle$
  %in Appendix B of \ci{Beccaria:2026ffm} (eq. (B.25) there), apart from the sign and $\rk_\theta$ corrections mentioned above, corresponds to the massive-kinetic smearing without  accounting for the measure, which explains why the coefficients found there did not match the dimensional reduction result.}

   %{\it Role of supersymmetry.}
 Let us  note that in  \ci{Beccaria:2026ffm} it was suggested that the $E^2$ terms (in (B.13) there)
  may cancel in a ``supersymmetric''
  regularization with $E_x =  E_y=2E_\theta$. Here we  see that with the measure contribution 
   included the relevant statement is different.
  If the measure is computed using the field-adapted $\delta(0)$    factors
%With the field-attributed measure
 then the  cancellation of the   propagator-dependent $E_\vp \rG_{\vp'}$  terms  in $f_2$ 
  holds for any  choice of individual  smearing kernels  %(see Appendix \ref{apA3})
  and the  finite result  given by \rf{999}  vanishes   provided 
    $3E_x + 5 E_y=8E_\theta$.
   % of the bosonic and fermionic mode densities in \rf{999}, which is the same condition that removes the logarithmic divergence proportional to the Euler number at one loop \ci{Drukker:2000ep}. 
      %This algebraic

         It is natural to expect that the common smearing is compatible with the residual \adst\
supersymmetry of the gauge-fixed \adss string action. There is, however, a more direct argument
that selects ${\rm Str}\, E=0$ within the present construction. As shown in Appendix \ref{apA3}
(see \rf{6002}), with the field-attributed measure \rf{a51} the measure density contains a
non-covariant term proportional to ${\rm Str}\, E$. Its vacuum expectation value vanishes, but it
induces the Lorentz-violating term \rf{6003} in the 1-loop effective action for the fluctuation
fields. The two consistency requirements, (i) independence of the 2-loop result of the choice of
variables, which requires the field-attributed $\delta(0)$'s (see the end of section \ref{s2}),
and (ii) 2d Lorentz covariance of the regularized measure, together imply ${\rm Str}\, E=0$
and hence $f_2=0$ in \rf{999}.

The absence of $\delta(0)$-dependent mass shifts in the 1-loop 2-point functions
(Appendix \ref{s44}) does not by itself discriminate between the values of ${\rm Str}\, E$:
with the field-attributed measure all $E_\vp$-dependent mass terms cancel for arbitrary $E_\vp$,
and it is the non-covariant term \rf{6003} that singles out ${\rm Str}\, E=0$.
It would still be important to confirm this from the supersymmetry Ward identities of the
regularized theory, or from a manifestly symmetric regulator (cf. Appendix \ref{apA2}).

%%%%%%%%%%%%%%%%%%%%%%%%%%%%%%%%%%%%%%%%%%%%%%%%%%%%
\section{2-loop correction  in \adsc\la{s4}}

%\subsection{String as the zero-level truncation of the M2 brane}

As discussed in the Introduction, the static and $\kappa$-symmetry gauge-fixed action of the IIA string near \adst\ in \adsc is the restriction of the M2 brane action  \ci{Beccaria:2026zjj} 
 in $\AdS_4 \times S^7/\ZZ_k$  to the $S^1$-independent ($n=0$) Fourier modes on the 11d Hopf-fiber circle  of $S^7/\ZZ_k$.
 The fluctuation fields at $n=0$ are  listed in  the table in section \ref{s24}:   %(see (1.28),(1.29) of \ci{Beccaria:2026zjj}):
  one complex scalar $x$ with $m^2=2$ (transverse directions in AdS$_4$), three complex massless scalars $w^a$ (${\rm CP}^3$ directions), and 16 real fermionic components. 
    The quadratic Lagrangian is (cf. \rf{27}) 
\al{ \la{44}
&\L_2 = |\del x|^2 + 2 |x|^2 + |\del w^a|^2 + \ha \bth \Dsl \theta \ ,  %\qquad \ \ \Dsl = \slashed \nabla + M \ , 
\\
 \Gb = \ii \G_{1234} \ , \qquad \Gb^2=-1,  & \qquad   \{\Gb, \g_\m\}=0\ , \qquad  \DF_\m =\hat  \nabla_\m + \ha \g_\m M \ , \qquad \ \ M = - \Gb P_\psi 
\ , }
where $\Gb$ plays the role of  the 2d chirality matrix combined with the internal structure, %($\Gb = \ii \G_{1234}$ in the 11d notation of \ci{Beccaria:2026zjj}; $\Gb^2=-1$, $\{\Gb, \g_\m\}=0$)
 and $ \DF_\m$   is  the Killing-spinor derivative  \ci{Beccaria:2026zjj}
(on the massless fermion  $P_\chi \theta$ it reduces to $\hat\nabla_\m$, so that $[\DF_\m, \DF_\n] = - \ha \g_{\m\n} P_\chi$).

  The  16 fermionic components $\theta$    can be  organized
   into 3 complex 2d Dirac fermions $\psi^a$ with $m=-1$ ($12$ real components)  and 
   one complex 2d Dirac fermion $\chi$ with $m=0$ ($4$ real components)
    represented by  $\theta$ with the internal projectors $P_\psi$ (of rank 12) and $P_\chi$ (of rank 4): $\psi=P_\psi \theta, \   \chi = P_\chi \theta$. 
The massless 2d  fermion $\chi$ is paired by the preserved supersymmetry with a massless scalar: $\chi$ and one of the $w$'s form the short displacement multiplet of the preserved $SU(1,1|3)$, with the boundary dimensions $\Delta = \ha$ and $1$ (the pair $(\bar w^3, \chi^c)$ in  \ci{Beccaria:2026zjj}; see also \ci{Bianchi:2020hsz}).

%In the computation described below we used the   2d string 
 %action  constructed directly from the level $n=0$   truncation 
 %  of the M2 brane action of \ci{Beccaria:2026zjj}. 
The  structure of the quartic Lagrangian is the same as in \rf{26}, i.e.  there are no cubic interaction 
 terms.
 %\foot{To connect with the real-component conventions of section \ref{s2} we  may set  $
% x = {x^1 + \ii x^2 \ov \sqrt 2} \ ,  w^a = {y^{2a-1} + \ii y^{2a} \ov \sqrt 2}, \   a=1,2,3  $.} 
Explicitly, the bosonic part of the 
 quartic Lagrangian follows from (3.7)--(3.10) of \ci{Beccaria:2026zjj} by dropping all terms with $\del_3$:
\ba
&\qquad \qquad \qquad \L_{4b}=    \L_{4x}  +   \L_{2x,2w}  +  \L_{4w} \ , \no \\
 \L_{4x}=& 2|x|^4+|x|^2|\del x|^2
 -\tfrac12\big|\del^\m x\,\del_\m x\big|^2 \ , \qquad\qquad 
 \L_{2x,2w}=|\del x|^2|\del w|^2
 -|\del x\cdot\del w^a|^2
 -|\del x\cdot\del\bar w^a|^2 \ , \la{45} \\
 \L_{4w}={}& -\tfrac12|w|^2|\del w|^2
 -\tfrac12(w^a\del_\m\bar w^a)(\bar w^b\del^\m w^b)
 +\tfrac12(|\del w|^2)^2
 -\tfrac12(\del_\m w^a\del^\m\bar w^b)
 (\del_\n\bar w^a\del^\n w^b)
 -\tfrac12|\del_\m w^a\del^\m w^b|^2  .\no 
\ea
%with $\L_{4\rm b} = \L_{4x} + \L_{2x,2w} + \L_{4w}$.
    % (including the Hopf-fiber off-diagonal components of the 11d metric, which contribute to %the bosonic terms quadratic in $w$ even at $n=0$).
   The mixed $\cL_{2b,2f}$ 
 and the four-fermion  $\cL_{4f}$  terms are  found   from the $n=0$ KK level truncation of  the expressions
  in (3.12) and (3.16) of \ci{Beccaria:2026zjj}
  (see also  Appendices C and D there).
    At $n=0$     the circle component of the Killing-spinor derivative, $D_3$, becomes an algebraic matrix, and the vertices reduce to 2d expressions of the same symbolic form as \rf{32},\rf{33}, with the field-dependent 
     matrices of the M2 brane action. 
     
    % and the bosonic part \rf{45},\rf{46} was checked against (3.7)--(3.10) of \ci{Beccaria:2026zjj}.

\subsection{Expectation value of the quartic Lagrangian}

The coincident limits of the propagators are the same as in section \ref{s24}, with the obvious changes. For the complex scalars
we have (cf. \rf{214}--\rf{220})
\al{ \la{47}
&\langle x \bar x\rangle = G_x \ , \qquad \langle w^a \bar w^b\rangle = \delta^{ab} G_w \ , \qquad G_w = G_y \ , \qquad \rG_x = \tfrac{1}{4\pi}(\LDR - 2) \ , \qquad \rG_w = \tfrac{1}{4\pi} \LDR \ ,
\\
& \la{48}
\langle |x|^2\rangle = \rG_x \ , \quad \langle |\del x|^2\rangle = E_x - 2 \rG_x \ , \qquad \langle |w|^2\rangle = 3 \rG_w \ , \qquad \langle |\del w|^2\rangle = 3 E_w \ . %\qquad E_x = E_w = \rE \ .
}
For the fermions, we may set $S(m)$ to be the regularized 2d fermion propagator on the block with mass $m$  (cf. \rf{2104})
\al{ &\la{555}
 S(m) = (\hat{ \slashed\nabla} + m \Gb)^{-1} K_\theta \ , \qquad \qquad E_\theta =[K_\theta] \ , \\
& [S(m)] = - \Gb\, \rG_{\theta}(m) \ , \qquad \qquad [\hat \nabla_\m S(m)] = \ha \g_\m \big[ E_\theta - m \rG_\theta(m) \big] = - [\hat \nabla'_\m S(m)] \ , \la{49} \\
& [\hat \nabla_\m \hat \nabla'_\n S(m)] = \big\{ \big[ - \tfrac{m}{2} E_\theta + ( \tfrac{m^2}{2} - \tfrac14) \rG_\theta(m) \big] g_{\m\n} + \tfrac14 \rG_\theta (m)\, \g_{\m\n} \big\} \Gb \ , \la{410}
}
which reduce to \rf{221},\rf{222} for $m=1$.\foot{In this notation the \adss fermionic constant is $\rG_\theta = \rG_\theta(+1)$.
Note also  that  as in section 2   all   $E_\vp$ are  defined  per real component, and the multiplicities of the fields   are accounted for directly.}
%The massless block of  the antisymmetric part of the mixed Killing spinor second derivative is $\tfrac14{\rm G}_\chi\gamma_{\m\n}\bar\Gamma$, cf. (3.10) but the text there already says that both the antisymmetric parts $\tfrac14{\rm G}_\chi\gamma_{\alpha\beta}$ and the mass-type parts are involved, so this is not essential.
 Here   $\rG_\theta(m)$ is  the same as  the coincident fermionic constant $\rG_f(m)$ of (1.37) of \ci{Beccaria:2026zjj} for $m\ne0$.
 
 For $m=0$ it is defined \ci{Beccaria:2026zjj} as the $m \to 0$ limit of $[S(m)]$, i.e. of
\be \la{4491}
\rG_\theta(m) = m\, \big[ (- \hn^2 - \tfrac12 + m^2)^{-1} K_\theta \big] \ ,
\ee
which follows from $S(m) = - (\hat{\slashed \nabla} + m \Gb)\, (- \hn^2 - \tfrac12 + m^2)^{-1} K_\theta$
and the vanishing of the coincident value of the first-derivative term (cf. Appendix \ref{apa2}).
The factor $m$ in \rf{4491} suppresses the log UV divergence of the coincident value of the
inverse of the squared operator and turns its infrared divergence \rf{413} into a finite,
sign-dependent limit, see \rf{442}.

 Indeed,  the massless 2d Dirac operator  on \adst\  has no gap in its spectrum, so that the coincident-point  value of the inverse of its square is infrared divergent,
\be \la{413}
 \big[ (- \hn^2 - \tfrac12 + m^2)^{-1} K_\theta \big]\Big|_{m\to 0} \  \to\   \te {1\ov 4 \pi |m|} \ .
\ee
 Then   using  the heat-kernel smearing  we define  (keeping $t\to 0$ dependence in  $\rG_\theta(m;t)$) 
\ba  &  \rG_\chi (t) \equiv \rG_\theta (0;t) = \lim_{m \to 0^\pm} m \, \big[ (- \hn^2 - \tfrac12 + m^2)^{-1}  K_\theta\big] 
\te = {1  \ov 4\pi}\,\eta \,  e^{-t/2} \ , \  \qquad \eta \equiv  \pm 1 \ , \la{442} \\
&    \rG_\chi \equiv \rG_\chi(0) =\tfrac {1}{4 \pi} \eta  \ ,  \qquad \qquad            \rG_\psi \equiv \rG_\theta(-1; t\to 0) = - \tfrac{1}{4\pi} (\LDR -1) \ , \la{411} \ea
where $\LDR$  appeared  in \rf{217}.
The  sign $\eta$ is determined by the choice of the boundary conditions for the massless fermion at the \adst\  boundary  \ci{David:2023btq} (see section 4 of \ci{Beccaria:2026zjj} for a detailed discussion).\foot{The same result follows without  going through the  $m\to 0$   limit if  we start with 
 the explicit massless inverse with the half-spinor boundary condition $\ha(1 + \eta \g_z \Gb)$ imposed at the boundary of \adst\   represented as 
 the upper half plane: the ``image'' part of the propagator solves the homogeneous massless Dirac equation, is an eigenfunction of $-\hn^2$ with eigenvalue $\ha$ (which is the origin of the factor $e^{-t/2}$ in \rf{442}), and has the coincident value $-  {1 \ov 4\pi}\eta \Gb$.}
 
  The limiting value $\rG_\chi = \rG_\chi(0)$ is thus  an ``infrared''  quantity,  not a local heat-kernel coefficient:
  % it is independent of the finite parts of the $E_\vp $ and of the propagator normalization constants,
   it is the only quantity in the present computation that is not of the $E_\vp$ or $\LDR$ (log UV divergence)  type.
    Note that its  square $\rG_\chi^2 = {1\ov 16 \pi^2}$  that enters the 2-loop coefficient below 
    does not depend on the choice of the boundary condition.

In terms of the $E_\vp$   constants   defined  per real component of the fields we get  (cf. \rf{223}) 
\be \la{414}
\langle \bth \Dsl \theta\rangle = - 12 E_\psi - 4 E_\chi \ , \qquad \langle \bth \Gb P_\psi\theta\rangle = - 12 \rG_\psi \ , \qquad \langle \bth \Gb P_\chi \theta\rangle = - 4 \rG_\chi \  . 
\ee
The expectation values of the three parts of the quartic  string Lagrangian and their sum  are then 
  \al{
\langle \L_{4\rm b} \rangle &= -\tfrac12 E_x^2-\tfrac32 E_w^2 +3 E_x \rG_x - 6 E_w \rG_w \ , \la{415} \\
\langle \L_{2\rm b, 2 f}\rangle & = -(6 E_\psi + 2 E_\chi)\, \rG_x+\big(3 E_x - 6 E_w\big)\rG_\psi  - 3 E_w \rG_\chi \ , \la{416} \\
\langle \L_{4\rm f}\rangle & =  \tfrac12 (3E_\psi  +  E_\chi)^2  - ( E_\psi + E_\chi) \rG_\psi + \big( 2 E_\psi + \tfrac43 E_\chi \big) \rG_\chi - \tfrac32 \rG_\chi^2  \ , \la{417}\\
\langle \L_{4}\rangle    &= \langle \L_{4\rm b}\rangle  + \langle \L_{2\rm b,2\rmf }\rangle  + \langle \L_{4\rm f}\rangle \no \\
&=
-\tfrac12\big[ E_x^2 + 3  E_w^2 -  (3E_\psi  +  E_\chi)^2\big] 
 + \big( 3 E_x - 6 E_\psi - 2 E_\chi \big) \rG_x - 6 E_w \rG_w \no \\
 &\ \  \ \ \ + \big( 3 E_x - 6 E_w - E_\psi - E_\chi \big) \rG_\psi
 + \big( - 3 E_w + 2 E_\psi + \tfrac43 E_\chi \big) \rG_\chi - \tfrac32 \rG_\chi^2 \la{4171}\ . 
}
 As in \rf{34}--\rf{36}, there are no $\rG_x^2$, $\rG_x \rG_\psi$ or $\rG_\psi^2$ terms (i.e. no double  UV  logarithms).

 All the terms in \rf{415}--\rf{4171} are proportional to the EOM terms  $E_\vp$, 
  with a single exception: the term $-\tfrac32 \rG_\chi^2$ in  \rf{417},\rf{4171}
   which is the square of the finite coincident constant  \rf{411} of the massless fermion.
The origin of this term can be seen more  directly   by restricting  $\L_{4\rm f}$ 
 to the four real components of the massless  field  $\chi \equiv P_\chi \theta$. 
The  $n=0$  part of the  four-fermion  term in the  M2 brane action of \ci{Beccaria:2026zjj}  then  contains  the following  terms 
quartic in $\chi$ (products of two $\chi$-bilinears)  % bilinear in $\chi$ terms
\al{ \la{4172}
 &\L_{4\rm f}\big|_\chi = \tfrac{1}{16} \big( {\rm e}^2 - {\rm u}_{\m\n} {\rm u}^{\n\m}\big) + \tfrac16\, \bb\, {\rm e} - \tfrac{5}{32}\, \bb^2 \ ,
\qquad\qquad  {\rm e} \equiv  \bar \chi \hat {\slashed\nabla} \chi \ , \quad  {\rm u}_{\m\n} \equiv  \bar\chi \g_\m\hat  \nabla_\n \chi \ ,\quad 
\bb \equiv  \bar\chi \Gb \chi \ . 
% \la{4171}
}
The part  of the expectation values  of  the three   structures  in    $\L_{4\rm f}\big|_\chi $  that  is     proportional to $\rG_\chi^2$  is then 
\be \la{4173}
 \langle  {\rm e}^2 - {\rm u}_{\m\n} {\rm u}^{\n\m} \rangle_{\rG_\chi^2} = - 4 \rG_\chi^2 \ , \qquad
 \langle \bb\, {\rm e}\rangle_{\rG_\chi^2} = 0 \ , \qquad  \langle \bb^2 \rangle_{\rG_\chi^2} = 8 \rG_\chi^2 \ ,\qquad\qquad 
 \langle \L_{4\rm f} \rangle_{\rG_\chi^2} = - \tfrac32 \rG_\chi^2 \ . 
\ee
%so that the finite term is $-(\tfrac14 + \tfrac54) \rG_\chi^2 = - \tfrac32 \rG_\chi^2$. Both the two-derivative structures (through the Majorana-type contractions in which the antisymmetric parts $\tfrac14 \rG_\chi \g_{\m\n}$ in \rf{410} and the mass-type parts $[S(0)] = - \Gb \rG_\chi$ are involved) and the algebraic $\bb^2$ structure are essential.
The  term  $\L_{4\rm f}\big|_\chi$   in 
 \rf{4172} was also  derived directly  from 
  the general type II GS action expanded  to quartic order in the fermions  \ci{Wulff:2013kga}  and  specialized to the \adsc background.
   It was then expanded near the \adst\  solution, projected on the four massless components and continued to Euclidean signature.\foot{The 
   auxiliary world-sheet metric  was  eliminated to the same order 
   which added  $\tfrac18 (h^{(2)\m}{}_\m)^2 - \tfrac14 h^{(2)}_{\m\n} h^{(2)\m\n}$ with $h^{(2)}_{\m\n} = \tfrac14(\bar\vartheta \g_\m \nabla_\n \vartheta + \bar\vartheta \g_\n \nabla_\m \vartheta)$.}
    In terms of the corresponding type IIA string variable $\vartheta$ one finds the same coefficients $\tfrac1{16}$ and $-\tfrac{5}{32}$
    as in  \rf{4172} but with  $\tfrac{1}{16}$ instead of $\tfrac16$ in front of the $\bb\, {\rm e}$ term. The difference is 
    accounted for by a field redefinition,
\be \la{4174}
 \chi = \vartheta - \tfrac{5 }{ 48}  T^{-1} \, \bb\, \vartheta \ , \qquad \ \ \Delta \L_4 = - \tfrac{5}{48}\, \bb\, {\rm e} \ , \qquad \ \ \log J =
  \tfrac{5 }{ 24 } T^{-1} \int d^2\xi \sqrt g\ \delta(0)\, \bb \ ,
\ee
whose Berezinian $J$ % (the trace of the derivative of $\bb \vartheta$ over the four odd components is $(4-2)\bb$) 
contributes a $\delta(0)$ term. The latter  can be shown to 
 cancel  against the expectation value of $\Delta \L_4$.\foot{The contributions to the 2-loop  coefficient 
   $f_2$ are $+\tfrac56 E_\chi \rG_\chi$ from $-\langle \Delta \L_4 \rangle$ and $- \tfrac56 \rE \rG_\chi$ from the Jacobian, 
   so that they  cancel assuming common smearing, $E_\chi=\rE$   (and  in general if  the measure is defined as in Appendix \ref{apA3}).
    Since $\langle \bb\, {\rm e}\rangle_{\rG_\chi^2} = 0$, the finite term $-\tfrac32 \rG_\chi^2$ is thus  the same.}

%%%%%%%%%%%%%%%%%%%%%%%%%%
\subsection{Measure  contribution and the total 2-loop coefficient}

The construction of the measure is the same as in section \ref{s24} and gives $\mc M$ 
of the form \rf{28} with the same constant $-6$ and 
\ba  \cJ_{}  &= \cJ_{\rm b} + \cJ_{\rm f} \ , \qquad \qquad  
\cJ_{\rm b} = \cJ_{\rm mom, b}  +  \cJ_{\rm sc, b} =   - |\del x|^2 - |\del w|^2 - 7 |x|^2 - 2 |w|^2 \ , \la{418} \\
\cJ_{\rm f} & =  \cJ_{\rm mom, f}  +   \cJ_{\rm sc, f} =  - \bth \Dsl \theta + \tfrac{5}{12} \bth \Gb P_\psi \theta - \tfrac{1}{12} \bth \Gb P_\chi \theta \ , \la{419}\\
 \cJ_{\rm mom, b} & = - \tfrac52 u^2 + \tfrac32 s^2 + |x|^2 - 2 |w|^2 \ , \quad \cJ_{\rm sc, b} = 2 u^2 - 2 s^2 - 8 |x|^2 \ , \quad u^2 = 2 |\del_0 x|^2 + 2 |\del_0 w|^2 \ ,\la{4193}  \\
 \la{4191}
 \cJ_{\rm mom, f} &= \bth \g_1 \DF_1 \theta - \bth \g_0 \DF_0 \theta
 + \tfrac14 \bth \Gb P_\psi \theta + \tfrac34 \bth \Gb P_\chi \theta \ , \quad
 \cJ_{\rm sc, f} = - 2\, \bth \g_1 \DF_1 \theta + \tfrac16 \bth \Gb P_\psi \theta
 - \tfrac56 \bth \Gb P_\chi \theta .
\ea
 Here there is the  same derivative structure as in the \adss case  in  \rf{29},\rf{209} 
  and the mass-type terms now
distributed between the massive and massless    fermion sectors  (their expectation values are given in \rf{a5}).
Then  we find that 
\be \la{421}
\langle \cJ \rangle = - E_x - 3 E_w + 12 E_\psi + 4 E_\chi - 5 \rG_x - 6 \rG_w - 5 \rG_\psi + \tfrac13 \rG_\chi \ .
\ee
Assuming that the measure is defined    using  common smearing  for all fields, i.e. $\delta(0) =\rE$    as  in  \rf{39}
the 
 total 2-loop coefficient 
\be \la{422}
 f_2 = - \langle \L_4\rangle +  f_{2\cal M} \ , \qquad \ \ \      f_{2\cal M} = \rE\langle \cJ\rangle - 6 \rE^2 \ ,
\ee
 takes the following form  when written in  terms of $\rho_\vp = E_\vp - \rE$  as in \rf{310},\rf{313}
\ba
f_2 ={}& \tfrac12\rho_x^2+\tfrac32\rho_w^2-\tfrac92\rho_\psi^2-3\rho_\psi\rho_\chi-\tfrac12\rho_\chi^2-(3\rho_x-6\rho_\psi-2\rho_\chi)\rG_x+6\rho_w\rG_w\no\\
 &-(3\rho_x-6\rho_w-\rho_\psi-\rho_\chi)\rG_\psi-\big(-3\rho_w+2\rho_\psi+\tfrac43\rho_\chi\big)\rG_\chi+\tfrac32\rG_\chi^2 \ . \la{423}
\ea
All the terms containing the $E_\vp$  and $\rE$  enter only through  the differences $\rho_\vp$
as in \rf{313}. 
 For the common massless  smearing, i.e. $E_x=E_w =\rE$,  
 we have 
 $\rho_x = \rho_w=0$, $\rho_\psi= \rho_\chi \equiv  r(t) = -{t\ov 96 \pi} + \OO(t^2)$, so that, as in \rf{314},
\be \la{4231}
 f_2(t) = \tfrac32 \rG_\chi(t)^2 - 8 r^2 + 8 r \rG_x + 2 r \rG_\psi - \tfrac{10}{3} r \rG_\chi(t) \ . 
\ee
Thus  in the limit of the removed cutoff $t\to 0$,   instead of 0 in \rf{314}  we end  up  with 
a  finite term\foot{The result \rf{424} was checked by evaluating the expectation value of the reduced ($n=0$)
action in two independent ways: (i) by a direct 2d computation with explicit Grassmann-polynomial
contractions, and (ii) by constructing the reduced action directly from the 11d form of the
fermionic part of the M2 brane action \ci{Beccaria:2026zjj}, with explicit $32\times 32$ gamma
matrices, and computing its expectation value by matrix methods. Both evaluations use the same
reduced action; its massless-fermion part was in addition checked against the 10d form of the
type II GS action, see \rf{4172}--\rf{4174}.}
 % the single finite term survives:
\be \la{424}
f_2 = \tfrac32 \rG_\chi^2 = \tfrac{3}{32 \pi^2} \ .
\ee
Let us note that  the same  value 
 is obtained  also if   one uses the  ``hard'' dimensional reduction  prescription as in \ci{Beccaria:2026ffm}
  in which   all propagators  are continued to $1+d= 2-2\epsilon$
  dimensions after doing all contractions in 2d.  In this case 
  all $\delta(0)$  factors  are set to zero (i.e. $E_\vp \to 0$)
  while  $\rG_\chi$ is still well defined:  for $m=0$ the 
   coincident fermionic propagator  in $\AdS_{1+d}$   is 
   given by  \ci{Beccaria:2026ffm}\foot{Here we use  sign convention opposite to the one in   \ci{Beccaria:2026ffm}, 
   i.e. $ \rG_\theta(m)  \to -  \rG_\theta(m) $.}
\be \la{425}\te 
  \rG_\theta(m) = { \G(\frac{1-d}{2})\, \G({1\ov 2}  + \Delta) \ov (4\pi)^{(d+1)/2}\,  \G({1\ov 2}  - d + \Delta)} \ , \qquad \Delta = \ha d  + m \ , \qquad \qquad
  \rG_\theta(0) = \frac{\G ({1+d \ov 2})}{(4\pi)^{(d+1)/2}} \ \to \ \te {1\ov 4\pi} 
  \ .
\ee
Thus $ \rG_\theta(0) =\rG_\chi$ has no pole,  and we reproduce  again \rf{424}. % $\rG_\chi^2 = {1\ov 16 \pi^2}$.

 If we instead of common smearing $\rE$ in the measure part   we 
   use  ``field-adapted''   prescription  discussed in the previous section and   Appendix \ref{apA3}  
   (i.e.  with $\rE\to E_\vp$ in the   corresponding parts of the measure contribution) 
   then  instead of \rf{423} we get (cf. \rf{999})
   \be \la{777}
   f_2 = \tfrac18 ({\rm Str}\, E)^2 + \tfrac32 \rG_\chi^2  \ ,  \qquad \qquad \qquad 
   {\rm Str}\, E \equiv  2 E_x + 6 E_w - 6 E_\psi - 2 E_\chi \ , 
   \ee
   where all $E_\vp$   are  constants  (with their  quadratically divergent parts cancelling in  the combination ${\rm Str}\, E$).
    The   choice  of their  finite parts defines the regularization prescription. 
   The  presence of the finite term $\tfrac32 \rG_\chi^2$  is universal  not depending on a  choice of UV regularization. 
   
   To summarize,  like in \adss   case the  result for 2-loop coefficient is  UV finite due to cancellation  between the 
   regularized action  and measure contribution.  This cancellation does not depend on regularization prescription 
    if the  measure part is defined in the  field-adapted way. 
    What is left, apart from the term $\tfrac18 ({\rm Str}\, E)^2$ in \rf{777} which vanishes for
${\rm Str}\, E=0$, is the contact-term free part of $\langle \L_{4\rm f}\rangle$, which in the \adss case was
     zero because all the fermions were massive, and which in the \adsc case is non-zero because of the
      massless fermion. 
      %The result \rf{424} is thus an infrared effect  insensitive to the finite parts of the propagators and of the $\delta(0)$'s.

 Like  in \adss   case the natural choice of the regularization prescription is the common  smearing one  where all $E_\vp$ are 
     fixed  to be the same  so that $\rho_{\vp}=0$ in \rf{423} or ${\rm Str}\, E =0$ in \rf{777}. 
     In this case the result agrees also  with the result  found in the ``hard'' dimensional
reduction regularization prescription.
As in the \adss case, this choice is required by the 2d Lorentz covariance of the
field-attributed measure: the bosonic part of the non-covariant term \rf{6002} is present also
here, with ${\rm Str}\, E$ given in \rf{777} (see Appendix \ref{apA3}).

%%%%%%%%%%%%%
\subsection{Comparison with the M2 brane result}

Starting with the M2  brane path integral in $\AdS_4 \times S^7/\ZZ_k$ expanded near $\AdS_2 \times S^1$
 minimal surface,   ref.\ci{Beccaria:2026zjj} found the 
 following result for the   corresponding 2-loop contribution to free energy  (for $k>2$)
 \al{& F_2 = \rT_2^{-1} f_{2\rm M}(k) \ , \qquad \qquad   f_{2\rm M}(k)  = - 64 \pi^2 k^{-1}  \langle \L_4\rangle_{_{\rM2}}\ , 
 \qquad 
 f_2 = \tfrac14  k^{-1}   f_{2\rm M}(k)  
 \ , \la{333} \\
 % organizes the 2-loop contribution as a ``perfect square'' (see (3.11), (3.13), (3.17) and (4.1) there)
 \la{426}
&\langle \L_4\rangle_{_{\rM2}} = - \tfrac38 \cX^2 \ , \qquad  \qquad \cX = \cX_\rmb - \cX_\rmf = \tfrac{1}{ 4 \pi} k  \sum_{n=-\infty}^\infty \X_n \ ,
\\ \la{624}
&   \X_n + \X_{-n} = - \te {1\ov \pi}   \  \ (n\ne 0)  \ , \qquad  \qquad \qquad \X_0 = - 2 \rG_\chi \ .
}
Here $\cX_\rmb$ and $\cX_\rmf$ are the bosonic and the fermionic combinations of the coincident 
3d propagators, and $\X_0$ is the contribution of the $n=0$ level in which only the massless fermion coincident constant $\rG_\theta(0) = \rG_\chi$  in  \rf{411} appears. 

The 3d computation  in  \ci{Beccaria:2026zjj} was done in the $\zeta$-function regularization, in which all the $\delta^{(3)}(0)$ terms vanish identically (there are no finite parts in odd dimensions, see Appendix \ref{apB}), so that neither the measure nor the contact terms from  the action contribute. 

Isolating  the $n=0$ level    contributions   to $\langle \L_4\rangle_{_{\rM2}} $ one finds, equivalently,  that 
\be \la{427}
\langle \L_{4\rm b}\rangle_{_{\rM2,0}} = 0 \ , \qquad \langle \L_{2\rm b, 2f}\rangle_{_{\rM2,0}} = 0 \ , \qquad \langle \L_{4\rm f}\rangle_{_{\rM2,0}}= - \tfrac38 \cX_{_{\rm f,0}}^2 = - \tfrac{3 k^2}{32 \pi^2}\, \rG_\chi^2 \ .
\ee
As a result,   using the relation between the  string and M2  brane  2-loop coefficients in \rf{333} 
we recover the expression $f_2= \tfrac32 \rG_\chi^2$  in \rf{424}.  The $\zeta$-function treatment of the 3d theory thus  corresponds, at the $n=0$ level, 
to setting all the $E_\vp$ to zero in \rf{415}--\rf{417} or \rf{423},\rf{777}, 
 in agreement with the 2d cutoff regularization 
  computation with the measure included.

  This is a non-trivial consistency check of the 2d computation: the 2d expectation values \rf{415}--\rf{417} were computed with the contact terms kept, and the $E_\vp$-independent part of their sum agrees with the $n=0$ truncation of the 3d perfect square.

As discussed  in \ci{Beccaria:2026zjj}, for  the  supersymmetry-consistent  choice of the value of   $\rG_\chi$
 the  result  for $\cX$ and thus for $f_{2\rm M}$ vanishes. 
In more detail, 
the difference between the string result \rf{424} and the vanishing M2 brane 2-loop  is due to the contribution of $n\ne 0$ 
M2-brane modes.  Using the Riemann $\zeta$-function to regularize the sum over $n$ \foot{The same value of the regularized sum is obtained for any smooth cutoff like $e^{-\epsilon |n|}$ or  $e^{-\epsilon n^2}$  on the level number. 
For example, 
  $\sum^\infty _{n=1} e^{-\epsilon n^2} = {\sqrt \pi \ov 2 \sqrt \epsilon} - \ha + \OO(\epsilon)$    which  gives  again $-\ha$ 
  after removing the   power divergence.}
\be \la{428}
\sum_{n \ne 0} \X_n = - \tfrac{1}{ \pi} \sum_{n=1}^\infty 1 = -   \tfrac{1}{ \pi}   \zeta(0) =  \tfrac{1}{ 2\pi}  \ , \qquad
\cX =\te  - {1 \ov 2\pi} k  \big( \rG_\chi - {1\ov 4\pi}\big) \ ,\qquad
f_{2\rm M} = 6 k  \big( \rG_\chi - {1\ov 4\pi}\big)^2 \ , 
\ee
we get   zero  for  $\rG_\chi = \rG_\theta(0^+) = {1\ov 4\pi}$  \ci{Beccaria:2026zjj},
\be  
\langle \L_4\rangle_{_{\rM2}} =0 \ , \qquad \qquad  f_2 = \tfrac14  k^{-1}   f_{2\rm M}(k)=0 \ .     \la{888}\ee
A heuristic reason why the higher-level M2 brane modes (with circle momentum $p_n = \ha k n$;
their masses are shifted relative to $p_n$, see \ci{Beccaria:2026zjj})
%A heuristic reason   why the higher-level  massive ($m_n = \ha k n $)  M2 brane  modes 
 do not decouple is the universality of the pair contribution $\X_n + \X_{-n} = - {1\ov \pi}$: each pair of levels contributes the same finite, $n$-independent amount, so that the sum over levels is a regularized ``$\sum 1$'', i.e. effectively a $\delta(0)$-type quantity on the  1d {circle}. In this sense the extra term  that cancels $\rG_\chi^2$ term in \rf{424} 
  is a 3d measure-type contribution, 
  which is seen from the 2d point of view as a finite local counterterm (see also Appendix \ref{apB}).
 % Similar non-commutativity of   taking the radius of $S^1$ circle  to zero (i.e. $k \to \infty$)  and 
 %  quantizing the M2  brane theory  was observed in \ci{Seibold:2024oyr,Beccaria:2025xry}. 

The  contribution of  massive KK  levels to \rf{426} may be  interpreted as  additional terms that  modify  the string theory result
 in \rf{422},\rf{424} to  (cf. \rf{333},\rf{428})
\be \la{431} 
f_2= \te  {3\ov 2} \rG_\chi^2  \ \  \to\ \  {3\ov 2} \rG_\chi^2    + \Delta f_2, \qquad  \ \  \Delta f_2 =  {3\ov 2} \big( -  \tfrac{1}{2\pi} \rG_\chi + \tfrac{1}{16 \pi^2} \big) \ . \ee
Then $\Delta f_2$ may be interpreted as   contribution  from  local terms  induced in the effective  type IIA 
 string Lagrangian    after 
integrating out   massive M2 brane modes 
\be \la{430}
\Delta \langle \L_4 \rangle =-  \Delta f_2 =  \te 
{3 \ov 4 \pi} \rG_\chi - {3\ov 32 \pi^2} = q_\chi \langle \bth \Gb P_\chi \theta \rangle + q_0 \ , \qquad \ \ q_\chi = - {3\ov 16 \pi} \ , \qquad q_0 = - {3\ov 32 \pi^2} \ .
\ee
This  relation by itself does not show that the non-zero levels actually induce  a local quadratic operator with the coefficient $q_\chi$.
 This stronger statement can be checked directly by keeping two external zero-level fermions $\chi= P_\chi \theta $ as external legs 
 and contracting a  pair of fluctuations with Fourier levels $(n, -n)$, $n>0$  in the M2 brane action
  with the regulator  factors $e^{-t(-\nabla^2 + p_n^2)}$  and  $e^{-t(-\hn^2 + p_n^2)}$, \ $p_n = \ha {k n }$ 
  and then summing over $n$.\foot{The 2-loop contribution in \rf{430} comes from   one loop of massive   KK modes and then  one loop of  massless $\chi$ modes.}
  %  extending the 2d prescription by the circle momentum. 
   The coefficient of $\bth \Gb P_\chi \theta$ in the resulting 1-loop   effective action   is  then 
   $ \hat q_\chi (t) =  {1\ov 4 k \sqrt{\pi t}} + { q}_\chi   + ..  , \ \ { q}_\chi  = - {3\ov 16 \pi} \ , $
% and append to the next sentence ("This leads to \rf{430} after subtracting power divergence."):
where the finite part is obtained by subtracting the displayed power divergence and no finite
constant (the  sum over $n$ is evaluated for $k > 3$, so that $p_n > \tfrac32$ for all $n \geq 1$; this
suffices for the string limit $k \to \infty$).
This  leads to \rf{430} after subtracting power divergence. 
%where the finite part is obtained by subtracting the displayed power divergence and no finite constant, and comes from the endpoint term of the Euler-Maclaurin summation of the level-dependent profile of the pair contribution (which reproduces the value $-{1\ov \pi}$ of \rf{426} at fixed level and the value $\ha$ of $-\zeta(0)$ for the finite part of the sum). 
Thus the non-zero levels induce a    local  quadratic term in the string  action 
\be \la{4302}
 \Delta S_2= { q}_\chi T^{-1}  \int d^2\xi \sqrt g\ \bth \Gb P_\chi \theta \ , \qquad \qquad  { q}_\chi  = -\te  {3\ov 16 \pi} \ . 
\ee
%with precisely the coefficient $c_\chi$ of \rf{430}, obtained without using the vacuum answer. 
The field-independent constant $q_0$ in \rf{430} is not reproduced by \rf{4302}: it comes from the
diagrams in which both contracted lines are  massive  KK modes, and from the 2d point of view it is
a vacuum ($[\delta(0)]^2$-type) contribution of the KK tower, in line with the interpretation of the
sum over levels as a regularized ``$\sum_n 1$'' given above. 
Note that the value of $q_0$ is fixed here by comparison with the M2 brane vacuum result
\rf{426}--\rf{428}; the diagrams with both contracted lines at non-zero levels have not been
re-evaluated independently in the proper-time prescription used for \rf{4302}.

%%%%%%%%%%%%%%%%%%%%%%%%%%%%%%%%%%%%%%%%%%%%%%%%%%%%
\section{Concluding remarks\la{s5}}

To summarize,  using a power (heat-kernel) cutoff  in 2 dimensions and including the contribution of the canonical measure in the gauge-fixed  path integral of  the GS string, we have found that the 2-loop correction to the  free energy   near \adst\   
surface  is UV finite in both the \adss and the \adsc cases. 
%The  non-vanishing UV pole  found   in \adss case   may thus be interpreted as  an artifact of a  direct 
% application of dimensional regularization. 

 In the same regularization (``common smearing'')    prescription for all the fields    the resulting expressions for 
the 2-loop coefficient in  the  string partition function are  given by 
\be \la{50}
 f_2 \Big|_{{\rm AdS}_5 \times S^5} = 0 \ , \qquad\qquad  \qquad f_2\Big|_{{\rm AdS}_4 \times {\rm CP}^3} = \tfrac32 \rG_\chi^2 = \te {3 \ov 32 \pi^2} \ .
\ee
These  agree  with the  expressions  found  using  the  ``hard''  dimensional regularization   prescription  under which 
all  contractions and summations are done directly  in 2 dimensions and only after that the Green's functions  are continued to $2-2\epsilon$ dimensions (so that, in particular,  all $\delta(0)$ terms are set to  zero as usual in dimensional regularization). 

The  2-loop result \rf{50}  is UV finite, so that the issue of the residual UV 
 pole found in \adss case in \ci{Beccaria:2026ffm} is resolved: that pole was an artefact of the dimensional 
  regularization, produced by evanescent coefficients multiplying double poles, i.e. by terms which in a 2d  cutoff  regularization are cancelled by the measure, and the ambiguous finite part accompanying it in \ci{Beccaria:2026ffm}  was not meaningful.

We have seen that all the terms depending on the regularized $\delta(0)$'s  that come from the action 
cancel against  similar  contributions coming from the  measure 
 if we use  the common smearing prescription  (cf.  \rf{313} and \rf{423}).

More generally, if the $\delta(0)$ in each term  in the measure is regularized in the same way as for  the corresponding field in the action,
 then  the  log divergent propagator-dependent  $\rG_\vp$ terms cancel for arbitrary choice of individual smearing kernels 
 and the 2-loop coefficient is   given simply   by  
 the square of the regularized supertrace of the identity  (see \rf{999} , \rf{777})
 \be \la{501}
 f_2 \Big|_{{\rm AdS}_5 \times S^5} =  \tfrac18 ({\rm Str}\, E)^2
  \ , \qquad\qquad  \qquad f_2\Big|_{{\rm AdS}_4 \times {\rm CP}^3} = \tfrac18 ({\rm Str}\, E)^2 + \tfrac32 \rG_\chi^2  \ .
\ee
 The only remaining scheme dependence is then 
  through ${\rm Str}\, E$, i.e.  the  difference between the  
   regularized  numbers of bosonic and fermionic modes. The latter  vanishes for the common smearing  when \rf{501} reduces to \rf{50}.
   The condition ${\rm Str}\,E=0$ is,  in fact,  required by the 2d Lorentz covariance of the
field-attributed measure, see \rf{6004}.
  
  % In \adss these are the only terms present in the expectation value of the quartic action, which gives $f_2 = 0$; in \adsc the finite term $\tfrac32 \rG_\chi^2$ of the massless fermion contribution is the single exception, and it survives.
 
 The vanishing of $f_2$ in the  \adss  case in \rf{50}  represents  a puzzle: 
 %On the negative side, $f_2=0$
it   is inconsistent with the expansion \rf{1} of the exact  Bessel function expression for the planar 
$\ha$ BPS Wilson loop in  the dual $\N=4$ SYM theory, which requires $f_2 = {3\ov 32\pi^2}$.
 Fine tuning ${\rm Str}\, E$ in \rf{501}  to match this value does not appear to be natural. 
 
One   heuristic   explanation for the vanishing 2-loop result
is that, like in the ABJM theory case \ci{Gautason:2025plx,vanMuiden:2026nsp,Bobev:2026gir},
 one may invoke the ensemble conjecture
  which explains why the 2-loop M2 brane correction to the Wilson loop in $\AdS_4 \times S^7/\ZZ_k$ 
   vanishes as   found  in \ci{Beccaria:2026zjj}. 
  The idea that the ensemble conjecture may apply also to the type IIB case was suggested in
   \ci{vanMuiden:2026lno}.\foot{The  discussion in \ci{vanMuiden:2026lno} 
   was  in the context of  the Schur-twisted background and the Schur index. It is not clear   if a similar reasoning 
    may  be   extended  to the  planar    circular Wilson loop  case.} 
  %The vanishing of the 2-loop correction for the \adss string found here may be an indication in this direction. Let us note that within the local 2d computation there is no candidate for a non-zero finite contribution in the \adss case: all the fermions are massive, so that no analog of the finite coincident constant $\rG_\chi$ exists, and the complete polynomial \rf{313} contains no finite part. 
  
  Another possible  explanation  is that the type II GS  string path integral  measure contains 
  an extra   contribution that  so far was not  accounted for. 
  An  indication of this was already observed  at 1-loop order: the string partition function  for \adst\ in $\AdS_n \times M^{10-n}$ 
   contains  \ci{Drukker:2000ep,Giombi:2020mhz}   a  universal    constant  log UV 
  divergent  term  ($\log  Z =    B_2 \log \Lambda, \  B_2={1\ov 4 \pi} \int d^2 \xi \sqrt g R^{(2)} =1$) 
    that should be cancelled 
   by a counterterm to   \ci{Giombi:2020mhz}  reproduce  the  1-loop   prefactor $\sqrt T$   in \rf{1}. 
  
   As was found in \ci{Giombi:2023vzu}   this prefactor is automatically  accounted for if one  defines  the \adsc  string path integral as 
   a  limit of the M2  brane path integral on $\AdS_2 \times S^1$. In the latter the  1-loop  contribution  represented as a sum over 
    KK modes  contains the \adst\  $\log \Lambda$ term with the   coefficient $B_2= \sum_n 1=  1 + 2 \sum^\infty _{n=1} 1 = 0$  so that 
    the  string ($n=0$)  contribution is cancelled   by  the  $2\zeta(0)$  contribution of the higher  M2  brane modes. 
    The non-trivial  part of the 1-loop M2  brane partition function  contains   the factor $[\sin {2\pi \ov k}]^{-1}$
    that  in the limit of large $k$  reproduces  the universal  \ci{Giombi:2020mhz}    string tension prefactor in \rf{41}.
    
     A related indication comes from the 1-loop correction to the current (tilt operator) 2-point function
    computed in \ci{Castellani:2026wwh}: as discussed in Appendix \ref{s44}, a 2d computation without
    evanescent terms would give the Bremsstrahlung coefficient ${\rm B}_1 = -{1\ov 2\pi^2}$ corresponding to
    a Wilson loop without the 1-loop  $\sqrt T$ prefactor, while the prescription $N_y = 5 + 2\eps$ used there to
    match the exact ${\rm B}_1 = - {3 \ov 8\pi^2}$ effectively restores it.

    Moreover, as we have seen above  (cf. \rf{426},\rf{888},\rf{430})
      the  M2  brane path integral also  supplies   extra    contributions that  added  to the 
    \adsc  2-loop  result computed directly from the  type IIA string action 
     make it  vanish, in agreement   with the localization   prediction in \rf{118}. 
    Indeed, as already 
 % The situation in the \adsc  case  
%We found that here the 2-loop correction is non-zero, $f_2 = {3 \ov 32\pi^2}$, while matching with the localization result expressed in terms of the ``shifted'' string tension $\sim \sqrt{\l - {1\ov 24}}$ in \rf{41} 
%requires it to vanish. 
 discussed in \ci{Beccaria:2026zjj}, the string result may be {\it defined} as the large $k$ limit of the M2 brane result, and then it indeed vanishes due to the extra regularized contribution of the KK modes.
 In general,  the limits of the truncation to the zero (string)  level  and of the 11d circle shrinking to zero size do not commute, as 
 observed  at the 1-loop level \ci{Giombi:2023vzu,Seibold:2024oyr}. 
The present 2-loop example is a new instance of this non-commutativity, 
with a   finite   extra  contribution  \rf{430},\rf{4302}  % of the measure type
 supplied by the KK tower.\foot{The two-point function computation  in  Appendix \ref{s44} shows that this term is not optional: without it the massless fermion acquires a finite mass while its scalar partner does not, in conflict with the residual supersymmetry. The term induced by the KK tower \rf{4302} 
  removes  this mass term  assuming that  $\rG_\chi=\rG_\theta(0^+)$.}
  %\foot{One may ask if, while in the M2 brane case one needs to set $\rG_\theta(0) = \rG_\theta(0^+) = {1\ov 4\pi}$ for consistency with supersymmetry as discussed in \ci{Beccaria:2026zjj}, in the string case one is to use instead a ``symmetric'' prescription, i.e. $\rG_\chi = \ha [ \rG_\theta(0^+) + \rG_\theta(0^-)]=0$, which would directly give $f_2=0$. But why? Such a prescription does not correspond to either of the two admissible boundary conditions for the massless fermion (cf. the discussion in section 4 of \ci{Beccaria:2026zjj}). Note also that the square of the averaged propagator is not the average of the squares, so that averaging the two 
  %branches is an additional prescription rather than a symmetrization of the result.}

We thus have the  evidence that the M2 brane path integral, where the issue of the local measure
 is less ambiguous than in the string case (recall that  $\delta^{(3)}(0)$ terms in 3 dimensions have no finite parts in the analytic regularization, cf.  Appendix \ref{apB}), automatically produces some extra measure-type factor in the type IIA 
 GS string path integral. 
Once this factor is included, it effectively restores the ``commutativity'' of the large $k$ and string-level  truncation limits. 
%Such a factor would then cancel the residual finite $\rG_\chi^2$ term in the \adsc case and 
%would restore the agreement with the ABJM localization result. 

%As we have seen in section \ref{s4}, this factor %is not of  a $\delta(0)$ (ultralocal)  type 
%regularized in the same way as the rest of the theory. It 
%is a finite local term  %with a 
%definite coefficient, 
%whose quadratic part is \rf{4302}, and its coefficient is not determined by the 2d theory alone
 %(see Appendix \ref{apA2}). 
 
 Unfortunately,  an analog of M2   brane  path integral  for the type IIB   GS string 
 is not known  at present. 
One may  speculate that if an additional    factor were 
  present also  in the type IIB in \adss  string path integral 
  it would then  provide an extra finite 2-loop contribution that would make the  \adss string
  result agree with the  dual SYM  theory result  in  \rf{1}, i.e. $f_2=\tfrac{3}{32\pi^2} $.\foot{Curiously, this is the same 
   value  as $f_2$   in \rf{50} found directly in \adsc   type IIA  string case. An extra  contribution 
      coming from the ``correct'' definition of type II string path integral is  then opposite in sign 
       in  IIA and IIB   cases:  the required  correction is  $\Delta f_2 = \tfrac{3}{32\pi^2} $ in IIB case and 
       $\Delta f_2 =- \tfrac{3}{32\pi^2} $ in IIA case.}

%We leave the interpretation of the  above  string 2-loop results open at the moment.  
To conclude, what was achieved  in this paper 
 is the resolution of the puzzle of the remaining 2-loop UV divergence in \adss case in \ci{Beccaria:2026ffm}, and also a novel understanding of the importance of the account of the measure contribution once one uses a power cutoff like the heat-kernel or the $\zeta$-function one.  
 The canonical measure of the gauge-fixed GS string is precisely the Jacobian for which the contact log divergent  part  of the 2-loop contribution 
 cancels, so that the 2-loop correction is UV finite and is given by the finite part of the combined
contribution of the action and of the measure, \rf{501}.

    It would be important to understand whether a consistent quantization of  type II  GS string 
 (e.g. based on a better understanding of the role of  $\kappa$-symmetry beyond the leading order 
   or on detailed  relation to the parent M2 brane path integral) 
   leads to an extra  finite local correction to the  IIA \adsc measure like \rf{4302} 
    and also 
     a similar  finite term in the  IIB \adss case.\foot{\la{f1}One may also  contemplate using 
another possible approach: to start with the pure-spinor formulation of the \adss superstring in \ci{Berkovits:2000fe},  based on a 
 sigma model on the same supercoset in which the fermionic momenta are independent fields, so that the measure of the matter sector is flat and manifestly covariant and there is no $\kappa$-symmetry gauge fixing issue 
  (cf.  Appendix \ref{apA2}). The 1-loop computations near 
   classical string solutions were done in this formalism in \ci{Cagnazzo:2012uq}, where the semiclassical equivalence with the results found from the  GS  action 
    was demonstrated. 
    A 2-loop comparison would require, in particular,  the knowledge of the 
     full ghost and measure contributions and  also require dealing   with  the 
      higher-derivative fermionic propagators,   which  appears to be   technically much harder than in the GS string approach. 
}

\section*{Acknowledgements}
We thank M. Beccaria, S. Kurlyand  and J. van Muiden for  the  collaboration on \ci{Beccaria:2026ffm,Beccaria:2026zjj}  and 
  important   discussions  and comments on the draft. 
This work was supported by the STFC grant ST/X000575/1  and  STFC Consolidated Award UKRI4230: Abdus Salam Centre for Theoretical Physics at Imperial College London.
Some of  the computations in  this paper were assisted  by 
 ChatGPT   and Claude models. 
%Some

\appendix
%%%%%%%%%%%%%%%%%%%%%%%%%%%%%%%%%%%%%%%%%%%%%%%%%%%%
\section{Details of the construction of the measure \la{apA}}

%\subsection{Derivation of the measure contribution  \la{apA}}

%\subsubsection*{Bosonic momenta}
In a local orthonormal frame on \adst\  let $u^I = \del_0 z^I$ and $s^I = \del_1 z^I$ for the 8 bosons $z^I$. The Euclidean Lagrangian 
has the form 
 \be \L = \ha u^2 + \ha s^2 + V(z) + T^{-1} \L_4(u,s,z,\theta) + ...\ . \ee
 The Hamiltonian path integral \rf{20} with the Liouville measure is converted into the Lagrangian one by integrating over the momenta 
  $p$. Expanding around the  Gaussian theory  as described in section \ref{s2}, one finds at order $T^{-1}$ the velocity Hessian \rf{24} multiplied by $\delta(0)$ and the constant $c$ in  \rf{203}, \rf{206} multiplied by $[\delta(0)]^2$. For the 
   \adss   Lagrangian  \rf{31},\rf{32}   we have %, with the other arguments set to zero,
\ba \la{a0}
 &\L_{4\rm b} (u, 0,0,0) = - \tfrac18 (u^2)^2 \ , \qquad  \qquad \Delta_u = \sum_{I=1}^8 {\del^2 \ov \del u^I\del u^I} \ ,\\
& \cJ_{\rm mom, b} = - \tfrac52 u^2 + \tfrac32 s^2 + \tfrac34 x^2 - \tfrac54 y^2 \ , \qquad \qquad    \Delta_u^2 \L_{4\rm b} = -{80}\ , \la{a1} \\
& \cJ_{\rm mom, f} = \ha \sum_I {\del^2 \L_{2\rm b, 2f} \ov \del u^I \del u^I} = \bth \g_1 \DF_1 \theta - \bth \g_0 \DF_0 \theta  - \tfrac14 \bth \Gb\theta \ , \la{a2}
\ea
where the first two terms in $\cJ_{\rm mom,f}$ have vanishing expectation value on the homogeneous \adst\ background.

%\subsubsection*{Fermionic constraints}
The fermionic Lagrangian is first order, $\L_{\rm f} = \theta^T A(\theta, z) \dot\theta + ...$, with $A = A_0 + T^{-1} A_2$, where $A_0 = \ha C \g_0$ ($C$ is the charge conjugation matrix) comes from the free term and $A_2$ from the $\dot\theta$-dependent terms in $\L_{2\rm b, 2 f}$  and in $\L_{4\rm f}$. % The bosonic variables are first partially Legendre transformed. 
The second-class constraints are \rf{204} 
and their matrix \rf{205} is $\mc C = A + A^T$ for the bosonic part and the corresponding Grassmann-differentiated expression for the fermionic part.
 The free odd block is $\mc C_0 = C \g_0$, acting on the sixteen physical components. As a result, 
 we get  $(\det \mc C)^{-1/2}$    factor per point, i.e. \rf{25} in the exponent.

For the \adss action, the bosonic part of $\mc C_2$ comes from $\tfrac14 A^{\m\n} \bth \g_\m \DF_\n \theta$ in \rf{32} with $\n=0$. From the Euclidean $A^{\m\n}$ of \rf{32} ($\tr\, {\bf 1} = 16$)
\ba \la{a3}
& A^{00} = x^2 - \ha u^2 + \ha s^2 \ , \qquad \mc C_{2 \rm b} = \ha A^{00} \mc C_0 +  {\rm traceless}   \ , \no \\
& \cJ_{\rm sc,b} = - \ha \tr(\mc C_0^{-1} \mc C_{2 \rm b}) = - 4 A^{00} = 2 u^2 - 2 s^2 - 4 x^2
\ea
%(the $\epsilon^{\m\n}$ and $A^{01}$ terms are traceless),
 in agreement with \rf{212}. The fermionic part comes from the terms with $\dot\theta$ in \rf{33}: the three mass-type structures
 cancel, and the two-derivative ones contribute 
\be \la{a4}
 \cJ_{\rm sc,f} = - 2\, \bth \g_1 \DF_1 \theta + \tfrac18 \bth \Gb \theta \ ,
\ee
which together with \rf{a2} gives \rf{2121} and hence \rf{209}. The absence of a $\dot\theta^2$ term in $A_2$, i.e. of new propagating fermionic variables at this order, follows from the identity for the two-derivative structures in \rf{33}: with $U_{\m\n} \equiv \bth \g_\m \DF_\n \theta$,
\be \la{a41}
 \tfrac{1}{16} (U_{00} + U_{11})^2 - \tfrac{1}{16} U_{\m\n} U_{\n\m} = \tfrac18 \big( U_{00} U_{11} - U_{01} U_{10} \big) \ ,
\ee
which is linear in $U_{00}$. % (the mixed terms, containing a single
%Killing-spinor derivative, cannot produce $\dot\theta^2$ terms, so that
%\rf{a41} is the complete check at quartic order). 
The same determinant
combination arises in the string ($n=0$)  IIA action, so that the fermionic Hessian
$\del^2 \L/\del \dot\theta \del \dot\theta$ vanishes at this order in both
cases, as assumed in \rf{2041}. 

The momentum dependence of $\cJ_{\rm sc,b}$ gives $-\tfrac12\Delta_u\cJ_{\rm sc,b}=-16$ per point, 
so that $c=10-16=-6$. The same construction for the AdS$_4\times$CP$^3$ action gives \rf{418}--\rf{421}
with the fermionic parts having the expectation values 
\be \la{a5}
\langle \cJ_{\rm mom, f}\rangle = - 3 \rG_\psi - 3 \rG_\chi \ , \qquad \qquad \langle \cJ_{\rm sc, f} \rangle = 12 E_\psi + 4 E_\chi - 2 \rG_\psi + \tfrac{10}{3} \rG_\chi \ ,
\ee
and the same constant $c=-6$.

\section{Measure with  field-specific  $\delta(0)$  factors  %in the measure and the supertrace identity
 \la{apA3}}

In  section \ref{s2}  the factor $\delta(0)$ in the measure \rf{206} was represented by a single regularized 
quantity $\rE$.

In general,  $\delta(0)$ is the number of integration variables per unit area of the field whose integration 
produces a given term, so that the $I$-th term of $\cJ_{\rm mom} = \sum_I \ha \del^2 \L_4/\del u^I \del u^I$  in \rf{24} 
 comes with the $\delta(0)$ of the boson $z^I$. Similarly, 
  each term of the fermionic trace in $\cJ_{\rm sc} = - \ha \tr(\mc C_0^{-1} \mc C_2)$ in \rf{25}
   comes with the $\delta(0)$ of the corresponding fermionic component.
   The double sums in the constant $c=-\tfrac18 \Delta_u^2 \L_4 - \ha \Delta_u \cJ_{\rm sc}$  in \rf{2006}
   come again with the corresponding products of $\delta(0)$'s. 
     If all the fields are smeared with the same kernel this distinction is immaterial, which is why it was ignored in section \ref{s2}. 
   
   \subsubsection*{Definition of the measure}
   
    If the fields are smeared with different kernels, however, the consistent prescription is
    to define the  measure  with  the field-attributed  regularized $\delta(0)\to E_\vp$  factors, i.e. 
\ba \la{a51}
& \delta(0)\, \cJ_{\rm mom} \to \sum_I E_I\, \cJ^{(I)}_{\rm mom} \ , \qquad \qquad
 \delta(0)\, \cJ_{\rm sc} \to \sum_\a E_\a\, \cJ^{(\a)}_{\rm sc} \ , \no \\
& [\delta(0)]^2 \, c \to - \tfrac18 \sum_{I,J} E_I E_J\, \del^2_I \del^2_J \L_4 - \ha \sum_{I,\a} E_I E_\a\, \del^2_I \cJ^{(\a)}_{\rm sc} \ ,
\ea
where $\cJ^{(I)}_{\rm mom} = \ha \del^2_I  \L_4$, $\del_I^2 = \del^2/\del u^I \del u^I$, $\cJ^{(\a)}_{\rm sc}$ is the $\a$-th diagonal element of $-\ha \mc C_0^{-1}\mc C_2$,  and $E_I$, $E_\a$ are the regularized $\delta(0)$'s of the individual
(bosonic and fermionic)  fields, i.e. the same quantities that enter the contact terms in  $\langle \L_4 \rangle$.

%\subsubsection*{\adss}
In the \adss  case   using  \rf{a51} one finds, for arbitrary values of $E_x, E_y, E_\theta$ and of the coincident propagators
$\rG_\vp$, the following  measure contributions $ f_{2\cal M}$ to  (cf. \rf{225},\rf{39})
\be f_2 = - \langle \L_4 \rangle +  f_{2\cal M} =  f_{2}^{\rm (b)}  +  f_{2}^{\rm (bf)}  +  f_{2}^{\rm (f)}  \ , \la{000}
\ee
from the three sectors in \rf{31}--\rf{33}. 
 In the bosonic sector, 
 \be 
 \sum_I E_I \langle \cJ^{(I)}_{\rm mom,b}\rangle \to  - \tfrac32 E_x^2 - \tfrac52 E_y^2 + \tfrac{21}{4} E_x \rG_x - \tfrac{25}4 E_y \rG_y \ ,\qquad \ 
 c\, [\delta(0)]^2 \to  \tfrac18 (3 E_x + 5 E_y)^2 + \tfrac14 (3 E_x^2 + 5 E_y^2)\ , \ee
% c\to  \tfrac18 (3 E_x + 5 E_y)^2 + \tfrac14 (3 E_x^2 + 5 E_y^2)\ , \ee
 so that together with    $\langle \cL_{4\rm b} \rangle$ in  \rf{34}  we get 
\be \la{a52}
 f_{2}^{\rm (b)} = \tfrac18 \big( 3 E_x + 5 E_y \big)^2 \ .
\ee
In the mixed sector,  we get 
\be \sum_\a E_\a \langle \cJ^{(\a)}_{\rm sc,b}\rangle \to 
 - 12 E_\theta \rG_x \ , \qquad 
 \sum_I E_I \langle \cJ^{(I)}_{\rm mom,f}\rangle \to  (- 6 E_x + 10 E_y) \rG_\theta\ , \qquad 
 c\, [\delta(0)]^2 \to  - 2 E_\theta (3 E_x + 5 E_y) \ ,   \ee
 so that together with \rf{35}
\be \la{a53}
 f_2^{\rm (bf)} = - 2 E_\theta \big( 3 E_x + 5 E_y \big) \ .
\ee
In the fermionic sector, $\sum_\a E_\a \langle \cJ^{(\a)}_{\rm sc,f}\rangle = E_\theta (16 E_\theta - 2 \rG_\theta)$, so that together with \rf{36}
\be \la{a54}
 f_2^{\rm (f)} = 8 E_\theta^2 \ .
\ee
All the propagator-dependent terms thus cancel, separately in each sector and for arbitrary regularized  values 
 $E_\vp$, and, remarkably,  the total is a perfect square  as in \rf{999}
\be \la{a55}
 f_2 = \tfrac18 \big( 3 E_x + 5 E_y - 8 E_\theta \big)^2 = \tfrac18 \big( {\rm Str}\, E \big)^2 \ , \qquad \qquad
 {\rm Str}\, E \equiv \sum_{I} E_I - \ha \sum_{\a} E_\a \ .
\ee
The weight $\ha$ of the fermionic  contribution 
 is the first-order counting (16 real components, 8 degrees of freedom), so that ${\rm Str}\, E$ is the regularized difference between the numbers of bosonic and fermionic modes per unit \adst\ area. 
 Its power-divergent part cancels ($8 - 8 = 0$), so that   the UV finiteness is automatic.
   The only scheme dependence of $f_2$ is thus  through the  single constant  ${\rm Str}\, E$. 
   It vanishes, in particular, for the  choice of 
   the common smearing, $E_x = E_y = E_\theta $.\foot{In the case of massless  smearing  when  $E_x=E_y= [e^{t \nabla^2}], 
   \ , E_\theta = [e^{t \hat \nabla^2}]$ we have 
   $E_\theta = E_x +  \OO(t)$, i.e. ${\rm Str}\, E=0$ in the limit of $t\to 0$ (cf. \rf{314}).}

   \subsubsection*{Lorentz covariance of the measure and  the value of ${\rm Str}\, E$}

The prescription \rf{a51} has a further consequence which is not visible in the vacuum value
\rf{a55}. The per-field velocity Hessian of the bosonic Lagrangian \rf{31} is
\be \la{6001}
\cJ^{(I)}_{\rm mom,b} = \ha \del_I^2 \L_{4\rm b}
= - \tfrac14 \big(u^2 - s^2\big) - \ha \big(u_I^2 + s_I^2\big) + \cJ^{(I)}_{0} \ ,
\ee
where $\cJ^{(x)}_{0} = \tfrac14 x^2$  and $\cJ^{(y)}_{0} = -\tfrac14 y^2$; the sum over the 8 bosons gives \rf{211}. The first term is not Lorentz covariant
($u^2 - s^2 = \del_0 z\cdot \del_0 z - \del_1 z\cdot \del_1 z$ in the local frame).
The bosonic part of the constraint matrix in \rf{a3} is proportional to the identity,
$-\ha \mc C_0^{-1} \mc C_{2 \rm b} = - \tfrac14 A^{00}\, {\bf 1} + {\rm traceless}$,
so that each of the 16 fermionic components contributes
$\cJ^{(\a)}_{\rm sc,b} = \tfrac18 (u^2 - s^2) - \tfrac14 x^2$.
Similarly, with $U_{\m\n} \equiv \bth \g_\m \DF_\n \theta$, the velocity Hessian of the mixed
term $\tfrac14 A^{\m\n} U_{\m\n}$ in \rf{32} gives $\tfrac18 (U_{11} - U_{00})$ per boson,
while the two-derivative part $-2 U_{11}$ of $\cJ_{\rm sc,f}$ in \rf{a4} comes with $E_\theta$.
Collecting the non-covariant terms in \rf{a51} we find
\be \la{6002}
\mc M_{\rm nonc} = - \tfrac14\, {\rm Str}\, E\, \big(u^2 - s^2\big)
 - \tfrac18\, {\rm Str}\, E\, \big(U_{00} - U_{11}\big) \ ,
\ee
with ${\rm Str}\, E$ as in \rf{a55}, while the remaining terms form a covariant density which
reduces to \rf{29},\rf{209} for equal $E_\vp$.

The vacuum expectation value of \rf{6002} vanishes by isotropy, which is why \rf{a55} holds
for arbitrary $E_\vp$. However, \rf{6002} contributes to the 1-loop effective action
\rf{4401} for the fluctuation fields. Since $\langle \L_4 \rangle_1$ in \rf{4401} is computed
with covariant propagators, \rf{6002} induces the Lorentz-violating term
\ba \la{6003}
\Delta \Gamma^{(1)}_2 = T^{-1}\, {\rm Str}\, E \int d^2\xi \sqrt g\, \Big[ & \tfrac14 \big(
\del_0 x\cdot\del_0 x - \del_1 x\cdot\del_1 x + \del_0 y\cdot\del_0 y - \del_1 y\cdot\del_1 y
\big) \no \\
& + \tfrac18 \big( \bth \g_0 \DF_0 \theta - \bth \g_1 \DF_1 \theta \big) \Big] \ .
\ea
The measure obtained by integrating out the momenta in the phase-space path integral, in which
the time direction is singled out, and regularized covariantly afterwards (see the discussion
after \rf{2006}) thus defines a Lorentz-invariant theory only if
\be \la{6004}
{\rm Str}\, E = 0 \ .
\ee
Together with the field-redefinition argument at the end of section \ref{s2}, which requires
the field-attributed $\delta(0)$'s in \rf{a51}, this fixes the remaining scheme dependence of
$f_2$ in \rf{a55}, giving $f_2=0$. In particular, smearing each field with its own kinetic
operator, for which ${\rm Str}\, E = - {1\ov 2\pi}$ (see \rf{a58} below), is not compatible with
the covariance of the canonical measure, while the common massless smearing \rf{213} is.%
\foot{One could, of course, add by hand a local Lorentz-violating counterterm cancelling
\rf{6003}, but such a term is not produced by the canonical construction of the measure.}

Similar expressions are found in the \adsc case. 
In the bosonic sector  starting with  the bosonic   Lagrangian  \rf{45} we get 
\be \la{a56}
 f_2^{\rm (b)} = \tfrac18 \big( 2 E_x + 6 E_w\big)^2 \ ,
\ee
again with all the $\rG_\vp$ tadpoles cancelled. 
The mixed and fermionic sectors  give the measure   contributions 
 with  the  same  tadpole structure  as in  \rf{416},\rf{417}.
 %\foot{The uniform trace of $\cJ_{\rm sc,b}$ gives
 % $-(6 E_\psi + 2 E_\chi) \rG_x$, the per-field Hessian of the mixed term gives $(3 E_x - 6 E_w)\rG_\psi - 3 E_w \rG_\chi$ with opposite sign, etc.}
 As a result, 
\be \la{a57}
 f_2 = \tfrac18 \big( 2 E_x + 6 E_w - 6 E_\psi - 2 E_\chi \big)^2 + \tfrac32 \rG_\chi^2 \ ,
\ee
which is again UV finite  and with the
term $\tfrac32 \rG_\chi^2$ independent of the smearing  choice.
The non-covariant part of the bosonic measure density is again
$-\tfrac14\, {\rm Str}\, E\, (u^2-s^2)$, now with ${\rm Str}\, E$ as in \rf{a57}: the
coefficients $-2$ and $+2$ of $u^2-s^2$ in $\cJ_{\rm mom,b}$ and $\cJ_{\rm sc,b}$ in \rf{4193}
are the same as in \rf{211},\rf{212}, being fixed by the induced-metric structure of the
quartic action (the bosonic part of $\mc C_2$ is again $\ha A^{00} \mc C_0$ up to traceless
terms). The covariance condition \rf{6004} thus applies also in the \adsc case and gives
$f_2 = \tfrac32 \rG_\chi^2$ in \rf{a57}.

% and of the sign $\eta$. The per-block split of the fermionic trace $\cJ_{\rm sc, f}$ between the $\psi$ and $\chi$ components in \rf{a57} was not recomputed separately; it is fixed by the mechanism above and by the totals \rf{a5}.

To summarize,  in the case when  the measure is defined as  in 
    \rf{a51} no UV  logarithms  appear  in the total 2-loop   expression for any choice of  the smearing kernels:
the logarithmically divergent  terms  like $E_\vp \rG_{\vp'}$  or   $\rE \rG_{\vp'}$ in \rf{313} (cf. \rf{2170}--\rf{218}) 
 appeared  only in the ``hybrid'' prescription in which the fields in $\langle \L_4 \rangle$
  are smeared with  kernels corresponding to  $E_\vp$ 
  while the measure  factors in \rf{39}   depend on a single  $\rE$.

%\subsubsection*{The value of ${\rm Str}\, E$}

Let us note for generality that in  the case of the ``massive'' smearing, i.e. 
when  each field is smeared with its own (massive) 
kinetic operator, 
 the finite parts of the $E_\vp$ are the  same as  the 
 $\zeta(0)$ values \rf{219} (and $ E_{\chi, \rm fin} = \tfrac{1}{24\pi}$ for the massless fermion), one finds in both \adss and \adsc cases
\ba \la{a58}
 &4 \pi \, {\rm Str}\, E_{\rm fin}\Big|_{\AdS_5 \times S^5}
  = 3 \big(-\tfrac73\big) + 5 \big(-\tfrac13\big) - 8 \big( - \tfrac56\big)
 = - 2 \ , \qquad \\
& 4 \pi \, {\rm Str}\, E_{\rm fin}\Big|_{\AdS_4 \times \CP^3}
  = 2 \big(-\tfrac73\big) + 6 \big(-\tfrac13\big) -  \big[ 6 \big(- \tfrac56\big)  + 2 \big( \tfrac16\big)\big]
 = - 2 \ .
\ea
Indeed,   $4 \pi E_{\vp, \rm fin}$ is given by $ \tfrac16 R^{(2)} - m_\vp^2$ for the scalar 
and $\tfrac16 R^{(2)} - \tfrac14 R^{(2)} - m^2_\vp$ for the spinor components,  so that 
the $\tfrac16 R^{(2)}$ terms cancel  due to the equal number of  bosonic and fermionic 
degrees of freedom  and thus 
\be \la{a59}
 4 \pi\, {\rm Str}\, E_{\rm fin} = - \big( \sum_{\rm b} m^2 - \sum_{\rm f} m^2 \big) + 2 R^{(2)} = 2 - 4 = - 2 \ .
 %\qquad \qquad
 %{V \ov 4\pi}\, {\rm Str}\, E_{\rm fin} = \zeta_{\rm tot}(0) = 1 = \chi \ ,
\ee
The universality  of this result follows from 
 the mass sum rule $\sum_{\rm b} m^2 - \sum_{\rm f} m^2 = - 2$ of the \adst\  string spectrum  (see \ci{Drukker:2000ep,Giombi:2020mhz}). 
  %(the fermionic sum being over the 8 degrees of freedom). 
  Thus in this ``massive'' smearing scheme  we have $V{\rm Str}\, E_{\rm fin}=\zeta_{\rm tot}(0)  $  (with $V=-2\pi$) 
  where   $\zeta_{\rm tot}(0)=1$    is the   familiar total 
  coefficient   %(times $4 \pi  V^{-1}= - 2$) 
    that appears in  the 1-loop computation  (it is equal to the Euler number $\chi = 1$ 
    for  the disk \ci{Drukker:2000ep}).  Hence in this scheme 
    \be \la{a60}
 f_2\Big|_{\AdS_5 \times S^5} 
   =\tfrac {1 }{ 32 \pi^2}  \ , \qquad \qquad\ \ \ \ \ \ 
 f_2\Big|_{{\rm AdS}_4 \times {\rm CP}^3}   = \te {1 \ov 32 \pi^2} + {3 \ov 32 \pi^2} = {1\ov 8 \pi^2}  \ .
\ee
This may be   compared with the results  ($f_2 = 0$ for  \adss and $f_2 ={3 \ov 32 \pi^2}  $ for \adsc)   in \rf{50} for the
  choice of common smearing  or   ${\rm Str}\, E = 0$. 
%Neither choice gives the \adss value $\tfrac{3}{32\pi^2}$ required by \rf{1}, which would need $4 \pi\, {\rm Str}\, E_{\rm fin} = \pm 2 \sqrt 3$. %  Unequal kernels do not induce finite mass shifts through the $\delta(0)$'s: all the propagator-dependent terms cancel, so that the vacuum-level counterterm interpretation of a mismatch between the $E_\vp$ does not arise; the two-point test of Appendix \ref{s44} is likewise unaffected (there the measure term $\tfrac{1}{12}\rE$ becomes $\tfrac{1}{12} E_\chi$ and the $E$-terms cancel exactly, leaving the finite $\tfrac34 \rG_\chi$). 
  
  The  remaining question is  which value of ${\rm Str}\, E$ in the  free energy 
   the correct quantization prescription 
   should actually select. 
   The 1-loop  result 
    \ci{Drukker:2000ep,Giombi:2020mhz}  
    suggests     to fix
     the normalization of the measure  so as to cancel the remaining  logarithmic divergence proportional to $\zeta_{\rm tot}(0) = 1$
     which  would effectively restore  the balance between the  bosonic and fermionic   contributions in \adst.
      %i.e. it effectively balances the bosonic and fermionic mode densities; 
       This may be  hinting at the choice of    ${\rm Str}\, E = 0$    at the 2-loop order;
      %within the 2-loop computation itself 
      this choice is required by the  condition 
      \rf{6004} of covariance of the 1-loop effective action.

     % This may also  be  hinting at the choice of    ${\rm Str}\, E = 0$    at the 2-loop order  as the preferred one
     % (see the end of section \ref{s3} for the other indications in favour of this choice).
%   is what gives $f_2 = 0$ in \adss and $f_2 = \tfrac32 \rG_\chi^2$ in \adsc at two loops. This is an argument for the common smearing, not a derivation of it.

%%%%%%%%%%%%%%
\section{Comments on  relation to invariant measure on  supercoset \la{apA2}}

The \adss GS action is the $\mathbb Z_4$-graded sigma model on the supercoset of $PSU(2,2|4)$ over $SO(4,1)\times SO(5)$ \ci{Metsaev:1998it}. One may ask whether this can be used to define the measure in a more covariant way. Three different things can be meant by a ``covariant measure''.

%{\it Formal covariance.} 

The Liouville measure on the physical phase space %, with the Senjanovi\'c factor,
 is a symplectic invariant: it does not change under canonical transformations. The superisometry group $PSU(2,2|4)$ acts on the physical phase space by canonical transformations (generated by the Noether charges), so the canonical measure of section \ref{s2} is formally $PSU(2,2|4)$ invariant as it stands. The supercoset formulation therefore adds no new invariance requirement at the unregularized level. What it does provide is an invariant characterization of the bosonic part: in a bosonic $G/H$ 
  coset sigma model  the Liouville measure $\prod_\xi\sqrt{\det G_{IJ}(X)}$ is the $G$-invariant (Haar) measure on $G/H$. In the coordinates used in  \ci{Beccaria:2026ffm},  one has $G_{ij}=\delta_{ij}(1-\tfrac14x^2)^{-2}$ for the three $AdS_5$ directions and $G_{ab}=\delta_{ab}(1+\tfrac14y^2)^{-2}$ for the five $S^5$ directions, so that the transverse part of the invariant measure is
\be \la{a13}
 \tfrac12\log\det G_\perp %=\tfrac12\big[3\cdot\tfrac{x^2}{2}-5\cdot\tfrac{y^2}{2}\big]+\dots
 =\tfrac34x^2-\tfrac54y^2+\dots \ ,
\ee
which is exactly the non-derivative part of $\cJ_{\rm mom,b}$ in \rf{211}. The derivative terms there are the static-gauge effect (the longitudinal momenta are solved from the constraints, so the Hessian of the Nambu Lagrangian depends on $\del z$), and the remaining bosonic terms \rf{212} come from the fermionic constraint matrix. Thus 
 the invariant coset measure is contained in the canonical one, with the gauge-fixing and fermionic pieces on top.

%{\it The invariant super-measure.} 
The natural target-space object on the supercoset is the $PSU(2,2|4)$ invariant Berezin measure, ${\rm sdet}\, E_M^A(Z)$ where   $Z^M=(X^I,\theta^\alpha)$. 
This is an algebraic function on superspace: expanded in $\theta$ it gives terms of the form $\delta(0)\int\bth M\theta$ with no worldsheet derivatives. The canonical fermionic measure is a different object. For first-order fermions it is the odd contribution $(\det\mc C)^{-1/2}$ to the square root of the Berezinian of the reduced symplectic form (with $\mc C$ the symmetric matrix of the second-class constraints), and $\mc C$ is built from the $\dot\theta$-coefficients of the action, which involve $\del_1\theta$  from  the $\theta\theta\theta\del\theta$ terms. 
This is why $\cJ_{\rm f}$ in \rf{209} contains the derivative bilinear $\bth\Dsl\theta$, which no supervielbein determinant can produce.

The reason is that in the GS string action \ci{Metsaev:1998it} the grade-one and grade-three
(odd) currents enter the Wess-Zumino term, which is first order in the derivatives of the odd
coordinates, while the metric term contains the grade-two currents only (see e.g. \ci{Vicedo:2009sn}).
The grade-two currents do contain fermionic bilinears, so that the metric term produces also
terms quadratic in the velocities of the odd coordinates, of the type $(\bth \G^a \dot\theta)^2$;
in the adapted $\kappa$-symmetry gauge these cancel at the quartic order. % see \rf{a41} and the
%discussion in section \ref{s21}. 
The gauge-fixed fermionic Lagrangian is therefore a first-order one,
and the corresponding measure is symplectic, not metric.\foot{The super-Haar measure would be the right object for a formulation in which the fermions are second order, or in which their momenta are independent fields, as in the pure-spinor formalism \ci{Berkovits:2000fe} where $(x,\theta,d_\alpha,\lambda,w)$ have a flat, manifestly covariant measure, the field-dependent kinetic
 structure of the fermions sits in the action (cf.  footnote \ref{f1}).}

%{\it Covariant regularization.}

The 2-loop result \rf{777} (or \rf{a55}), obtained with the field-attributed measure of
Appendix \ref{apA3}, holds for any smearing of the propagators. %; the value $f_2=0$ in \adss requires
%${\rm Str}\, E=0$.
%The 2-loop result  \rf{313} holds for any smearing of the propagators.
  The value $f_2=0$ requires ${\rm Str } E=0$. %   assuming 
  % that the regularized $\delta(0)$'s
  % of all fields coincide with that of the measure. 
 That is the step a supercoset symmetry  may  justify. %, and it is where the supercoset structure is useful. 
 A  symmetry-consistent formulation  of   ``common smearing''
  would be to regulate all fluctuations with the same operator built
   from the Casimir of the superalgebra preserved by the background: $OSp(4^*|4)$ for the \adss  case 
    and $SU(1,1|3)$ for the \adsc one.
    
     The fluctuations with scaling dimensions 
     $\Delta=2,1,\tfrac32$ (and, in the IIA case, also the massless fermion with $\Delta = \ha$) sit in supermultiplets of that algebra, so such a regulator would relate the bosonic and fermionic kernels in the way the ``supersymmetry relation'' $\rG=-{\rm S}$ of \ci{Beccaria:2026ffm} anticipated. Whether it reproduces the prescription 
     of section \ref{s23}, i.e. $E_x=E_y=E_\theta$ 
     is a  check that  may be interesting to do. 
     One may  start with  the covariant background-field expansion of the coset itself, $g=g_0\,e^{\ru}$ with $\ru$ in the complement of the stability subalgebra\foot{It was  used in the earlier 2-loop \adss coset computation  in  \ci{Roiban:2007jf},  which, however, used dimensional regularization.}
      in which the invariant measure is flat in $\ru$ up to the ultralocal Jacobian
%\be \la{a14}
$ \rm  J= {\rm sdet} \frac{1-e^{-{\rm ad}\,\ru}}{{\rm ad}\,\ru}\big|_{\rm coset} $.   It is 
the super-analogue of $\sqrt{G}$ in normal coordinates.\foot{Here the Berezinian is taken over the coset tangent directions, after the projection that separates the coset vielbein from the stability-group connection \ci{Metsaev:1998it}; the unrestricted group determinant is not the quotient-space measure.}
 Comparing $ \rm J$  (after the change of variables to the canonically normalized 2d spinors) 
 with the canonical measure density $\cJ$ in \rf{2006}
  would be a covariance check of the bosonic and the mass-type fermionic terms.
   The derivative term in $\cJ_{\rm f}$ would then have to be accounted for by the (regularized) Jacobian
of the change of variables, which remains to be exhibited (which involves the background vielbein, the $\kappa$-projector and the free constraint matrix $\mc C_0$).

%{\it The missing finite term.} 

One may ask if 
 an invariant measure  may supply an  extra finite term  needed to match localization result?
 The invariant target-space Berezinian alone does not determine such a term: any extra ultralocal
algebraic term $\delta(0)\int\bth M\theta$ contributes $\rE\,\langle\bth M\theta\rangle\sim  \rE\,\rG_\theta$,
which is logarithmically divergent through the massive blocks and would also break the tadpole
cancellations of section \ref{s3}. A  finite term of the type \rf{430} can only come from a
non-ultralocal Jacobian, which would have to be computed with a covariant regulator together with
its accompanying terms.
 \iffa The answer is no: 
  any extra ultralocal algebraic term $\delta(0)\int\bth M\theta$ 
  gives  $\rE\,\langle\bth M\theta\rangle\sim  \rE\,\rG_\theta$, which is logarithmically divergent due to  the massive 
  fermion  contribution
   in both \adss and \adsc cases.  
   \fi
    %and  breaks the tadpole cancellations 
      The term in  \rf{430} is finite precisely because it involves the  massless 
      sector  only and comes without a $\delta(0)$.\foot{In the string theory  setting 
      % Its coefficient is therefore not determined by the target-space Berezinian;
        it could arise from a non-ultralocal Jacobian  of the change of variables from coset fermions to 2d spinors
         evaluated with a covariant regulator  (cf. \ci{Wiegmann:1989md}), 
      while in the M2 brane description it  originates from integrating out   higher  KK modes. }
        %The supercoset is useful here too, but in a different way: it tells one which regulator to use in computing such a Jacobian, not what the Jacobian is.

%%%%%%%%
\section{Measure in the M2 brane case \la{apB}}

The general  discussion of the measure  in  section \ref{s2}  applies to a theory in  any number of  dimensions. 
 For the 3d case of the M2 brane  action in $\AdS_n \times  M^{11-n}$ 
  in the static and $\kappa$-symmetry gauges     one obtains for the 2-loop   measure  contribution (cf.\rf{206})\foot{$\rT_2$ is dimensionless 
          M2  brane tension   and here  we   assume that   $\L=  \L_2+  \rT_2^{-1} \L_4+...$.}
%     cf.  \ci{Beccaria:2026zjj}.}
\be \la{b1}
\mc M_3 = 1 +  { \rT_2^{-1}} \int d^3 \s \sqrt g \Big[ \delta^{(3)}(0)\, \cJ_3 +  \cc \, [ \delta^{(3)}(0)]^2\Big] + \OO(\rT_2^{-2})  \ .
\ee
Here  $\cJ_3$  is again a local combination quadratic in the 3d fields, built from the 3d quartic Lagrangian by the same two operations (the velocity Hessian, and the field dependence of the matrix of the second-class fermionic constraints), and $\cc$ is a constant as in \rf{2006}. 

The crucial difference with the 2d case is in the value of  regularized 
 $\delta(0)$. The coincident-point  value of the 3d heat kernel is
\be \la{b2}
\delta^{(3)}(0) \to  e^{-t(-\nabla^2_3 + m^2)}(\s,\s) =\te  {1\ov (4\pi t)^{3/2}} \big( 1 + a_1 t + a_2 t^2 + ...\big) \ ,
\ee
i.e. it contains only half-integer powers of $t$.  In odd dimensions there is no $t^0$ term, and the $\zeta$-function regularized value of $\delta^{(3)}(0)$ (the finite part) vanishes identically, $\zeta(0; \s, \s) = 0$, for the smooth AdS$_2 \times S^1$ geometry with no normalizable zero modes and no boundary contributions.
 In the $\zeta$-function (or dimensional) regularization used in \ci{Beccaria:2026zjj} and in \ci{Beccaria:2025ahf} the measure \rf{b1} thus does   not contribute. 
 
 \iffa 
 Note that this concerns the individual contact traces in the analytic prescription; it does not extend to arbitrary products of cutoff-regularized contacts, for which the finite part is a different operation. For example, with
\be \la{b21}
 K(t) = (4\pi t)^{-3/2} e^{-(1+m^2) t} \ , \qquad \qquad {\rm FP}\, K(t) = 0 \ , \qquad {\rm FP}\, [K(t)^2] = - {(1+m^2)^3 \ov 48 \pi^3} \ ,
\ee
the finite part of the square of a cutoff contact does not vanish even though that of the contact itself does.
\fi

The same regularization sets to zero all the ``equation of motion'' contact terms in the expectation value of the quartic action,
%(EOM  parts  dropped in Appendices F--H of \ci{Beccaria:2026zjj}), 
i.e. precisely the terms that the measure would have cancelled. 
Setting ``$\delta(0)=0$'' is thus a consistent treatment of the total path integral. 
This is why the measure was not important in the M2 brane computation of \ci{Beccaria:2025ahf} for AdS$_3$ in AdS$_7\times S^4$, where the 2-loop expectation value was found to be proportional to the EOM  terms and thus to vanish.
Similarly,  in the computation of \ci{Beccaria:2026zjj}  only the contact-free part (the perfect square \rf{426}) survived. 

By contrast, 
in 2 dimensions,   the $t^0$ term in \rf{23} is present, and the $\zeta$-function or heat-kernel regularized $\delta(0)$ is a non-zero finite quantity which differs from field to field. This  is why the measure had to be included explicitly in the string computation.

  %, with world-volume coordinates $(\s^\mu = \xi^\m, \s^3=y)$ where $\xi^\m$ are the \adst\ coordinates and $y$ is the coordinate identified with that of the  the Hopf circle of $S^7$ 
 % of length $L = {4\pi \ov k}$,

It is instructive to discuss  in more detail the 3d  case with explicit power cutoff  and compare it to the 2d   case. 
 Let us   consider, e.g.,  the 
M2 brane  action on $\AdS_2 \times S^1$  in    $\AdS_4 \times S^7/\ZZ_k$ 
 (with the world-volume coordinates $\s^\mu = \xi^\m, \s^3=y$\,  ($y\equiv y+ L)$, 
  where $\xi^\m$ are the \adst\ coordinates and $y$ is the coordinate identified with that of   the Hopf circle of $S^7/\ZZ_k$)  and  reduce the 3d measure to  the 2d string theory one. 
 % of length $L = {4\pi \ov k}$,
Let  us  expand  the 3d fields as
\be \la{b3}
 \vp(\xi, y) = L^{-1/2} \sum_{n=-\infty}^{\infty} \vp_n(\xi)\, e^{i p_n y} \ , \qquad \qquad  L=   4 \pi k^{-1} \ , \qquad 
 p_n =\ha  {k n } \ ,
\ee
Then in \rf{b1}  we will have 
$\int dy\, \cJ_3 = \sum_n \cJ^{(n)}(\vp_n )$  where $\cJ^{(n)}(\vp_n )$    has the structure   $\vp_n  \OO_n \vp_{-n}$.
Then $n=0$ term  here $\cJ^{(0)}(\vp_0 )$     corresponds to   the 2d combination $\cJ$ in \rf{418}.
%\foot{Note that 
%the circle component of the Killing-spinor derivative becomes  the algebraic matrix that survives  upon  2d reduction.}
% The coefficient, however, is not the 2d one.
  With a massless 3d smearing kernel $e^{t \nabla^2_3}$, where $\nabla^2_3 = \nabla^2 + \del_y^2$, the 3d coincident kernel is a sum over the KK levels of the 2d ones,
\be \la{b4}
\delta^{(3)}(0) \ \to \ [e^{t \nabla^2_3}]={1\ov L} \sum_{n=-\infty}^{\infty} \rE^{(n)} \ , \qquad \qquad \rE^{(n)} \equiv \rE\, e^{-t p_n^2} \ ,
\ee
where $\rE = \rE^{(0)}$ is the massless  2d coincident kernel (i.e. \rf{23}   with $m=0$) 
and $\rE^{(n)}$ is the one for the $n$-th KK mode. %Using the relation between tensions 
%$8 \rT_2^{-1} L^{-1} = \rT_1^{-1}$, 
The zero-mode combination $\cJ^{(0)}(\vp_0)$ in \rf{b1} is then 
multiplied by the total KK mode density $\sum_{n} \rE^{(n)}$ rather than by $\rE$ alone as in \rf{28},\rf{422}.

 In the M2 brane  2-loop computation the contact terms in  the contribution of the 
  action have the same structure: in the double-bubble diagram one of the two loops can be the one for 
  any mode $n$ while the other is the zero mode, producing $(\sum_n \rE^{(n)})\, \rG_{\vp}$ type terms. One therefore expects the cancellation between the action and the measure to take place with the coefficient 
  $\sum_n \rE^{(n)}$  playing  the role  of   what $\rE$  does  in 2d case.
  Then   only the contact-free part survives,
   the difference between the 3d and 2d results being not in the measure sector but in the contact-free part, where the integrated-out  massive   KK modes  leave the finite local term   \rf{430}.\foot{This expectation has been verified explicitly only in 2 dimensions and, for the quadratic term in
the external massless fermions, in the level-by-level computation leading to \rf{4302}. The full 3d
cutoff computation of the vacuum contribution has not been carried out; note that for a cutoff
regularization in 3 dimensions the analogs of the differences $\rho_\vp$ are not $\OO(t)$ but power
divergent, $\rho^{(3)}_\vp \propto m_\vp^2 t^{-1/2}$, so that a 3d heat-kernel scheme is not automatically
equivalent to the $\zeta$-function one.
% and, for the \iffa block with eq. b21 (lines 2470-2480), one may keep just the sentence:
Note that the vanishing of the analytic value of an individual 3d contact term does not imply the
vanishing of the finite part of a product of cutoff-regularized contacts; the analytic prescription
assigns zero to the individual contacts before forming their products.
}
   
%   This expectation has been verified explicitly only in 2 dimensions and, for the quadratic term in the external massless fermions, in the level-by-level computation of section \ref{s4}, see \rf{4302}. The full 3d cutoff comparison is a further computation: for a cutoff regularization in 3 dimensions the analogs of the differences $\rho_\vp$ are not $\OO(t)$ but power divergent, $\rho^{(3)}_\vp \sim m_\vp^2 t^{-1/2}$, so that a 3d heat-kernel scheme is not automatically equivalent to the $\zeta$-function one, and whether the corresponding remainders cancel by the mass sum rules at each KK level  remains to be seen. 

\section{Coincident limit of  fermionic propagator\la{apa2}}

With $S = \Dsl^{-1} K_\theta = - \Dsl\, (-\Dsl^2)^{-1} K_\theta$ 
(where $K_\theta$ is a regularization of the  fermionic $\delta^{(2)}(\xi,\xi')$ in the fermionic sector)
and $-\Dsl^2 = - \hn^2 + \ha$ on the unit-radius  \adst\  (using $[\nabla_\m, \nabla_\n]\theta = \tfrac14 R^{(2)} \g_{\m\n}\theta = - \ha \g_{\m\n}\theta$), one has for the coincident-point value 
\be \la{a6}
 [S] = - \Gb\, [ (-\Dsl^2)^{-1} K_\theta] \equiv - \Gb \rG_\theta \ ,
\ee
where  the coincident value of the first derivative of the (bispinor) kernel vanishes by homogeneity. For the first derivative, isotropy gives $[\hat\nabla_\m S] =\ha  a \g_\m$ and contracting with $\g^\m$
\be \la{a7}
  a = [\hat{\slashed\nabla} S] = [\Dsl S] - \Gb [S] = E_\theta - \rG_\theta \ .
\ee
For the mixed second derivative the symmetric part is fixed by its trace,
\be \la{a8}
 [\hat \nabla_\m \hat\nabla'^\m S] = - [\hat \nabla^2 S] = [(-\Dsl^2) S] - \ha [S] = - \Gb E_\theta + \ha \Gb \rG_\theta \ ,
\ee
and the antisymmetric part by the curvature,
\be \la{a9}
 [\hat\nabla_\m \hat\nabla'_\n S] - [\hat\nabla_\n\hat\nabla'_\m S] = [[\hat\nabla_\m, \hat\nabla_\n] S] = - \ha \g_{\m\n}[S] = \ha \g_{\m\n} \Gb \rG_\theta \ ,
\ee
giving \rf{222}. Converting to the Killing-spinor derivatives, with $\overline{\DF_\n \theta} =\hat  \nabla_\n \bth - \ha \bth \Gb \g_\n$,
\be \la{a91}
 [\DF_\m S \overleftarrow{\DF}{}'_\n] = [\hat\nabla_\m \hat\nabla'_\n S] - \ha [\hat\nabla_\m S] \Gb \g_\n + \ha \g_\m \Gb [\hat\nabla'_\n S] - \tfrac14 \g_\m \Gb [S] \Gb \g_\n
 = \ha E_\theta \, \g_{\m\n} \Gb \ .
\ee
The extra terms cancel both the symmetric part of \rf{222} and all the $\rG_\theta$-dependence, leaving a 
purely antisymmetric $E$-term, i.e. proving that  $\rk_\theta = \ha E_\theta$, as quoted in \rf{222}.\foot{Note that the vanishing of $[\DF_\m, \DF_\n]$ on the background does not imply the vanishing of the antisymmetric part of $[\DF_\m S \overleftarrow{\DF}{}'_\n]$: the coincident limit of a derivative acting on each of the two points of $S$ is not the commutator of two derivatives acting on one point. This is the origin of the non-zero value $\rk_\theta = \ha E_\theta$, which was assumed to vanish in Appendix B of \ci{Beccaria:2026ffm}.}
 For a general mass $m$ the same steps give \rf{49},\rf{410}.

%\subsubsection*{Spectral values}

The coincident  values of the propagators in \rf{2170} follow from the \adst\  spectral representations \ci{Camporesi:1991nw,Camporesi:1995fb}: for a scalar of dimension $\Delta$, with $-\nabla^2$ having the eigenvalues $\nu^2 + \tfrac14$,
\be \la{a11}
\rG(t)= [G_\Delta](t)=\tfrac1{2\pi}\int_0^\infty d\nu\,\nu\tanh(\pi\nu)\,{e^{-t(\nu^2+\frac14)} \ov \nu^2+(\Delta-\frac12)^2} \ ,
\ee
which gives $\rG=\tfrac1{4\pi}(\LDR-2) + \OO(t \LDR)$ and $\tfrac1{4\pi}\LDR + \OO(t \LDR)$ for $\Delta=2,1$.

For the spinor %, with the density $\nu\coth(\pi\nu)$,
with  $(\Delta-\frac12)^2=m^2$  and $-\hn^2$ having the eigenvalues $\nu^2 + \ha$ (cf. \rf{2009})
\be \la{a12}
 \rG_\theta(m;t) = {m \ov 2\pi}\int_0^\infty d\nu\,\nu\coth(\pi\nu)\,{e^{-t(\nu^2 + \frac12)} \ov \nu^2+m^2} \ .
\ee
This  gives $\rG_\theta=\tfrac1{4\pi}(\LDR-1) + \OO(t\LDR)$ for $m=1$.
For $m\to 0^\pm$, 
since $\nu \coth(\pi\nu)\to {1\ov\pi}$ as $\nu\to0$,  we get   $ \rG_\theta(m\to 0^\pm;t) = {1 \ov 4\pi}\eta\, e^{-t/2}$ 
 in agreement with \rf{442}. 
 
% The finite parts of the scalar propagators are $-2\psi(\Delta)+2\psi(1)$; the $E$ values \rf{218} are the Seeley coefficients of the scalar and spinor Laplacians on \adst. The finite-cutoff identities \rf{313} and \rf{423} are algebraic; in the \adsc case the integrals \rf{a11},\rf{a12} were in addition evaluated numerically at finite $t$ and the cancellation of the $E$-dependent terms in \rf{423} was confirmed numerically (with 50-digit arithmetic in the direct 2d computation, at several values of $t$).

%\subsubsection*{}

Let us comment on the Majorana Wick contraction  rule.
For the product of two bilinears of a Majorana fermion $\theta$, let $\lambda_1 = \mc D_1 \theta$ and $\lambda_2 = \mc D_2 \theta$ be two (possibly differentiated) spinors built from $\theta$ at the points $1$ and $2$, and $A, B$ two matrices. Then
\be \la{a10}
 \big\langle(\bth_1 A\lambda_1)(\bth_2 B\lambda_2)\big\rangle=\tr(A S_{11})\tr(B S_{22})-\tr(A S_{12} B S_{21})-\tr(A T_{12}\wt B \wt U_{12}) \ ,
\ee
with $S_{ij}=\langle\lambda_i\bth_j\rangle$, $T_{12}=\langle\lambda_1\bar\lambda_2\rangle$, $U_{12}=\langle\theta_1\bth_2\rangle$, and $\wt B=C^{-1}B^TC$, $\wt U_{12} = C^{-1} U_{12}^T C$, where the traces are over the 16-dimensional space of the physical components.

 The last term in \rf{a10}  is special to Majorana fermions and is the one through which the antisymmetric coefficient $\rk_\theta $  
 in \rf{222} enters the four-fermion contributions \rf{36} and \rf{417}. For the two-derivative structures $\tfrac1{16} (\bth \Dsl \theta)^2$ and $-\tfrac1{16}(\bth\g^\m\DF^\n\theta)(\bth\g_\n\DF_\m\theta)$ it gives $- E_\theta \rG_\theta$ each in the \adss case
 (using $\g^\m \g_{\m\n} \g^\n = 2$), 
 while in the \adsc case the same contractions in   the massless sector part   produce the term $-\tfrac32 \rG_\chi^2$ in \rf{417}.

\np

%%%%%%%%%%%%%%%%%%%%%%%%%%%%%%%%%%%%%%%%%%%%%%%%%%%%
\section{1-loop   2-point   functions in  \adsc and \adss     % in \adsc  case 
\la{s44}}
%comments on 2-point function

The 2d prescription of sections \ref{s2}--\ref{s4} can be tested beyond the free energy or vacuum expectation value
 by computing the 1-loop correction to the quadratic part of the 
 effective action for the fluctuation fields. 
 Since there are no cubic vertices, this is obtained by contracting one pair of fields in $S_4$ and subtracting the
  quadratic measure term (cf. \rf{28})
\be \la{4401}
 \Gamma^{(1)}_2 = { T}^{-1} \int d^2 \xi \sqrt g\, \big[ \langle \L_4 \rangle_{1} - \rE\, \cJ \big] \ ,
\ee
where $\langle\ \ \rangle_1$  indicates 
 that just  one Gaussian contraction is done.\foot{In quartic fermionic terms 
  the terms with two derivatives on the external fermions are  integrated by parts using 
  that $\int d^2 \xi \sqrt g\, (\hat\nabla_\a \bth) \g^{\m\n} \Gb P_\chi \hat\nabla_\n \theta = - \ha \int d^2 \xi \sqrt g\, \bth \Gb P_\chi \theta$, etc.}
  
   In the AdS$_5\times S^5$ case all the terms in \rf{4401} 
   are proportional to the free quadratic operators of the fields, i.e.
    they amount to wave-function renormalizations with no mass shifts (see below).

    \subsubsection*{\adsc}
%%%%%%   
   In the \adsc case the same is true for all the fields but one: the coefficient of the mass-type bilinear $\bth \Gb P_\chi\theta$ of the massless fermion $\chi$  is (for  $t\to 0$   and with  $\rE$ in the measure  and    $E_\vp$  set equal   at the end)  %kept  generic)
\be \la{4402}
 {\cal Z}_\chi = \tfrac{1}{12} \rE - \ha E_\psi - \tfrac13 E_\chi + \tfrac34 E_w + \tfrac34 \rG_\chi \  
% = - \tfrac56\, r(t) + \tfrac34 \rG_\chi (t) 
 \ \to \ \tfrac34 \rG_\chi \ . 
\ee
Here   $ \rG_\chi$ is  the constant in \rf{442},\rf{411}.  The five terms in \rf{4402} 
 come, respectively, 
   from the measure, from the massive-fermion contraction,
 from the massless-fermion contraction, from the massless-scalar contraction, and from the finite part of the massless propagator. 
 
 In the case when the measure is defined with the field-attributed $\delta(0)$'s of
Appendix \ref{apA3}, the first term in \rf{4402} is replaced by the sum of three contributions:
$\tfrac34 E_w$ from the velocity Hessian of the mixed term (in which the bilinear $\bth \Gb P_\chi\theta$
multiplies $(\del w)^2$, cf. \rf{4191}), and $-\tfrac12 E_\psi - \tfrac13 E_\chi$ from the $\psi$- and
$\chi$-blocks of the trace in $\cJ_{\rm sc,f}$, i.e.
 $ %\be \la{4405}
 \tfrac{1}{12}\, \rE \ \to \ \tfrac12 E_\psi + \tfrac13 E_\chi - \tfrac34 E_w
$ %\ee
which reduces to $\tfrac{1}{12}\rE$ for $E_\vp = \rE$. The $E_\vp$-terms in \rf{4402} then cancel
identically for any choice of the smearing kernels, and one is left with the same finite term
$\tfrac34 \rG_\chi$.

 %In the case when  the measure is defined with  field-attributed $\delta(0)$ as in 
  %Appendix \ref{apA3} the first term is $\tfrac{1}{12} E_\chi$ instead of $\tfrac{1}{12} \rE$; 
  %the $E_\vp $-terms then cancel   in general and one ends  just with the same   finite term $\tfrac34 \rG_\chi$.
  
  Thus we get a finite 1-loop induced  mass-type  term $\bth \Gb P_\chi\theta$ 
   for $\chi=P_\chi \theta$ (cf. \rf{430}). After dividing out the kinetic normalization $\ha$ 
  in the free  Lagrangian \rf{44}, the  fermion $\chi$ acquires a mass
\be \la{4403}
 \delta m_\chi = 2\,  {\cal Z}_\chi  T^{-1} = \tfrac{3}{ 2}\rG_\chi { T^{-1} } \ .
\ee
The  scalar partner of $\chi$, however,  remains massless (there is no mass shift for the $w^a$   fields  in \rf{44}).
% The same 1-loop coefficient \rf{4402}  follows from the truncated  action \rf{4172}. 
 Since $\chi$ and its scalar partner belong to the same short multiplet of the preserved $SU(1,1|3)$,
  the mass shift  \rf{4403} 
   is  thus not compatible with the residual supersymmetry. 
   This indicates  an incompleteness  of the  approach   based   on the 2d regularization prescription  for the type  IIA GS action  itself. 
   
  % This is a necessary (not sufficient) supersymmetry test, which the 2d prescription, 
 %  taken as the complete definition of the quantum theory, does not pass on the fixed branch $\eta$; it limits the inference that the cancellation of the contact terms alone selects a supersymmetric prescription.

The finite term \rf{4302} induced by integrating out 
 the non-zero  modes  of the M2 brane is precisely what  resolves the problem: adding it to \rf{4402} gives
\be \la{4404}
 {\cal Z}_\chi  +  { q}_\chi  = \tfrac34 \big( \rG_\chi - \tfrac{1}{4 \pi}\big) = 0\ , 
  \qquad  \qquad \rG_\chi = \rG_\theta(0^+) = \te {1\ov 4 \pi} \ .
\ee
Thus the massless fermion stays massless
for the same   definition of $\rG_\chi$  for which the 2-loop correction to M2  brane free energy 
in \rf{333},\rf{426},\rf{428}  vanishes. 

This  two-point test thus confirms, independently of the  vacuum  partition function  (Wilson loop) value, that the finite 
term produced  by integrating out  the KK tower is required for the consistency of the type IIA 
string regularization prescription with the residual  supersymmetry. % and that it is absent in the purely 2d computation.

   \subsubsection*{\adss}
In the \adss case one finds, for the common smearing (or, with the field-attributed measure, for
${\rm Str}\, E = 0$, cf. \rf{6003}), that all the terms in \rf{4401} are proportional to the original
free quadratic operators of the fields, i.e. they amount to (logarithmically divergent)
wave-function renormalizations, with no mass shifts.

 %In the \adss case one finds that all the terms in \rf{4401} are proportional to the original free quadratic operators of the fields, i.e. they amount to finite %renormalizations of the kinetic terms, with no mass shifts. 
Let us make this explicit for the massive scalar, for which the 1-loop correction to
the 2-point function was computed in dimensional regularization  in Appendix C of
\ci{Beccaria:2026ffm}. Contracting  2 fields  in \rf{31},\rf{32}  keeping  two $x$-fields as external
 and subtracting the  corresponding measure  contribution  we get\foot{Here fermion loop contributes $-4 E_\theta x^2$ from the $A^{\m\n}$ term in \rf{32} and
$-2 \rG_\theta [(\del x)^2 + 2 x^2]$ from the mass-type term.} 
\ba\la{c10}
&\qquad \Gamma^{(1)}_2 = \ha T^{-1}  \int d^2\xi \sqrt g\, \big[ \cZ_x (\del x)^2  + \td \cZ_x\, x^2 \big] \ , \ \ \qquad \\
\cZ_x= & -  E_x
 +  \rE + \tfrac72 \rG_x -  4 \rG_\theta \ , \ \ \qquad \qquad 
\td \cZ_x = \tfrac32 E_x - 8 E_\theta + \tfrac{13}{2} \rE +  7 \rG_x
 - 8 \rG_\theta  \ , \la{c100}
\ea
Assuming the common smearing $E_\vp=\rE$ 
the   contact $\delta(0)$ terms cancel  and we get  (cf. \rf{27})
\be \la{c11}
 \Gamma^{(1)}_{x^2} = T^{-1}   \cZ_x 
 \int d^2\xi \sqrt g\, \L_{2x} \ , \qquad  \L_{2x} = \ha (\del x)^2 + x^2 \ , 
      \qquad
 \cZ_x =  \tfrac72 \rG_x - 4 \rG_\theta = -\tfrac{1}{8\pi} (  \LDR + 6)
\ ,
\ee
i.e. a wave-function renormalization of $x$   with  no mass shift, in accordance with the
protected dimension $\Delta=2$ of the displacement multiplet. 
This is an indication that the common smearing prescription is  
 consistent with the underlying supersymmetry.
The same result is obtained with the field-attributed measure \rf{a51} for any $E_\vp$, up to
the Lorentz-violating term \rf{6003}, which is absent only for ${\rm Str}\, E=0$.

In the notation of
\ci{Beccaria:2026ffm}, where the correction is reduced to the coefficient $P$ of
$\int G(\omega_1,\omega) G(\omega,\omega_2)$ after integrating by parts, this means setting 
$P=0$, which is the ``hard''  dimensional reduction $d \to 1$ limit of (C.4) there, $P \sim 1-d$. 
The finite
mass counterterm $C_1 = - P\big|_{d\to 1}  = {1\ov 2\pi}$ found in the ``soft''  dimensional regularization 
version is
produced by the evanescent factor $1-d$ multiplying the poles of $\rG$ and $\rm S$, i.e. it
is the 2-point analog of the single pole in $f_2$ discussed in section \ref{s3}, and it is
absent in a 2d regularization with the measure contribution  included.

 The comparison with the
subleading term of the Bremsstrahlung function, $\rB_1/\rB_0 = - {3 \ov 4\pi }T^{-1}$, noted as a
puzzle in \ci{Beccaria:2026ffm}, should therefore not be made with a mass shift but with
the normalization of the boundary limit of the $x$ propagator, $\langle x x \rangle_1 =
\big(1 + {\LDR + 6 \ov 8 \pi T}\big) \langle x x\rangle_0$.  Since the finite part of a
wave-function renormalization is scheme dependent, this requires a renormalization of the
boundary operator which  we  will not consider here.
%\foot{Heat-kernel regularization is likely to be useful also 
%in the context discussed in  \ci{Castellani:2026wwh}.}

The same computation for the massless scalars $y^a$, with two external $y$'s and one contraction in
\rf{31},\rf{32}, gives\foot{Here the $(\del z)^4$ terms in \rf{31} give $-\ha E_y (\del y)^2$ (the terms involving the total contact
sum cancel as in the $x$ case), the term $-\tfrac14 y^2 (\del y)^2$ gives $-\tfrac54 E_y y^2 - \tfrac54 \rG_y (\del y)^2$
(using $\langle \del y \cdot \del y\rangle = 5 E_y$, $\langle y^2 \rangle = 5 \rG_y$), the fermion loop of the
$A^{\m\n}$ term in \rf{32} vanishes (the $y$-dependent part of $A^\m{}_\m$ is zero), the mass-type term
$-\tfrac18 (\del y)^2 \bth \Gb\theta$ gives $2 \rG_\theta (\del y)^2$, and the measure contributes
$-\rE \cJ_{\rm b}|_y = \ha \rE (\del y)^2 + \tfrac54 \rE y^2$.} 
\be 
 \Gamma^{(1)}_2 =\ha  T^{-1}  \int d^2\xi \sqrt g\, \big[ \cZ_y (\del y)^2  + \td \cZ_y\, y^2 \big] \ , \qquad 
 \cZ_y = -  E_y + \rE - \tfrac52 \rG_y + 4 \rG_\theta \ , \quad 
 \td \cZ_y = - \tfrac52 E_y + \tfrac52 \rE \ . \la{c13}
\ee
For the common smearing (and  for any
smearing with the field-attributed $\delta(0)$'s in the measure  of Appendix \ref{apA3}
satisfying ${\rm Str}\, E=0$; for ${\rm Str}\, E\neq 0$ there is in addition the
non-covariant term \rf{6003}) we get
\be \la{c14}
 \Gamma^{(1)}_{2} = T^{-1}\,  \cZ_y \int d^2\xi \sqrt g\, \L_{2y} \ , \qquad \L_{2y} = \ha (\del y)^2 \ ,
 \qquad  \cZ_y = - \tfrac52 \rG_y + 4 \rG_\theta = \tfrac{1}{4\pi} \big( \tfrac32 \LDR - 4 \big) \ .
\ee
Thus no mass term is generated for the Goldstone fields $y^a$, as required by the non-linearly
realized broken $SO(6)/SO(5)$ symmetry (the unbroken $SO(5)$ alone would allow the invariant
$y^a y^a$), and there is a logarithmically divergent wave-function renormalization.

%Thus  no mass term is generated for the Goldstone fields $y^a$ %(as required by the unbroken part of the
%$SO(6)$ symmetry)
% and there is a logarithmically divergent wave-function renormalization.
%\be \la{c15}
% \langle y^a y^b \rangle_1 = \Big( 1 - { \cZ_yT^{-1} } \Big) \langle y^a y^b \rangle 
% = \Big( 1 - {3 \LDR - 8 \ov 8 \pi T}\Big) \langle y^a y^b \rangle_0 \ .
%\ee
The  log divergence $\sim \ell$ in \rf{c14} is not a problem: $ \cZ_y$ is not a physical quantity.
Under a local
redefinition $y^a \to y^a +  \alpha\, T^{-1} y^a y^2$ it changes, at this order, by
$-2 \alpha (N_y + 2) \rG_y = - 14 \alpha \rG_y$ (the accompanying Jacobian contributes only $E$-terms,
which cancel against the change of the contact terms as discussed in section \ref{s2}).

 A direct check
of this is provided by comparing with the computation of \ci{Castellani:2026wwh}, in which the same 1-loop
correction was computed in dimensional regularization with the $S^5$ fluctuations parametrized by
 normal coordinates ($\hat y$) rather than by the stereographic-type coordinates of
\ci{Beccaria:2026ffm} used here.
The $y$-loop contribution found there is $\tfrac13 (N_y - 1)\rG_y = \tfrac43 \rG_y$,
while the fermion-loop contribution $-4 \rG_\theta$ is the same (the mixed term $-\tfrac18 (\del y)^2 \bth\Gb\theta$ is
parametrization independent at this order). 
The two parametrizations are related by\foot{For the unit sphere the stereographic
radius is $\rho = 2 \tan \tfrac{r}{2} = r + \tfrac{1}{12} r^3 + \dots$ in terms of the geodesic one.}
$y = \hat y\, \big(1 + \tfrac{1}{12} T^{-1}  \hat y^2\big) + \dots$, i.e.
$\alpha = \tfrac{1}{12}$ in the above formula, which shifts $ \cZ_y$ by $ - \tfrac76 \rG_y$,
i.e. the coefficient of $\rG_y$ in the 1-loop renormalization  factor  $  1 - { \cZ_yT^{-1} }$    from $\tfrac43$ to $ \tfrac52$,
in agreement with \rf{c14}. This confirms
 that  (i) 
 the $\delta(0)$-independent part of the 1-loop correction to $\langle y y \rangle$ is the same in
the two regularizations, and  (ii) its finite part has no invariant meaning.

%============================================================================================
The physical quantity is instead the normalization of the 2-point function of the boundary tilt
operator, $\langle \rt^a(\tau) \rt^b(\tau') \rangle = C_\rt \delta^{ab} |\tau - \tau'|^{-2}$  ($\tau$  is coordinate along the  straight WL)
which is related to the
Bremsstrahlung function by $C_\rt = 2 \rB(\l)$ \ci{Correa:2012at} and is thus known exactly (cf. \rf{1})
\be \la{c16}
 \rB(\l) =\te  {1\ov 2\pi^2}\, \l \del_\l \log \langle W \rangle = {\sql \ov 4 \pi^2} - {3 \ov 8 \pi^2} + {3 \ov 32 \pi^2 \sql} + \dots
 = {1\ov 2\pi} T- {3 \ov 8 \pi^2} + {3 \ov 64 \pi^3 } T^{-1} + \dots \ .
\ee
As emphasized in \ci{Castellani:2026wwh}, $C_\rt$ should be defined not through
the boundary limit of $y^a$ (whose normalization is the parametrization-dependent and divergent
$ \cZ_y$) but through the boundary limit of the Noether current $J^a_\m$ of the broken $SO(6)$, whose
normalization is fixed by the Ward identity. 
Here $a = 1,\dots,5$ labels the five generators $M_{a6}$ of $SO(6)$ broken to $SO(5)$.\foot{On the gauge
theory side this breaking is due to
 the scalar coupling $\Phi^6$ in the Wilson-Maldacena loop $W = {\rm tr}\, P\, e^{\int d\tau\,(i A_\tau + \Phi^6)}$,
and on the string side -- due to  the choice of the point on $S^5$ where \adst surface is localized. }
The tilt
operators $\rt^a(\tau)$ are the corresponding protected ($\Delta = 1$) boundary operators (cf. \ci{Giombi:2017cqn})
 normalized by the
broken Ward identity $\del^\m J^{a6}_\m = \delta^3(x_\perp)\, \rt^a$ of the $SO(6)$ R-symmetry current. 

On the
worldsheet, the $SO(6)$ symmetry acts non-linearly on the five  Goldstone massless $S^5$ fluctuations $y^a$
and $J^a_\m$ denotes the corresponding Noether current of
the gauge-fixed string action \rf{27},\rf{31}--\rf{33} (see \ci{Castellani:2026wwh}). 
In the normalization of \ci{Castellani:2026wwh}, with the action 
$T \int d^2\xi \sqrt g\, (\L_2 + \L_4 + \dots)$ in terms of unrescaled fields, it reads
$
 J^a_\m = - T\, \del_\m y^a + \OO(y^3,\, y\,\theta^2) , $     with  %\qquad \qquad
$ \langle J^a_\m(\xi)\, J^b_\n(\xi') \rangle_0 = T\, \del_\m \del'_\n G_y(\xi,\xi')\, \delta^{ab}$. 
The tilt operator is defined by its $z\to 0$  boundary limit, $J^a_z(z,\tau) \to \rt^a(\tau)$
(with $\xi = (\tau, z)$  being the Poincar\'e coordinates of \adst). 
%In our conventions, with the fields
%rescaled by $T^{-1/2}$, the leading term is $-\sqrt T\, \del_\m y^p$ and the 1-loop corrections to
%\rf{c20} are of relative order $T^{-1}$, as below  in \rf{c17}.

In \ci{Castellani:2026wwh} the 1-loop correction to
$\langle J^a_\m J^b_\n\rangle$ was computed in dimensional regularization  with the result (in our notation their $S_0 = - \rG_\theta$)
\be \la{c17}
 \langle J^a_\m J^b_\n\rangle_1 = \langle J^a_\m J^b_\n \rangle_0 \te \Big( 1 -  T^{-1}  \big[ (N_y - 1)\, \rG_y - {1-d \ov 2(1+d)}\, N_x m_x^2 \rG_x - \tfrac14 N_\theta\, \rG_\theta \big] \Big) \ .
\ee
Here \rf{c17} is the correlator at separated points; the contact term proportional to
$g_{\m\n}\delta^{(2)}(\xi,\xi')$ required by the Ward identity (eq. (3.54) of \ci{Castellani:2026wwh})
is a different object from the coincident-point contacts $E_\vp$ discussed in this paper and does not
affect the boundary coefficient.
The poles cancel between the $y$ and $\theta$ loops for
$N_y=5$, $N_\theta=16$, and $C_t$ is obtained from the finite part of the bracket. It was found there that
the  direct  application of dimensional regularization  does not reproduce \rf{c16},  but  the agreement is obtained if the
number of massless scalars is continued as $N_y = 5 + 2 \eps$ (so that $d + 1 + N_x + N_y = 10$ is kept
fixed). 

In the  context  of the present paper  this result  may be interpreted as follows.  The three
finite contributions to the bracket in \rf{c17} are
$
 y: \  {2 \eps \ov 4\pi \eps} = {1\ov 2\pi} $, \ 
 $x: \   - {\eps \ov 2} \cdot {6 \ov 4\pi \eps} = - {3 \ov 4\pi}$, \ 
 $\theta: \  4\, (\rG_y - \rG_\theta)_{\rm fin} = {1\ov \pi} $, 
with the total $\tfrac{3}{4\pi}$, giving $\rB_1 = - \tfrac{3}{8\pi^2}$. 
The first two are ``evanescent'' terms of
 the type discussed in section \ref{s3}: the $x$-loop contribution vanishes identically in 2
dimensions, for any $E_x$ and $\rG_x$ (it comes from the contraction of a traceless tensor with
$\langle \del_\m x \del_\n x \rangle \sim  g_{\m\n}$), and the $y$-loop one is the continuation of the number of
fields times the pole of $\rG_y$. 
Only the third contribution  is a genuine finite part: the difference
$\rG_y - \rG_\theta = \tfrac{1}{4\pi}$ between the coincident massless-scalar and massive-fermion propagators,
which is regularization independent (cf.\ \rf{2170}). 

In a 2d cutoff  regularization with the measure contribution  included,
where the $\delta(0)$ terms cancel as in section \ref{s3} and no evanescent terms exist, one therefore
expects the bracket in \rf{c17}  to be just $4(\rG_y - \rG_\theta) = \tfrac{1}{\pi}$, i.e.\foot{This is assuming that the measure does not contribute a finite term to the current correlator beyond the
cancellation of the contact terms. We have not carried out this computation explicitly: the Noether
current in the regularized theory should include the contribution of the measure term
$\delta(0) \int \cJ$ in the effective action. }
\be \la{c19}
 C_t^{\rm (2d)} 
 =2\rB^{\rm (2d)} = \te  {1 \ov \pi} T - {1\ov \pi^2} \ ,
\ee
The  2d regularization value   \rf{c19} differs from the  localization one  
\rf{c16} by $\Delta \rB_1 = - \tfrac{1}{2\pi^2} + \tfrac{3}{8\pi^2} = - \tfrac{1}{8 \pi^2}$.

This difference has a suggestive interpretation. According to \rf{c16}, the coefficient $\rB_1$ is determined
by the $\l$-dependence of the 1-loop prefactor of the Wilson loop, $\l \del_\l \log  {N \ov \l^{3/4}}  =
- \tfrac34$, which in the string theory  language is the product of the disk normalization $ g_s^{-1} \sim {N\ov \l}$
 and of the measure factor $\sqrt T \sim \l^{1/4}$ in $Z_1$ in \rf{2}.
The 2d value \rf{c19}, $\rB_1^{\rm (2d)} = - \tfrac{1}{2\pi^2} $
 is precisely the value
that would follow from \rf{c16} for a Wilson loop {\it without} the $\sqrt T$ factor, i.e. from the
direct 1-loop string partition function  \ci{Drukker:2000ep} in which this factor is absent and has
to be supplied by the measure normalization \ci{Giombi:2020mhz} (or, in the IIA case, by the parent M2
brane path integral \ci{Giombi:2023vzu}). 

%In this sense the direct  computation of the current-current  correlator and
%the direct computation of the 1-loop partition function are mutually consistent, and what is missing
%in both is the same measure factor $\sqrt T$ as  discussed in section \ref{s5}.

In this sense the direct computation of the current-current correlator (under the assumption
 that the measure contributes to it only through the cancellation of the contact terms)
and the direct computation of the 1-loop partition function would be mutually consistent, the
tentative value \rf{c19} differing from the exact one by precisely the amount generated by the
$\sqrt T$ factor in the coupling-derivative relation \rf{c16}. Whether the two mismatches have a
common origin requires the direct 2d computation of $\langle J J \rangle$ with the measure and the
properly normalized current included.
The continuation
$N_y = 5 + 2\eps$ of \ci{Castellani:2026wwh}, together with the evanescent $x$-loop term, shifts the  value in the bracket in \rf{c17} 
by $-\tfrac{1}{4\pi}$ and thus $\rB_1$ by $+\tfrac{1}{8\pi^2} $: it effectively
reinstates the contribution of the $\sqrt T$ factor. 

It would be interesting to verify this interpretation
by computing $\langle J J\rangle_1$ at one loop directly in the 2d heat-kernel regularization with the measure contribution included,
and to see whether the extra measure-type part required to reproduce $\rB_1$ is the same as the
one required to reproduce the 2-loop   coefficient  $f_2 = \tfrac{3}{32\pi^2}$.\foot{Note that the latter is the coefficient
of the next term, $\tfrac{3}{64 \pi^3} T^{-1} = {1\ov 2\pi} f_2$, in \rf{c16}: the two coefficients test different terms in
$\log \langle W \rangle$, the 1-loop prefactor and the 2-loop correction respectively.}

\np

 \small
\baselineskip 11pt
\bibliography{2loop}
\small
\bibliographystyle{JHEP-v2.9}
\end{document}